\documentclass[11pt,a4paper]{article}
\usepackage{jheppub}
\usepackage{appendix}
\usepackage{ifthen} 
\usepackage{overpic} %
\usepackage[utf8]{inputenc}
\usepackage[normalem]{ulem}
\newboolean{pdflatex}
\setboolean{pdflatex}{true} 

\newboolean{articletitles}
\setboolean{articletitles}{true} 

\newboolean{uprightparticles}
\setboolean{uprightparticles}{false} 
\usepackage{bm}
\usepackage{lineno}

\usepackage{placeins}
\usepackage{subcaption}

\usepackage{tabto} 
\usepackage{siunitx}
\usepackage{mathtools}
\usepackage{soul}

\usepackage[utf8]{inputenc}
\usepackage[T1]{fontenc}

\numberwithin{equation}{section}
\makeatletter
\let\oldtheequation\theequation
\renewcommand\tagform@[1]{\maketag@@@{\ignorespaces#1\unskip\@@italiccorr}}
\renewcommand\theequation{(\oldtheequation)}
\makeatother

\usepackage{xspace} 
\usepackage{upgreek}

\def\lhcb   {\mbox{LHCb}\xspace}

\def\MagUp {\mbox{\em Mag\kern -0.05em Up}\xspace}

\ifthenelse{\boolean{uprightparticles}}%
{

 \def\PDelta      {\ensuremath{\Delta}\xspace}                 
 \def\PXi         {\ensuremath{\Xi}\xspace}                 
 \def\PLambda     {\ensuremath{\Lambda}\xspace}                 
 \def\PSigma      {\ensuremath{\Sigma}\xspace}                 
 \def\POmega      {\ensuremath{\Omega}\xspace}                 
 \def\PUpsilon    {\ensuremath{\Upsilon}\xspace}
 \let\oldPi\Pi
 \def\PPi         {\ensuremath{\oldPi}\xspace}

 \def\PB      {\ensuremath{\mathrm{B}}\xspace}                 
 \def\PD      {\ensuremath{\mathrm{D}}\xspace}                 

 \def\PK      {\ensuremath{\mathrm{K}}\xspace}                 
 \def\thebaroffset{0.0em}
}
{

 \mathchardef\PDelta="7101
 \mathchardef\PXi="7104
 \mathchardef\PLambda="7103
 \mathchardef\PSigma="7106
 \mathchardef\POmega="710A
 \mathchardef\PUpsilon="7107
 \mathchardef\PPi="7105
 \def\PB      {\ensuremath{B}\xspace}                 
 \def\PD      {\ensuremath{D}\xspace}                 

 \def\PK      {\ensuremath{K}\xspace}                 
 \def\thebaroffset{0.18em}
}
\makeatletter
\ifcase \@ptsize \relax
  \newcommand{\miniscule}{\@setfontsize\miniscule{4}{5}}
\or
  \newcommand{\miniscule}{\@setfontsize\miniscule{5}{6}}
\or
  \newcommand{\miniscule}{\@setfontsize\miniscule{5}{6}}
\fi
\makeatother

\DeclareRobustCommand{\optbar}[1]{\shortstack{{\miniscule (\rule[.5ex]{1.25em}{.18mm})}
  \\ [-.7ex] $#1$}}

\def\Kbar    {{\ensuremath{\offsetoverline{\PK}}}\xspace}

\def\KorKbar {\kern \thebaroffset\optbar{\kern -\thebaroffset \PK}{}\xspace}

\def\D       {{\ensuremath{\PD}}\xspace}

\def\DorDbar {\kern \thebaroffset\optbar{\kern -\thebaroffset \PD}\xspace}

\def\Dp      {{\ensuremath{\D^+}}\xspace}
\def\Dm      {{\ensuremath{\D^-}}\xspace}

\def\DpDm    {\ensuremath{\Dp {\kern -0.16em \Dm}}\xspace}

\def\B       {{\ensuremath{\PB}}\xspace}
\def\Bbar    {{\ensuremath{\offsetoverline{\PB}}}\xspace}

\def\BorBbar {\kern \thebaroffset\optbar{\kern -\thebaroffset \PB}\xspace}

\def\Bd      {{\ensuremath{\B^0}}\xspace}

\def\BdorBdbar {\kern \thebaroffset\optbar{\kern -\thebaroffset \Bd}\xspace}
\def\Bu      {{\ensuremath{\B^+}}\xspace}

\def\Bp      {{\ensuremath{\Bu}}\xspace}

\def\Bs      {{\ensuremath{\B^0_\squark}}\xspace}

\def\BsorBsbar {\kern \thebaroffset\optbar{\kern -\thebaroffset \Bs}\xspace}

\def\Y#1S{\ensuremath{\PUpsilon{(#1S)}}\xspace}

\def\LorLbar     {\kern \thebaroffset\optbar{\kern -\thebaroffset \PLambda}\xspace}

\def\to                 {\ensuremath{\rightarrow}\xspace}

\def\AT#1     {\ensuremath{A_{\mathrm{T}}^{#1}}\xspace}           

\def\C#1      {\ensuremath{\mathcal{C}_{#1}}\xspace}                       
\def\Cp#1     {\ensuremath{\mathcal{C}_{#1}^{'}}\xspace}                    
\def\Ceff#1   {\ensuremath{\mathcal{C}_{#1}^{\mathrm{(eff)}}}\xspace}        
\def\Cpeff#1  {\ensuremath{\mathcal{C}_{#1}^{'\mathrm{(eff)}}}\xspace}       
\def\Ope#1    {\ensuremath{\mathcal{O}_{#1}}\xspace}                       
\def\Opep#1   {\ensuremath{\mathcal{O}_{#1}^{'}}\xspace}                    

\newcommand{\aunit}[1]{\ensuremath{\text{\,#1}}}       

\newcommand{\tev}{\aunit{Te\kern -0.1em V}\xspace}
\newcommand{\gev}{\aunit{Ge\kern -0.1em V}\xspace}
\newcommand{\mev}{\aunit{Me\kern -0.1em V}\xspace}
\newcommand{\kev}{\aunit{ke\kern -0.1em V}\xspace}
\newcommand{\ev}{\aunit{e\kern -0.1em V}\xspace}
 
\newcommand{\mevc}{\ensuremath{\aunit{Me\kern -0.1em V\!/}c}\xspace}
\newcommand{\gevc}{\ensuremath{\aunit{Ge\kern -0.1em V\!/}c}\xspace}
\newcommand{\mevcc}{\ensuremath{\aunit{Me\kern -0.1em V\!/}c^2}\xspace}
\newcommand{\gevcc}{\ensuremath{\aunit{Ge\kern -0.1em V\!/}c^2}\xspace}

\def\gsim{{~\raise.15em\hbox{$>$}\kern-.85em
          \lower.35em\hbox{$\sim$}~}\xspace}
\def\lsim{{~\raise.15em\hbox{$<$}\kern-.85em
          \lower.35em\hbox{$\sim$}~}\xspace}

\def\evtgen     {\mbox{\textsc{EvtGen}}\xspace}

\def\geant      {\mbox{\textsc{Geant4}}\xspace}

\def\photos     {\mbox{\textsc{Photos}}\xspace}

\def\pythia     {\mbox{\textsc{Pythia}}\xspace}

\def\tell1  {TELL1\xspace}
\def\ukl1   {UKL1\xspace}

\newcommand{\lhcborcid}[1]{\href{https://orcid.org/#1}{\hspace*{0.1em}\raisebox{-0.45ex}{\includegraphics[width=1em]{figs/orcidIcon.pdf}}}}

\title{Towards replacing detector simulation with heterogeneous GNNs in flavour physics analyses}

\author[3]{G. Hijano}
\author[2]{D. Lancierini}
\author[1]{A. Marshall}
\author[2]{A. Mauri}
\author[3]{P. Owen}
\author[2]{M. Patel}
\author[1]{K. Petridis}
\author[3]{S. R. Qasim}
\author[3]{N. Serra}
\author[3]{W. Sutcliffe}
\author[2]{H. Tilquin}

\affiliation[1]{University of Bristol, Bristol, UK}
\affiliation[2]{Imperial College London, London, UK}
\affiliation[3]{University of Zürich, Zürich, Switzerland}

\emailAdd{alex.marshall@cern.ch}

\abstract{
\noindent 
Driven by the increasing volume of recorded data, the demand for simulation from experiments based at the Large Hadron Collider will rise sharply in the coming years. Addressing this demand solely with existing computationally intensive workflows is not feasible. This paper introduces a new fast simulation tool designed to address this demand at the LHCb experiment. This tool emulates the detector response to arbitrary multibody decay topologies at LHCb. Rather than memorising specific decay channels, the model learns generalisable patterns within the response, allowing it to interpolate to channels not present in the training data. Novel heterogeneous graph neural network architectures are employed that are designed to embed the physical characteristics of the task directly into the network structure. We demonstrate the performance of the tool across a range of decay topologies, showing that the networks can correctly model the relationships between complex variables. The architectures and methods presented are generic and could readily be adapted to emulate workflows at other simulation-intensive particle physics experiments.

\vfill
\begin{center}
\copyright 2025 CERN for the benefit of the LHCb collaboration. CC BY 4.0 licence.
\end{center}

}

\usepackage{microtype}
\usepackage{lineno}  
\usepackage{xspace} 

\usepackage{graphicx}  
\usepackage{color}
\usepackage{colortbl}
\graphicspath{{./figs/}} 
\DeclareGraphicsExtensions{.pdf,.PDF,png,.PNG}

\usepackage{amsmath} 
\usepackage{amssymb}
\usepackage{amsfonts}
\usepackage{upgreek} 
\usepackage{makecell}
\newcommand*\patchAmsMathEnvironmentForLineno[1]{%
\expandafter\let\csname old#1\expandafter\endcsname\csname #1\endcsname
\expandafter\let\csname oldend#1\expandafter\endcsname\csname
end#1\endcsname
 \renewenvironment{#1}%
   {\linenomath\csname old#1\endcsname}%
   {\csname oldend#1\endcsname\endlinenomath}%
}
\newcommand*\patchBothAmsMathEnvironmentsForLineno[1]{%
  \patchAmsMathEnvironmentForLineno{#1}%
  \patchAmsMathEnvironmentForLineno{#1*}%
}
\AtBeginDocument{%
\patchBothAmsMathEnvironmentsForLineno{equation}%
\patchBothAmsMathEnvironmentsForLineno{align}%
\patchBothAmsMathEnvironmentsForLineno{flalign}%
\patchBothAmsMathEnvironmentsForLineno{alignat}%
\patchBothAmsMathEnvironmentsForLineno{gather}%
\patchBothAmsMathEnvironmentsForLineno{multline}%
\patchBothAmsMathEnvironmentsForLineno{eqnarray}%
}

\usepackage{hyperxmp}

\usepackage[colorinlistoftodos,textsize=scriptsize]{todonotes}

\usepackage[all]{hypcap} 

\usepackage{xspace} 
\usepackage{upgreek}
\newcommand{\offsetoverline}[2][0.1em]{\kern #1\overline{\kern -#1 #2}}%

\ifthenelse{\boolean{uprightparticles}}%
{

 \def\PDelta      {\ensuremath{\Delta}\xspace}                 
 \def\PXi         {\ensuremath{\Xi}\xspace}                 
 \def\PLambda     {\ensuremath{\Lambda}\xspace}                 
 \def\PSigma      {\ensuremath{\Sigma}\xspace}                 
 \def\POmega      {\ensuremath{\Omega}\xspace}                 
 \def\PUpsilon    {\ensuremath{\Upsilon}\xspace}

 \def\PB      {\ensuremath{\mathrm{B}}\xspace}                 
                  
 \def\PD      {\ensuremath{\mathrm{D}}\xspace}

 \def\PK      {\ensuremath{\mathrm{K}}\xspace}

 \def\Pi      {\ensuremath{\mathrm{i}}\xspace}

}
{

 \mathchardef\PDelta="7101
 \mathchardef\PXi="7104
 \mathchardef\PLambda="7103
 \mathchardef\PSigma="7106
 \mathchardef\POmega="710A
 \mathchardef\PUpsilon="7107
                  
 \def\PB      {\ensuremath{B}\xspace}                 
                  
 \def\PD      {\ensuremath{D}\xspace}

 \def\PK      {\ensuremath{K}\xspace}

 \def\Pi      {\ensuremath{i}\xspace}

}

\makeatletter
\ifcase \@ptsize \relax
  \newcommand{\miniscule}{\@setfontsize\miniscule{4}{5}}
\fi
\makeatother

\DeclareRobustCommand{\optbar}[1]{\shortstack{{\miniscule (\rule[.5ex]{1.25em}{.18mm})}
  \\ [-.7ex] $#1$}}

  \def\Kbar    {{\kern 0.2em\overline{\kern -0.2em \PK}{}}\xspace}

\def\KorKbar {\kern 0.18em\optbar{\kern -0.18em K}{}\xspace}

\def\B       {{\ensuremath{\PB}}\xspace}
\def\Bbar    {{\ensuremath{\kern 0.18em\overline{\kern -0.18em \PB}{}}}\xspace}
\def\BorBbar    {\kern 0.18em\optbar{\kern -0.18em B}{}\xspace}

\def\Bu      {{\ensuremath{\B^+}}\xspace}

\def\Bp      {{\ensuremath{\Bu}}\xspace}

\def\Bd      {{\ensuremath{\B^0}}\xspace}

\def\Bs      {{\ensuremath{\B_{s}^0}}\xspace}

\def\to                 {\ensuremath{\rightarrow}\xspace}

\def\C#1      {\ensuremath{\mathcal{C}_{#1}}\xspace}                       
\def\Cp#1     {\ensuremath{\mathcal{C}_{#1}^{'}}\xspace}                    
\def\Ceff#1   {\ensuremath{\mathcal{C}_{#1}^{\mathrm{(eff)}}}\xspace}        
\def\Cpeff#1  {\ensuremath{\mathcal{C}_{#1}^{'\mathrm{(eff)}}}\xspace}       
\def\Ope#1    {\ensuremath{\mathcal{O}_{#1}}\xspace}                       
\def\Opep#1   {\ensuremath{\mathcal{O}_{#1}^{'}}\xspace}                    

\renewcommand{\gev}{\, \mathrm{GeV}}
\renewcommand{\tev}{\, \mathrm{TeV}}

\renewcommand{\C}[1]{{\cal C}_{#1}}
\renewcommand{\Cp}[1]{{\cal C}_{#1'}}

\usepackage{mciteplus}
\usepackage{multirow}
\usepackage{booktabs}

\newcommand{\dira}{\ensuremath{\rm{DIRA}}\xspace}
\newcommand{\ipChiSq}{\ensuremath{\chi^2_{\rm{IP}}}\xspace}

\newcommand{\vtxChiSqNdf}{\ensuremath{\chi^2_{\rm{DV}}/\rm{ndof}}\xspace}

\newcommand{\ghostProb}{\ensuremath{ \rm{prob}_{\rm{ghost}}}\xspace}
\newcommand{\chiSqTrack}{\ensuremath{\chi^2_{\rm{TrackFit}}}\xspace}

\begin{document}

\newcolumntype{L}[1]{>{\raggedright\let\newline\\\arraybackslash\hspace{0pt}}m{#1}}
\newcolumntype{C}[1]{>{\centering\let\newline\\\arraybackslash\hspace{0pt}}m{#1}}
\newcolumntype{R}[1]{>{\raggedleft\let\newline\\\arraybackslash\hspace{0pt}}m{#1}}

\renewcommand{\thefootnote}{\fnsymbol{footnote}}
\setcounter{footnote}{1}

\renewcommand{\thefootnote}{\arabic{footnote}}
\setcounter{footnote}{0}
\maketitle
\flushbottom

\cleardoublepage

\pagestyle{plain}
\setcounter{page}{1}
\pagenumbering{arabic}

\def\rex {\mbox{\textsc{Rex}}\xspace}

\section{Introduction}\label{sec:introduction}
    
    Experiments at the Large Hadron Collider~\cite{Evans:2008zzb} (LHC) are now collecting data at an unprecedented rate, and with the planned upgrades, this rate will only increase~\cite{CERN-LHCC-2021-012}. To maximise the scientific output from the data taken, one requires a detailed understanding of detector efficiencies for a wide range of decay topologies. This is achieved by generating vast quantities of simulated data, which currently requires repeatedly running the same computationally-expensive simulation software for millions of CPU hours annually. This forms a large fraction of each experiment's computing budget~\cite{Bozzi:2024hev,collaboration2022atlas,Software:2815292}. Despite the significant allocation of resources, the limited size of available simulation samples remains a major source of uncertainty in high-profile published analyses~\cite{LHCb:2024jll, LHCb:2023zxo, LHCb:2023uiv, LHCb:2021zwz}. 
    As the volume of recorded data grows, the amount of simulated data needed to maintain a comparable level of understanding of both signal and background processes increases proportionally. The development of fast simulation tools is therefore essential to ensure the precision of future analyses is not constrained by this limit~\cite{HEPSoftwareFoundation:2017ggl}.

    The tools being developed across the experiments at the LHC to simulate detector responses broadly fall into three categories. The first category involves tools which make simplifications to existing full simulation frameworks; these often include parameterising or entirely avoiding expensive elements of the simulation. Examples include ReDecay~\cite{Muller:2018vny} and FastSim~\cite{abdullin2011fast}. These strategies have been in use for years and are well-established. However, their continued reliance on full simulation frameworks limits the overall speed-up possible, which generally proves modest relative to future requirements. The second category encompasses fully parametric tools such as DELPHES~\cite{deFavereau:2013fsa} and RapidSim~\cite{Cowan:2016tnm} that offer extremely fast performance by approximating the detector response using simplified parametrisations. Although highly efficient, these tools are limited in accuracy and lack the level of detail required for precision studies. Finally, the third category covers a growing class of tools that leverage machine learning techniques to attempt to reproduce the accuracy of the full simulation at the speeds of the fully parametric tools. These tools either model parts of the detector response to overcome specific bottlenecks or, in some cases, even attempt to model the entirety of the detector response. The full detector response is often modelled by generating the low-level energy deposits or hits, and then passing this information to the full reconstruction algorithms to produce analysis-level variables. A diverse range of such efforts, in various stages of development, exists across the LHC experiments; some examples are available in Refs.~\cite{Maevskiy:2019vwj, Rogachev:2022hjg, Barbetti:2023bvi, CMS:2024jdl, Vaselli:2024hml, ATLAS:2021pzo, atlas2024deep}. Of particular interest is the approach presented by the CMS collaboration in Ref.~\cite{Vaselli:2024hml}, which offers the high-fidelity generation of analysis-level variables in a single step. This so-called end-to-end approach has the most potential for speed, albeit with some trade-off in flexibility and interpretability. There is significant demand for such a tool to be developed for the requirements of the LHCb experiment. A well-designed tool would prove invaluable in reducing the load on existing simulation frameworks and reducing the existing dependence on parametric tools\footnote{Note that despite its approximate nature, RapidSim is still widely used within LHCb analyses in cases where selection effects can be estimated or are less important, see Refs.~\cite{LHCb:2025shu, LHCb:2024ett, LHCb:2024rkp, LHCb:2024wve, LHCb:2024hfo, LHCb:2024oco}.}. 
    Additionally, it would enable novel studies that are currently impossible due to the computational demands of full simulations, such as exhaustive grid searches for relevant background processes. 

\begin{figure}[t]
        \begin{center}
        \centering
        \includegraphics[width=0.99\textwidth,keepaspectratio]{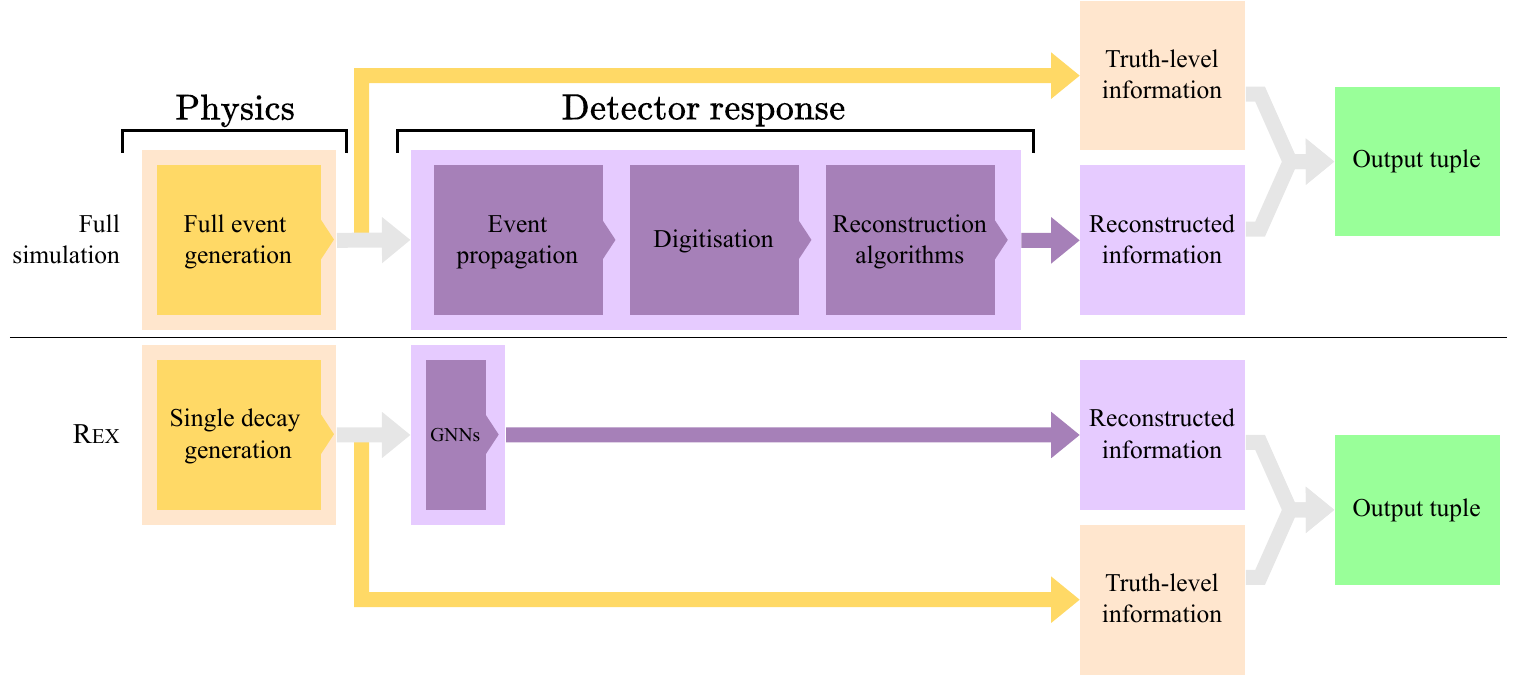}
        \vspace*{-0.5cm}
        \end{center}
        \caption{
        \small Overview of the stages of the full simulation and those of the approach of \rex.}
        \label{fig:intro:flow_compare}
    \end{figure}

    This paper introduces \rex, a Graph Neural Network (GNN) based tool that generates realistic sets of analysis-level variables on an event-by-event basis. These variables include reconstructed kinematic quantities, particle identification (PID) variables, and variables associated with track reconstruction and vertex quality. This generation is conditioned on the kinematics of each decay, which are directly available from existing event generators. These networks learn stochastic mappings with which they can generate a realistic detector output from these conditions, along with vectors of randomly generated noise. At no point does the model replicate the internal steps of the LHCb reconstruction algorithms; instead, it learns to directly produce the required high-level output. Therefore, in deployment, this tool is isolated from the LHCb simulation and reconstruction software. The contrast between this approach and the full simulation is illustrated in Fig.~\ref{fig:intro:flow_compare}.

    Trained in a manner agnostic to any specific topologies or final states, \rex is designed to be as generalised as possible and is capable of modelling a broad range of decay modes commonly encountered in analyses of heavy flavour decays. This includes fully reconstructed modes, partially reconstructed decays with missing particles, modes involving misidentified tracks, and those affected by significant radiative losses. Whilst the model is trained on a sample of heavy flavour decays, \rex is able to produce representative samples for any multi-body decay\footnote{For this initial implementation, reconstructed topologies are limited to having four, or fewer, final state tracks.} at LHCb. 

    The approach taken is founded on the premise that the detector response across a wide range of decay topologies is already well represented in existing fully simulated samples. While the set of explicitly simulated modes within these samples is limited, missing modes will typically differ only by minor structural variations, such as an extra particle in the final state or an alternate resonance structure. Crucially, as the characteristics of the target high-level variables are governed by the same principles across all examples, by learning these principles, a model can accurately interpolate between examples. Whilst there is a stochastic nature to the full simulation, over large samples, the distributions of the target high-level variables follow reproducible, learnable patterns.
    
    This paper is structured as follows: the description of the training sample is provided in Sec.~\ref{sec:simulation}; an outline of \rex is given in Sec.~\ref{sec:fast_framework}; Sec.~\ref{sec:architecture} describes the various network architectures employed. This is followed by examples of performance in Sec.~\ref{sec:performance}, and then, finally, conclusions are presented in Sec.~\ref{sec:conclusion}.
    
\section{LHCb simulation framework}\label{sec:simulation}

    This tool has been developed for the requirements of the LHCb experiment; therefore, a brief outline of the standard LHCb simulation framework is provided below.
    
    In the LHCb simulation, $pp$ collisions are generated using \pythia~\cite{Sjostrand:2006za,Sjostrand:2007gs} with a specific \lhcb configuration~\cite{LHCb-PROC-2010-056}. In a typical simulation workflow, for example, for that of $B^+\to K^+e^+e^-$ decays, $pp$ collisions are simulated until a $B^+$ meson is generated. In this case, the $B^+$ particle, the head of the decay, is then forced to decay into the target signal final state, $K^+e^+e^-$, using a pre-defined decay model. The kinematics of particle decays are described by \evtgen~\cite{Lange:2001uf}. Final-state radiation is generated using \photos~\cite{Golonka:2005pn, Davidson:2010ew}. Each generated so-called event is made up of the $B^+$ meson decay products of interest and any other particles generated in the collision, which are all then passed to the detector response stage. The kinematics and identification of the involved particles are stored at this point; this is the so-called truth-level information. 
    
    The propagation of generated particles through the detector material and any interaction with the detector is then computed using the \geant toolkit~\cite{Allison:2006ve, Agostinelli:2002hh}, as described in Ref.~\cite{LHCb-PROC-2011-006}. This produces a simulated response in all detector sub-systems. This is then digitised and passed through the same trigger, particle identification, and reconstruction algorithms as used during online data-taking; here, reconstructed track objects are combined to form reconstructed candidates. It is the reconstruction-level output quantities from these algorithms that form the final detector response that can be used to compute the derived quantities, such as efficiencies, required for physics analyses. It is these high-level quantities that are extracted from simulated events to create training samples for this tool. 

    \subsection{Training Data Preparation and Preprocessing}\label{sec:simulation:data}
    
        Training samples are extracted from the extensive archive of simulated data produced for LHCb analyses, covering the full range of previous simulation requests. For each simulated event, all final-state tracks are truth-matched~\cite{Tolk:2014llp} to ensure they are associated with single particles and originate from heavy meson decays; ghosts, tracks containing hits from multiple particles, and clones, multiple tracks matched to the same particle, are excluded. Valid tracks, and various sub-sets of these tracks, are systematically combined to form candidates representing every possible decay topology. These topologies include both direct multi-body decays and configurations with reconstructed intermediate resonances. Tracks are reused as appropriate to cover all topologies and unique combinations. At no point does the original true topology of a given simulated event affect the candidates constructed. Such a process creates an exhaustive sample of fully- and partially-reconstructed candidates. For the studies presented in this paper, only charged tracks from pions, kaons, muons, and electrons are included in the sample; neutral particles and protons are therefore excluded from the first iteration of \rex. Additionally, for this iteration, only final states of four tracks or fewer are considered; however, by employing graph representations for decays, this tool can naturally accommodate higher multiplicity final states with no change to the network architecture. For this study, only a small subset of the available simulated data is employed. Events are processed until there are 1.5 million training samples for each option of 2, 3 and 4 final state particles; these are evenly balanced across the topologies for each option. 

        \noindent The first prototype of \rex produces the following variables for each track:
        \begin{itemize}
            \item \ghostProb: the probability that a track is a fake track that doesn't correspond to the trajectory of a true particle, but originates from a mismatch of hits from separate particles or from detector noise, a so-called ghost~\cite{DeCian:2017ytk}.
            \item \chiSqTrack/ndof: the fit quality of the Kalman filter-based track fit~\cite{LHCb:2018zdd,Billoir:2021srr,DeCian:2017ytk}. 
            \item \ipChiSq: a measure of how likely it is that a given track originated from the primary vertex (PV), computed as the difference between the $\chi^2$ of the PV fit when including and excluding the track in question.  
            \item PID$i$: a set of particle identification variables computed by taking the difference in the log likelihood of a track being of a specific type (e.g., $K$, $e$, $\mu$, $p$~) relative to the $\pi$ hypothesis. These variables incorporate information from the calorimeter, RICH, and muon systems~\cite{LHCb:2014set}. 
            \item ProbNN$i$: a set of optimised PID variables obtained using multivariate techniques that combine tracking and PID information from each sub-system into a single probability value for each particle hypothesis, $i$ is one of $\pi$, $K$, $e$, $\mu$, $p$. 
        \end{itemize}
        \noindent Then, for each reconstructed candidate, whether that is an intermediate resonance or the parent candidate at the head of the decay chain, the following variables are generated: 
        \begin{itemize}
            \item $\chi^2_{to~\textrm{PV}}$: a measure of how well the decay vertex of the reconstructed particle in question can be separated from the primary vertex. 
            \item $\cos(\dira)$: the cosine of the direction angle (\dira), the angle between the momentum vector of the reconstructed particle in question and the vector connecting the associated PV and the reconstructed decay vertex. 
            \item \ipChiSq: a measure of how likely it is that the reconstructed particle in question originated from the primary vertex (PV), computed as the difference between the $\chi^2$ of the PV fit when including and excluding the particle in question. 
            \item \vtxChiSqNdf: a measure of the fit quality of the secondary vertex fit~\cite{LHCb:2018zdd,Billoir:2021srr}. 
        \end{itemize}
        Finally, for the parent candidate only, there are the vertex isolation variables ($VtxIso$), which employ event-wide information to determine whether or not any unaccounted for charged tracks also originate from the vertex in question. This set of target variables includes the majority of the high-level variables used in most of the physics analysis at LHCb. For any analysis pipeline requiring the generation of bespoke variables, retraining the network will be necessary; however, the software is designed to make this process as straightforward as possible. 
        
       Additionally, a set of conditional variables is extracted from each event, corresponding to a set of properties that guide the response of each network within \rex, as discussed in Sec.~\ref{sec:fast_framework}. The conditional inputs include:
        \begin{itemize}
            \item \textbf{Per-track variables} such as the true identification of the particle that formed each track, the true and reconstructed momenta and pseudorapidity of each track.
            \item \textbf{Track-pair variables} describing relationships between tracks — for example, the angle between their momentum vectors, and the angles between their reconstructed momentum vectors.
            \item \textbf{Head candidate-level variables} computed for the reconstructed head particle. These include quantities such as the \dira, calculated from both true and reconstructed momenta, the total true and reconstructed momenta of the candidate, and the difference between the true momentum of the head particle and the sum of the true momenta of its daughter particles. 
            \item \textbf{Intermediate candidate-level variables} are constructed similarly to those of the head candidate. However, since an intermediate particle in the reconstruction topology may not correspond to a physical particle, no truth-level quantities are available, and only reconstructed information is used. 
        \end{itemize}
        The chosen set of conditional variables is designed to capture as much information as possible upon which the target high-level variables may depend.

        The raw values of both the target and conditional variables in the training samples are preprocessed such that the distributions are easier for the network to handle. This constitutes mapping distributions into smoother representations, normalising them to align with the chosen activation space, and reducing the density of events near the activation space boundaries. Quantile transformations are used to achieve processed distributions with these properties~\cite{Gavranovic:2023oam}. The distribution of each variable is mapped onto a truncated normal distribution that is subsequently normalised to the range [-1, 1] to align with the $tanh$ activation function. This is a monotonic and reversible transformation that preserves correlations between variables while being robust to outliers and large differences in scale between the core and the tails of distributions. This preprocessing is found to improve training stability and facilitate faster convergence. 

\section{Fast simulation framework}\label{sec:fast_framework} 

    \begin{figure}[t]
        \begin{center}
        \centering
        \includegraphics[width=0.99\textwidth,keepaspectratio]{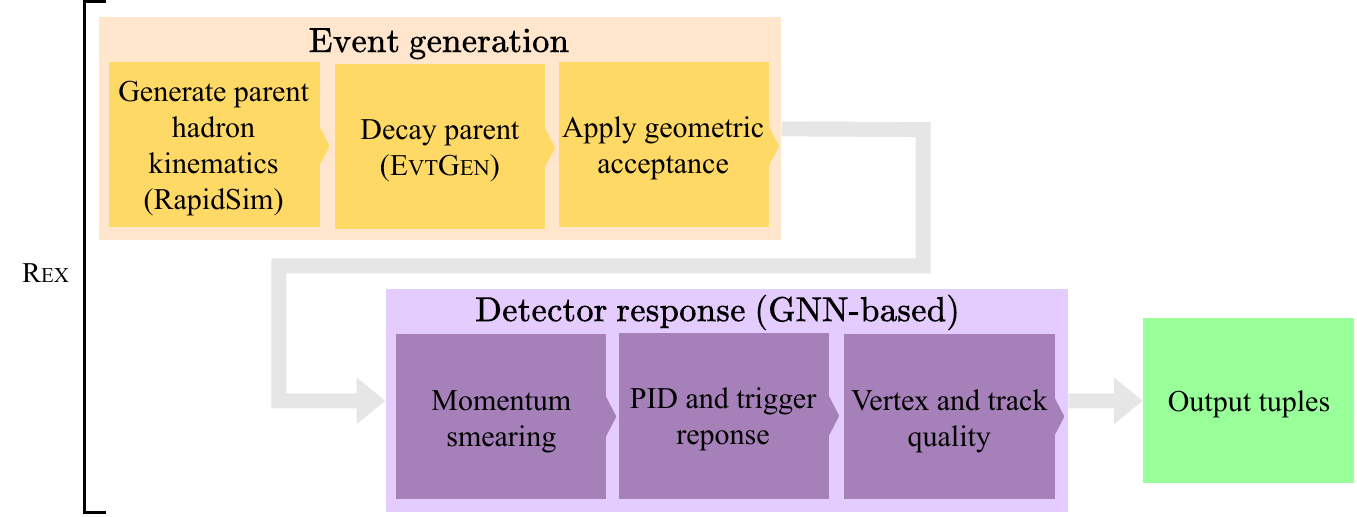}
        \vspace*{-0.5cm}
        \end{center}
        \caption{
        \small Overview of the structure of \rex.}
        \label{fig:intro:flow}
    \end{figure}

    An overview of the structure of the tool is provided in Fig.~\ref{fig:intro:flow}, and described in the following paragraphs. 
    
    \paragraph{Event generation} 

    This approach entirely bypasses \pythia and instead directly samples the kinematics of single heavy-flavour hadrons from Fixed Order Next to Leading Order predictions of charm and bottom production at the LHC~\cite{Cacciari:1998it, Cacciari:2012ny}. In collisions at the LHC, heavy-flavour hadrons are often produced in pairs. Any influence of the corresponding partner hadron or of particles from the rest of the event is implicitly included on average by training on fully simulated events, which were generated with \pythia and within which event-wide effects are present. This is demonstrated in Sec.~\ref{sec:performance:vtx}, where the generated reconstruction performance is shown to closely match that of the full simulation, rather than resembling an idealised case. The location of the PV is generated using a dedicated network trained on fully simulated events, which is introduced in Sec~\ref{sec:architecture:PVsmearing}, and the lifetimes of any particles, along with any flight distances, are modelled within the RapidSim package.
    
    Particle decays are handled with the same generator as in the full simulation, \evtgen. This ensures a physical description of decay amplitudes and allows users to generate events with standardised physics models. Final-state radiation is also included with the same library as the full simulation, \photos.
    
    In $pp$ collisions at LHCb, heavy-flavour hadrons and their decay products are produced in $4\pi$ steradians. In the full simulation, the geometrical acceptance is naturally captured. However, in this fast approach, the geometrical acceptance must be approximated. This is achieved by requiring that all final-state particles fall within a specified pseudorapidity range. Events are only retained in cases where all final-state particles are within this range. A more accurate modelling of this acceptance will be included in a later version \rex.
    
    \paragraph{Detector response} 
    
    The modelling of the detector response is decomposed into three interdependent components: momentum smearing, generation of particle identification and trigger responses, and vertexing. While each component is interdependent during inference, during training, each network is trained independently. Each component is handled with a GNN, the architecture of each is described in Sec.~\ref{sec:architecture}. This modular decomposition is motivated by considerations of both training efficiency and flexibility during inference. With this treatment, it becomes possible to update the network weights of specific stages as needed. For instance, one could substitute a different model for momentum smearing without retraining the entire system. The networks are implemented using \texttt{PyTorch Geometric}~\cite{Fey:2019wpv}.

    \paragraph{Inference and output} 

    \rex is available as a lightweight Python package with minimal dependencies and is isolated from the extensive LHCb software stack. The package interfaces with RapidSim automatically, executing queries in the background before running the required networks and then finally organising the output. The generated output is formatted such that it mirrors that of the full simulation and can therefore act as a drop-in replacement. The package handles the computation of the reconstructed masses of all intermediate and parent particles, automatically accounting for different mass hypotheses in the case of misidentified tracks.
    
    To provide a sense of timing performance, generating analysis-ready output for a sample of ten million $B^+ \to K^+ \mu^+ \mu^-$ decays takes approximately one hour on a 64-core AMD EPYC 7702P CPU. This time includes the RapidSim and EvtGen-based event generation, which accounts for a similar portion of the total time taken as the detector response emulation. 
    This is a speed-up of $\mathcal{O}(10^5)$ over the full simulation using a typical approximation that each event takes one minute~\cite{Muller:2019ogi}. 
    Such a significant speed-up enables on-demand generation, which has the potential to reduce the pressure of long-term storage of simulated samples significantly. 
    Furthermore, such a workflow of generating on demand could enable a partial decentralisation of the simulation process, helping to reduce administrative overhead for the collaboration.

\section{Network Architecture}\label{sec:architecture} 

    During inference, the networks used within this tool are the generators of conditional Generative Adversarial Networks (GANs)~\cite{goodfellow2020generative}. In the standard GAN framework, a generator network is trained to produce synthetic samples that are indistinguishable from the training samples. To enable this, during training, updates to the weights of the generator network are driven by a second network, the discriminator network. The goal of the discriminator network is to distinguish between real and generated samples. The two networks are trained in tandem in a technique known as adversarial training. Updates to each network are performed iteratively via backpropagation, with losses being computed over mini-batches of training data. In the simplest GAN setup, the generator learns to produce samples using only random noise vectors as inputs. However, the GANs employed here are all conditional, that is, they have additional input vectors containing values that represent physical characteristics, as described in Sec.~\ref{sec:simulation:data}, that can be used to guide the generation. These variables are also shown to the discriminator so that it can flag generated samples with characteristics that are incompatible with the conditional inputs. 

    To emulate the detector response, these networks are structured as heterogeneous graph neural networks~\cite{yang2023simple} (HGNNs). Graph neural networks (GNNs) are a broad class of models designed to operate on graph-structured data, where nodes represent entities (such as particle tracks) and edges represent interactions between those entities. Information is propagated across a graph via message-passing operations, in which each node exchanges messages with its connected neighbours, then updates its internal representation based on any received information. These updates often include pooling operations, aggregating incoming information at nodes connected to multiple neighbours. Details of the message-passing operations employed in these networks are provided in the following sections. HGNNs extend the framework of GNNs, adding flexibility that allows for multiple types of nodes and edges, each with its own learnable update mechanism. For instance, different node types may represent tracks formed by different species of particles. In this case, edge types would encode directed, typed interactions between those tracks—for example, an edge from a kaon track to an electron track may be treated differently from one in the reverse direction. This increased flexibility of HGNNs enables them to capture complex dependencies for a wide variety of graph structures, making them well-suited to the task at hand. These qualities of HGNNs have been exploited in other particle physics applications, see Refs.~\cite{Sutcliffe:2025arn, Huang:2023ssr, Caillou:2024dcn}. The structure of each network and the construction of each graph are detailed in the following sections. 

    Training was performed on NVIDIA Tesla T4 (16 GB) GPUs. The HGNN-based networks were each trained for approximately 48 hours.

    \subsection{Primary vertex smearing}\label{sec:architecture:PVsmearing}

        The process outlined in Sec.~\ref{sec:fast_framework}, with which the kinematics of the head particle of each decay are generated, does not produce a generated location of the PV. As the reconstruction quality has a dependence on the location of the PV, a realistic distribution of PVs is required in the generated sample. 
        We train a very simple GAN constructed of fully connected layers to generate the true location of PVs using a training sample extracted from fully simulated events. This GAN is conditioned on the true momenta of the decaying meson to account for boosting effects. When queried, the generated smearing is passed on to all downstream vertices such that any separation between vertices is maintained. 

    \subsection{Momentum Smearing}\label{sec:architecture:smearing}
    
        Momentum smearing constitutes the first stage of the detector response model. The structure of this network is designed to handle two effects. Firstly, the scale of resolution effects varies across particle species, most notably for electrons, whose tendency to emit bremsstrahlung radiation significantly degrades their reconstruction quality at LHCb. Secondly, global properties of an event can impact the momentum resolution. For example, in the case that two particles are emitted with very small angles between them, they can influence each other's reconstruction. Consequently, to treat each track in isolation with independent queries of a network would be an incomplete solution. 

        \begin{figure}[h]
        \begin{center}
        \centering
        \includegraphics[width=\textwidth,keepaspectratio]{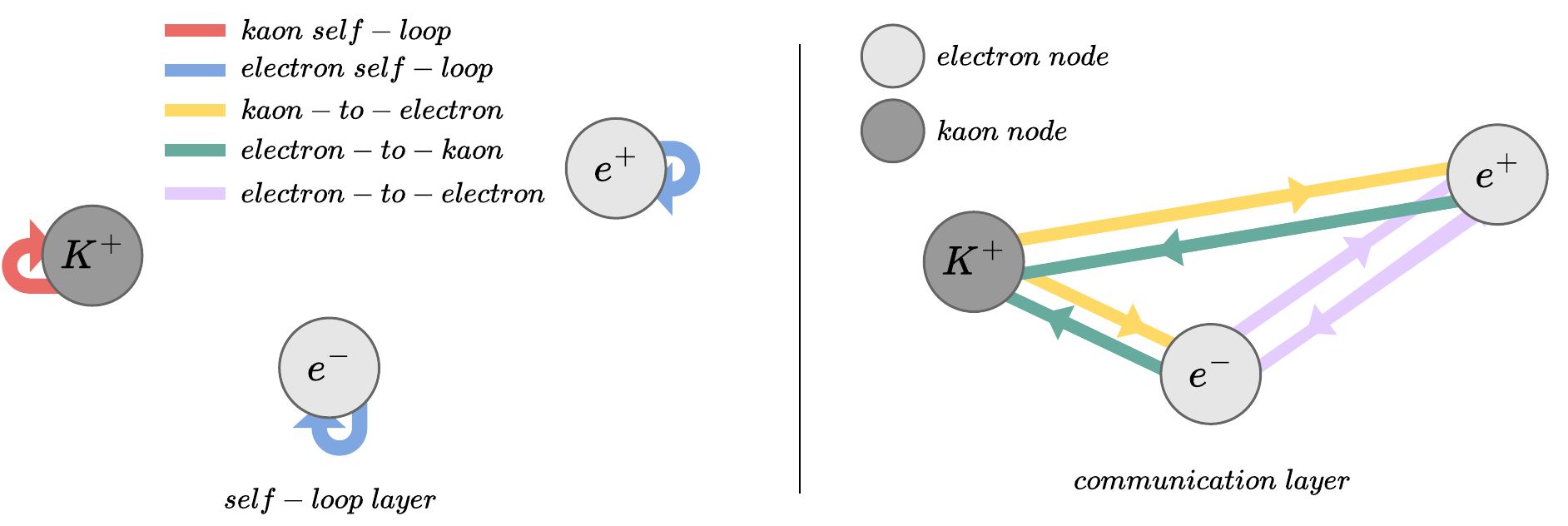}
        \vspace*{-0.5cm}
        \end{center}
        \caption{
        \small Example of the edge connection structure of a graph employed during a communication layer of the smearing and PID networks.}
        \label{fig:architecture:unique_node_graph}
    \end{figure}
    
        To address both these challenges, we employ a heterogeneous graph-based approach using multiple node types. Each event is structured into an all-to-all connected graph within which each node represents a reconstructed track and is assigned a type corresponding to its particle species, independently of charge. The graph incorporates multiple edge types that facilitate track-to-track communication. 
        In this setup, a distinct edge type is defined for each ordered pair of node types. For example, \textit{kaon-to-electron} and \textit{electron-to-kaon}, and these are used to connect each node to every other. This structure gives the network flexibility to capture any differences that may exist in the various species-specific interactions. Each graph is initialised with vectors of conditions and vectors of uniquely generated random noise at each node; additionally, the angles between particles are provided as conditional edge features in the initialisation. Each graph structure is then passed through the network, resulting in vectors of output at each node.
        
        During a pass of the network, the hidden representations of each graph are updated multiple times by a sequence of self-loop layers and communication layers. Illustrations of the structure of the nodes and edges in these layers are presented in Fig.~\ref{fig:architecture:unique_node_graph}. In the self-loop layers, each node type is processed by a distinct and isolated linear transformation. In the communication layers, information is exchanged between nodes along the edges via Graph Attention Convolution layers~\cite{Velickovic:2017lzs} (GATs) in the process of message-passing. In a standard graph convolution, for each target node, the hidden states of its neighbours (and optionally the node itself) are first linearly transformed to create a collection of incoming messages; these are then aggregated by pooling to update the state of the target node. GATs extend this operation to include a learnable attention mechanism that assigns a weight to each message to scale its relative importance before pooling. 
        Since inter-track correlations are expected to introduce only subdominant modifications to the smearing, we restrict this message passing so that only a fixed subset of the hidden representation of each node is shared with its neighbours. The rest of each hidden representation is not updated by the communication layers. The subset of nodes used in the communication is defined before the training process.

        This heterogeneous node-based approach removes the need for the particle species labels to be included in the conditional input and instead encodes this information into the graph structure itself. This approach automatically handles imbalances in the training sample, with the weights of particle-specific nodes and edges never being updated unnecessarily.

\begin{figure}[h]
            \begin{center}
            \centering
            \includegraphics[page=4, width=0.4\textwidth,keepaspectratio]{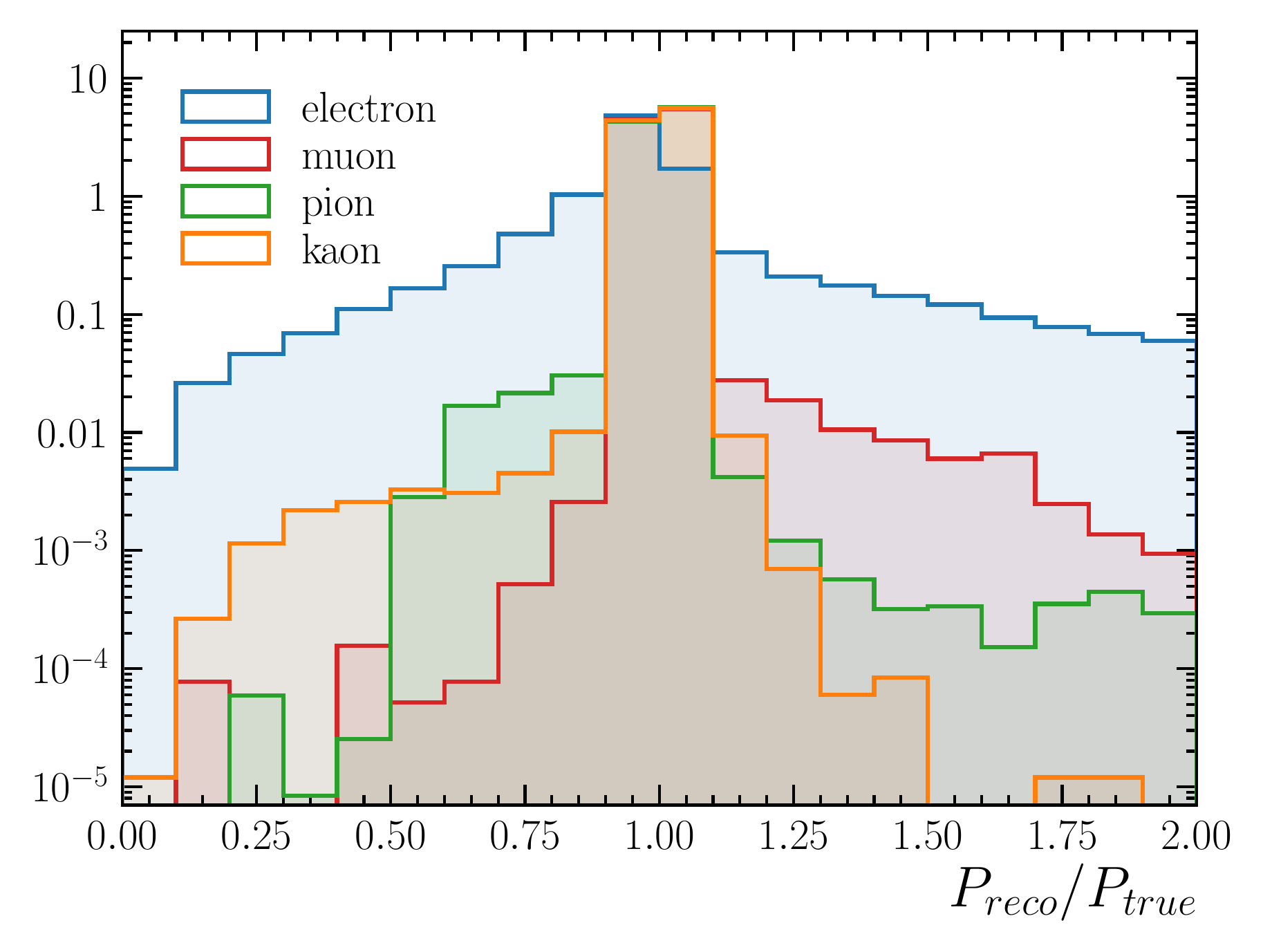}\\
            \includegraphics[page=7, width=0.4\textwidth,keepaspectratio]{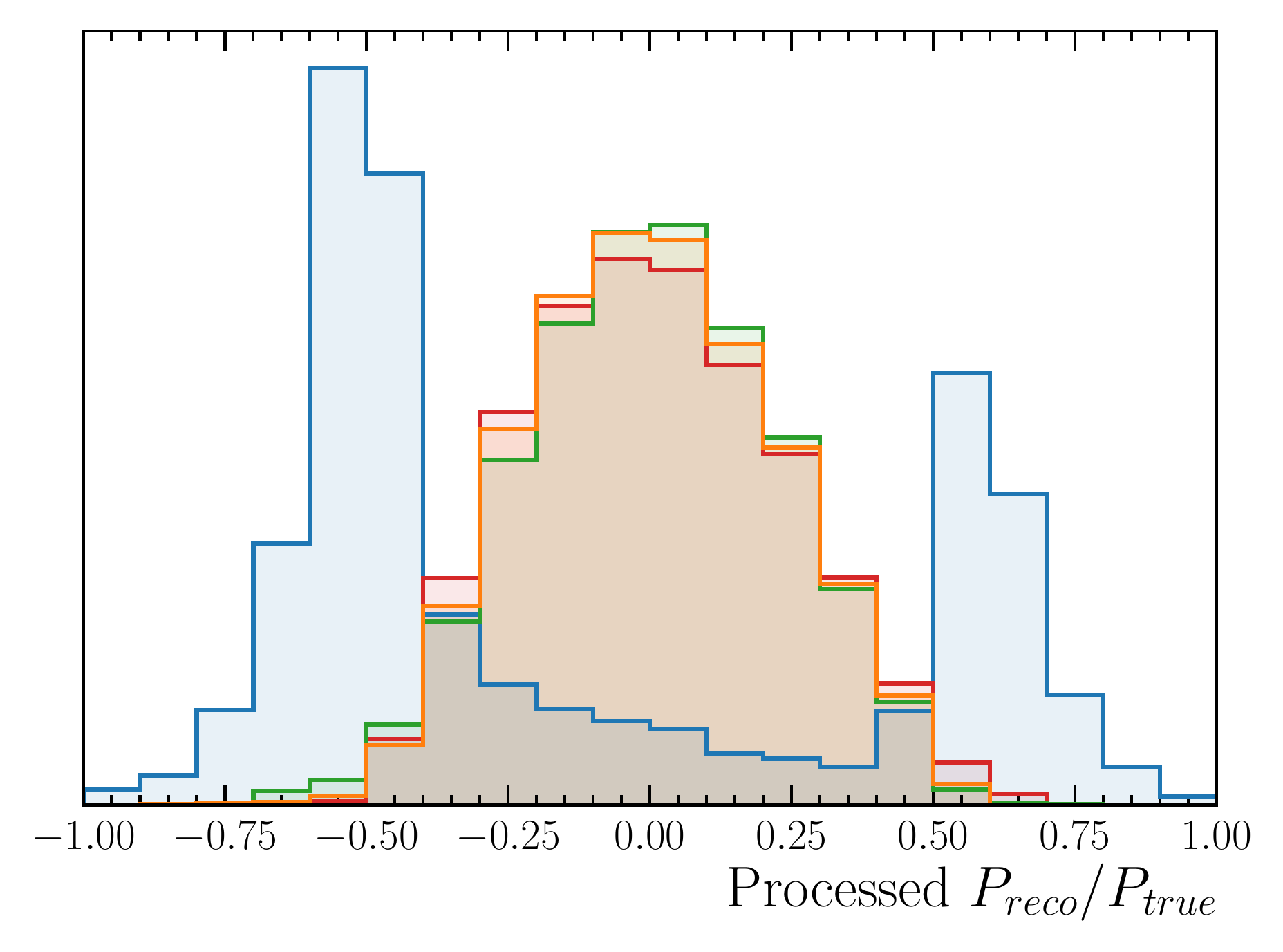}
            \includegraphics[page=11, width=0.4\textwidth,keepaspectratio]{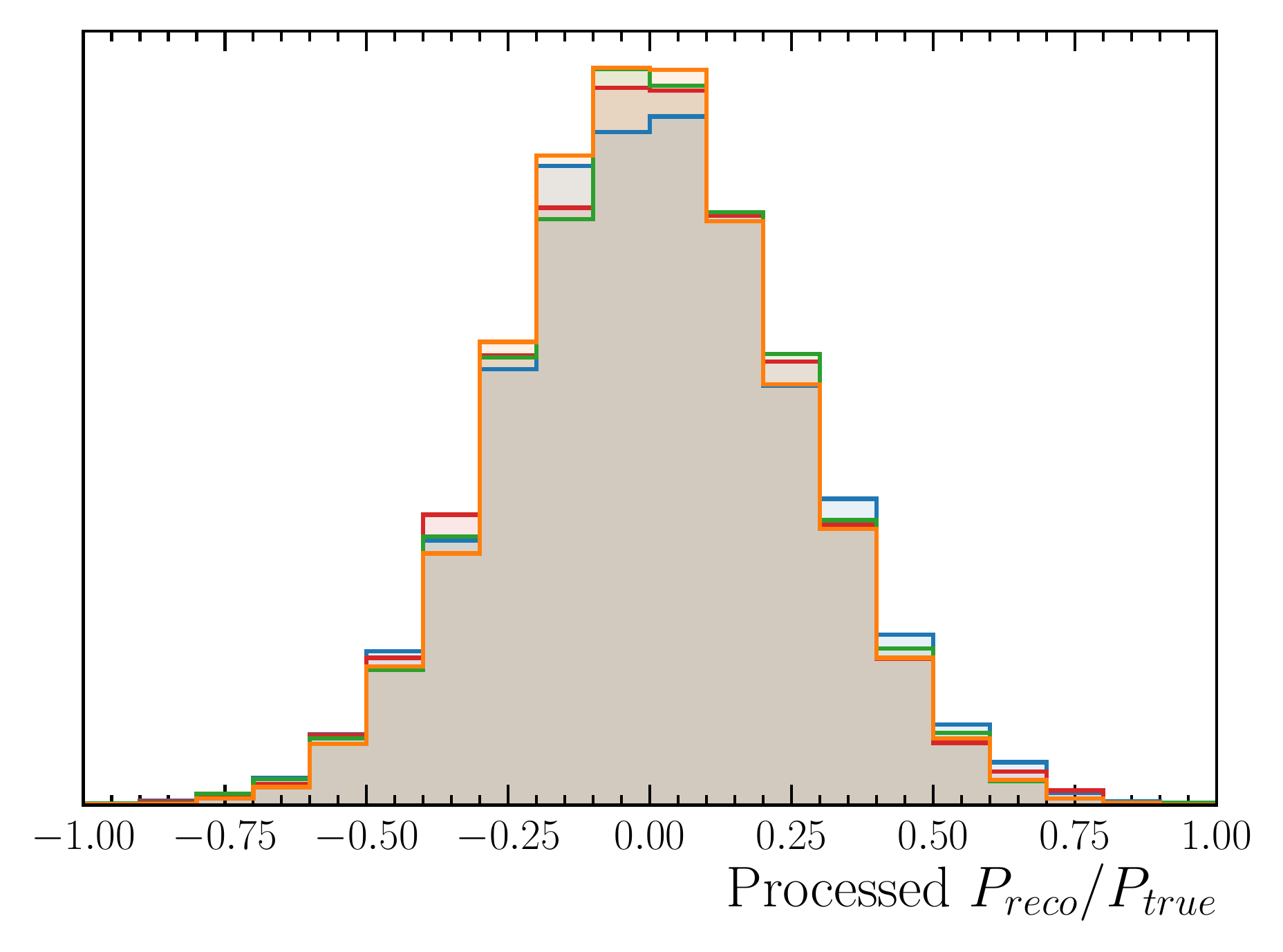}
            \vspace*{-0.5cm}
            \end{center}
            \caption{
            \small Examples of pre-processing for a variable describing the level of momenta smearing, the ratio of the reconstructed momenta $P_{reco}$ to the true momenta $P_{true}$. The physical distribution of the variable (top), the distribution using shared pre-processing (left), and unique per-particle-type pre-processing (right) are all provided.}
            \label{fig:architecture:unique_processing}
        \end{figure}

        Additionally, this heterogeneous graph formulation enables species-specific preprocessing. This is helpful when the scale of input variable distributions differs significantly across particle types. This is true for the distributions of momentum residuals, where electrons, being more poorly reconstructed, exhibit a much broader spread than other species. An illustration of this effect and both a common and a species-specific preprocessing strategy is shown in Fig.~\ref{fig:architecture:unique_processing}. In the case of a single global preprocessing, the majority of the electron distribution exists in the tail regions of the processed distribution, which is undesirable. The species-specific preprocessing avoids this, significantly improving the stability of the network during training and speeding up convergence.

    \subsection{PID}\label{sec:architecture:PID}

        The network architecture employed here is the same as that of the momentum smearing, with a unique node type per particle species. The target variables here are the PID variables PID$i$ for each of $K$, $e$, $\mu$, and $p$, and ProbNN$i$ for each of $\pi$, $K$, $e$, $\mu$, and $p$. The conditional input variables used to query this network are the same as those of the smearing network; however, they now also include the generated reconstructed momenta, such that the generated PID information is correctly correlated to the momentum reconstruction quality. Again, unique pre-processing per node type is advantageous, as the distribution of each target variable is very different, being variables designed to separate different particle species. 

    \subsection{Vertexing}\label{sec:architecture:vertexing}

        \begin{figure}[h]
            \begin{center}
            \centering
            \includegraphics[width=0.99\textwidth,keepaspectratio]{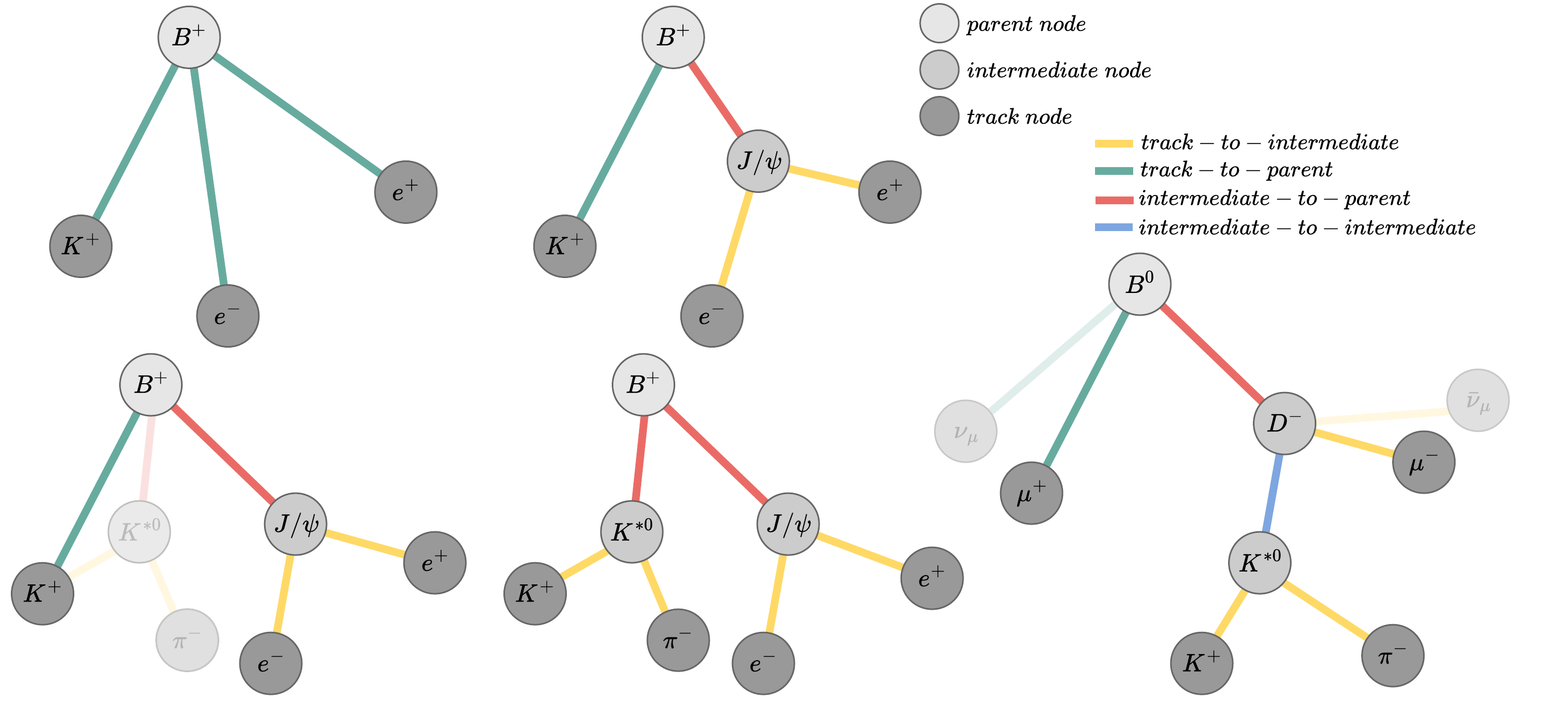}
            \vspace*{-0.5cm}
            \end{center}
            \caption{
            \small Example edge structures within graphs passed through the vertexing network, the examples cover various 3- and 4-body reconstruction topologies. Faded nodes indicate particles missed; these are not included in the graph structures that encode the reconstruction topology.}
            \label{fig:architecture:example_decays}
        \end{figure}
        
\begin{figure}[h]
            \begin{center}
            \centering
            \includegraphics[width=0.99\textwidth,keepaspectratio]{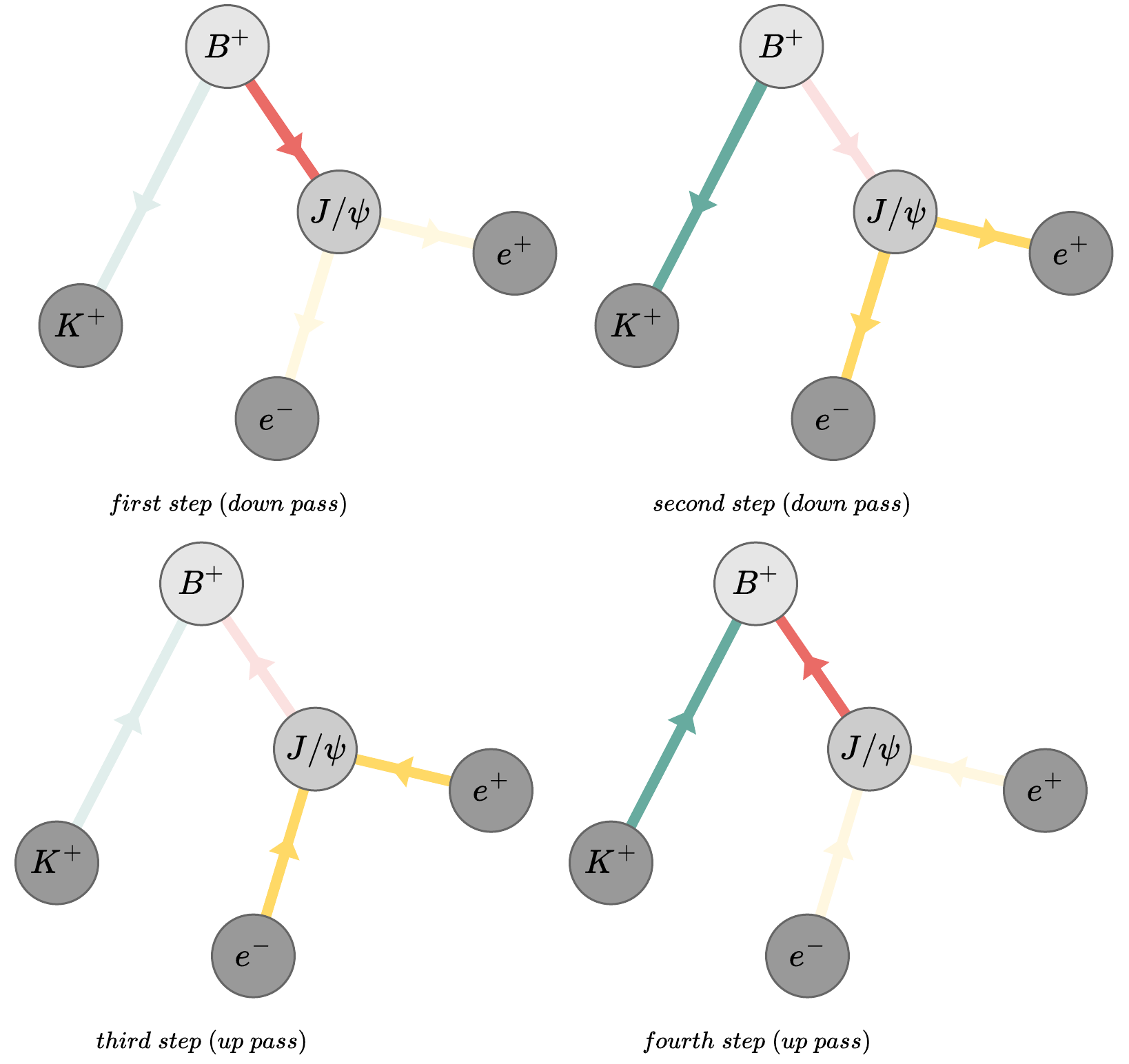}
            \vspace*{-0.5cm}
            \end{center}
            \caption{
            \small The stages of the hierarchical propagation of information during a single layer of the generator of the vertexing network, for the example of $B^+\to K^+ J/\psi(\to e^+ e^-)$. Faded edge connections indicate inactive edges in a given step.}
            \label{fig:architecture:sequential}
        \end{figure}

        The architecture of the vertexing network is significantly different. Here, a graph represents the reconstruction of a decay tree from track objects. Within each graph, there are now only three possible node types: track nodes, nodes representing reconstructed intermediate candidates and nodes representing parent candidates. Each graph is initialised with vectors of conditions and of uniquely generated random noise at each node. 
        Within each graph, nodes and their connecting edges are organised in a hierarchical arrangement in accordance with the requested reconstructed topology.  
        Some examples are provided in Fig.~\ref{fig:architecture:example_decays}; these include cases of partially reconstructed decays in which one or more final state particles are not reconstructed; missing particles are indicated in the graphs with faded nodes. There is an edge type for all the possible connections, such as track-to-intermediate or parent-to-track, and each edge type has a unique set of weights. All edges in the network are directional, and information is propagated within a graph sequentially, as illustrated in Fig.~\ref{fig:architecture:sequential}. Within each layer of the generator network, two passes are made: first, information flows \textit{down} the graph — from the reconstructed parent candidate, through any reconstructed intermediate objects, to the track nodes. This \textit{downward} pass allows the network to correlate track-level features with properties of the reconstructed objects. This is particularly important for variables such as the track $\chi^2_{\textrm{IP}}$, which encode how the parent candidate’s $\chi^2_{\textrm{IP}}$ changes when a track is excluded. Then, a second pass proceeds \textit{up} the graph. This \textit{upward} pass is motivated by the existing LHCb reconstruction algorithms, which build candidates having access solely to reconstructed track momenta. The final pass being \textit{upward} is stylistically important, as it ensures that any downstream nodes remain unchanged and the representations left behind are the exact representations used in any pooling operations. For the discriminator network, the order of these passes is reversed.

        On top of the primary edges connecting nodes in accordance with the reconstructed hierarchy, additional skip connections were trialled. These were added to allow direct communication between otherwise disconnected nodes, for example, the $K^+$ and the $J/\psi$ nodes of the graph in Fig.~\ref{fig:architecture:sequential}. There was no significant improvement in the performance, so these are avoided for simplicity. The hierarchical structure without skip connections is also philosophically closer to how the existing LHCb reconstruction algorithms build up reconstructed candidates. 
                
        \begin{figure}[h]
            \begin{center}
            \centering
            \includegraphics[width=0.75\textwidth,keepaspectratio]{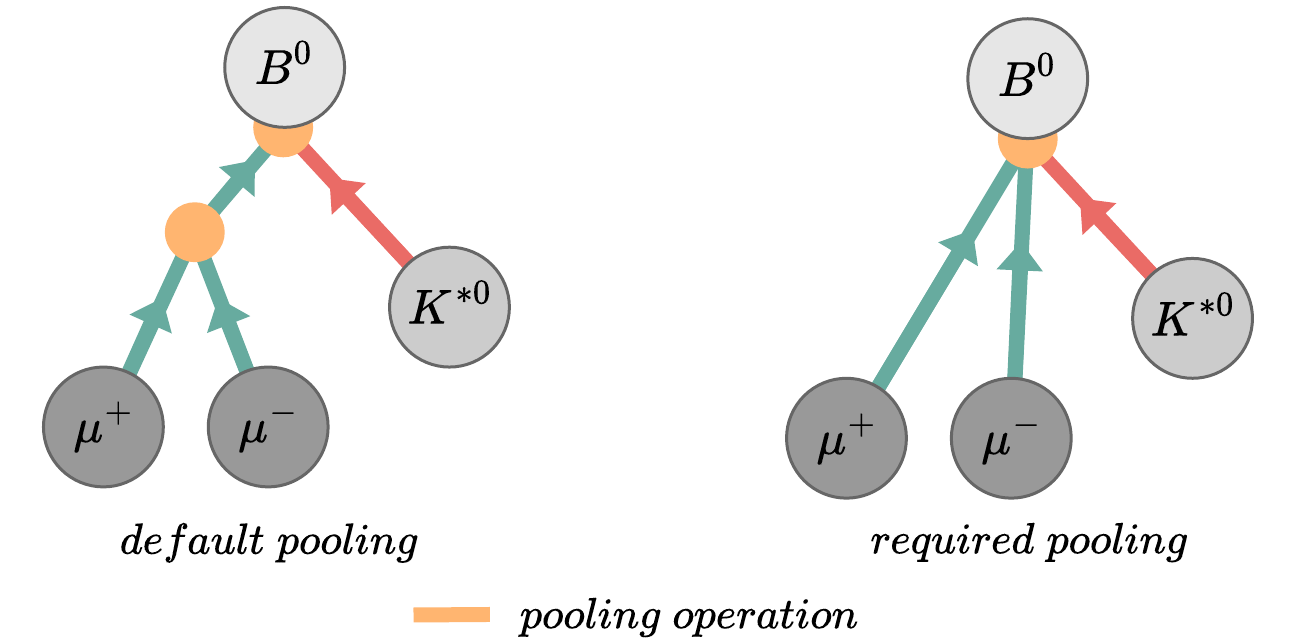}
            \vspace*{-0.5cm}
            \end{center}
            \caption{
            \small Example of a single vertex within a wider graph indicating two possible approaches to pooling hidden representations.}
            \label{fig:architecture:pooling}
        \end{figure}

        For this network, we implemented a customised version of the \texttt{PyTorch Geometric} heterogeneous graph class that allows for node-independent pooling during the propagation up each graph, as illustrated in Fig.~\ref{fig:architecture:pooling}. In the standard heterogeneous graph class, pooling at nodes occurs in two stages: firstly, hidden representations from any edges of each edge type are pooled, then secondly, these resulting representations are pooled across edge types. This approach is problematic for this application as it would change the relative importance of each edge across different cases. We avoid this by combining representations in a single step, agnostic of edge type. Three pooling operations are used simultaneously: min, max and sum. A learnable attention mechanism then allows the network to adapt the importance of each operation for each dimension of the hidden representation. 
        
        This architecture incorporates minibatch discrimination (MBD) layers~\cite{Salimans:2016kpc}, which enables the discriminator to employ information from across a small batch of samples rather than just assessing each example independently. This was found to improve the diversity of the generated outputs. This is particularly valuable when modelling variables that are only weakly correlated with the conditioning inputs, such as \chiSqTrack/ndof and \ghostProb. MBD layers are applied both at the level of individual track nodes and on the graph-level pooled representation before its final compression into the discriminator output.

        The conditional inputs at each track node type include information about the true kinematics, the generated reconstructed kinematics, and trigger and particle identification (PID) information. As described in Sec.~\ref{sec:simulation:data}, different conditions are given as inputs to nodes of each type. Importantly, all conditional variables used are derived exclusively from lab-frame observables; quantities such as reconstructed invariant masses are deliberately excluded. This design choice aims to improve the network's ability to generalise across different decay modes and more easily cover modes not shown during training. This avoids learning about specific resonant structures present in the training data. For example, in decays such as $B^+\to K^+J/\psi(\to \mu^+\mu^-)$, providing the invariant mass of the dimuon system as a condition would implicitly encode knowledge of the $J/\psi$ resonance. The network may then associate large deviations from the $J/\psi$ mass with poor reconstruction quality. If the network were subsequently queried on non-resonant $B^+\to K^+\mu^+\mu^-$ decays, it would incorrectly generate variables indicating poor reconstruction quality for events with dilepton invariant mass far from the $J/\psi$ resonance. This approach also mirrors the LHCb vertexing algorithms, which only have access to lab-frame variables when constructing candidates.
        
        One condition worth highlighting is that which encodes the physical separation between the origin vertices of final state particles and the decay of the parent particle. Separation is possible in cases where a decay has occurred via a long-lived intermediate state, such as a $D^0$.
        In the example of $B^+\to \bar{D}^{0}(\to K^+e^-\bar{\nu}_e)\pi^+$, the $\bar{D}^{0}$ meson has a long enough lifetime to create a physical separation between the $B^+$ decay vertex and the origin vertex of the $K^+$ and $e^+$.
        
        Regarding training convergence, no formal stopping criterion has yet been implemented — a limitation acknowledged in Sec.~\ref{sec:conclusion} as a target for future development. Training is currently stopped based on a qualitative assessment of when the performance on an unseen validation subsample of the training data ceases to improve.

\section{Physics performance}\label{sec:performance}

    \begin{figure}[h]
        \begin{center}
        \centering
        \includegraphics[page=1, width=0.32\textwidth,keepaspectratio]{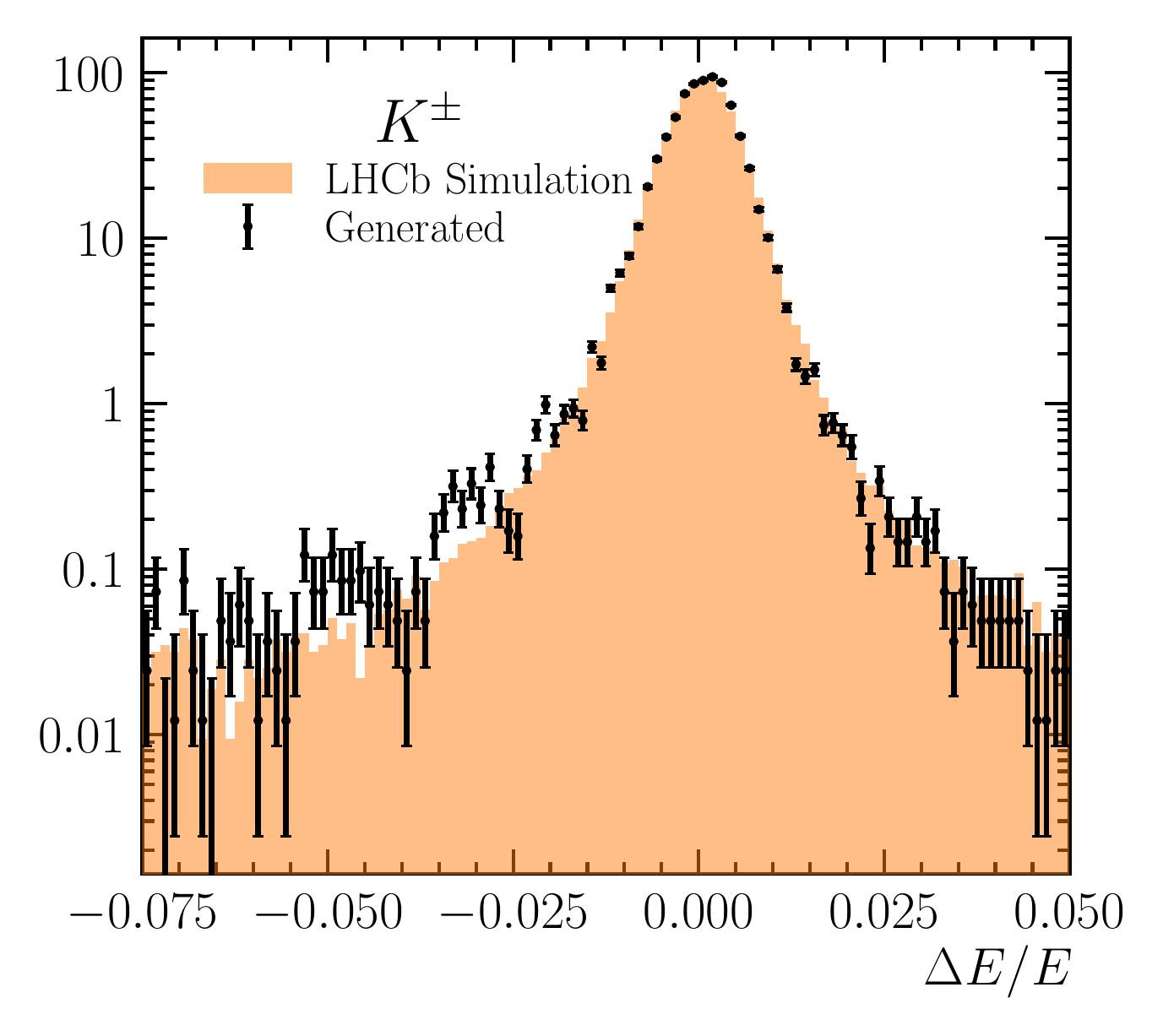}
        \includegraphics[page=3, width=0.32\textwidth,keepaspectratio]{figures/smearing_plots.pdf}
        \includegraphics[page=6, width=0.32\textwidth,keepaspectratio]{figures/reco_mass_plots.pdf}\\
        \includegraphics[page=5, width=0.32\textwidth,keepaspectratio]{figures/smearing_plots.pdf}
        \includegraphics[page=7, width=0.32\textwidth,keepaspectratio]{figures/smearing_plots.pdf}
        \includegraphics[page=11, width=0.32\textwidth,keepaspectratio]{figures/reco_mass_plots.pdf}
        \vspace*{-0.5cm}
        \end{center}
        \caption{
        \small Distributions (left and centre) of $\Delta E/E$ for each particle type, and (right) reconstructed-$B$ mass distributions for $B^+\to K^+ e^+ e^-$ and $B^+\to K^+ \mu^+ \mu^-$. Both simulated and generated distributions are provided in each case.}
        \label{fig:performance:smearing}
    \end{figure}

    Within this section, the performance of each component of \rex is presented, followed by an example analysis case study and a discussion on uncertainties. 
        
    \subsection{Momentum smearing}\label{sec:performance:smearing}

        \begin{figure}[h]
            \begin{center}
            \centering
            \includegraphics[page=2, width=0.48\textwidth,keepaspectratio]{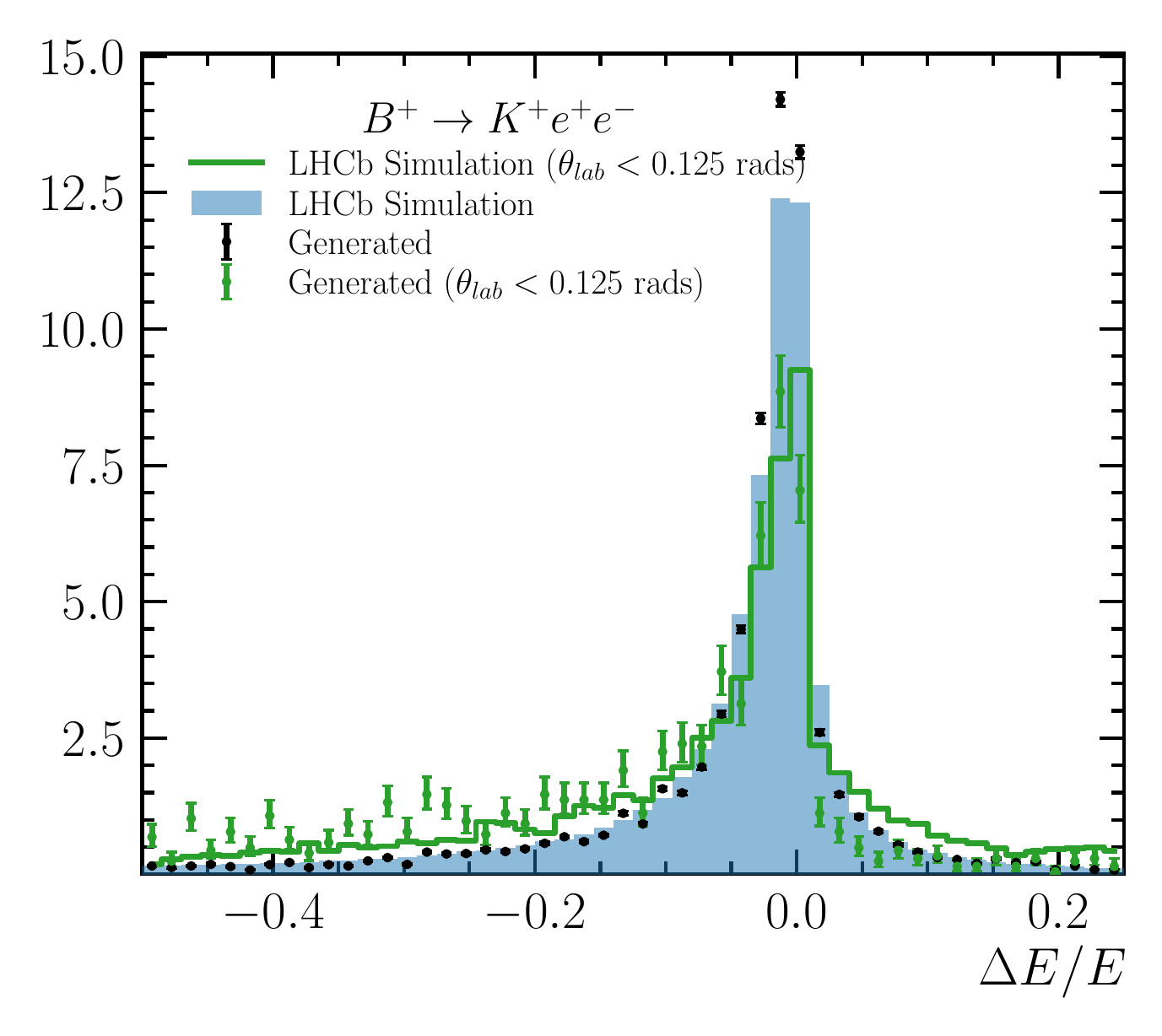}
            \includegraphics[page=4, width=0.48\textwidth,keepaspectratio]{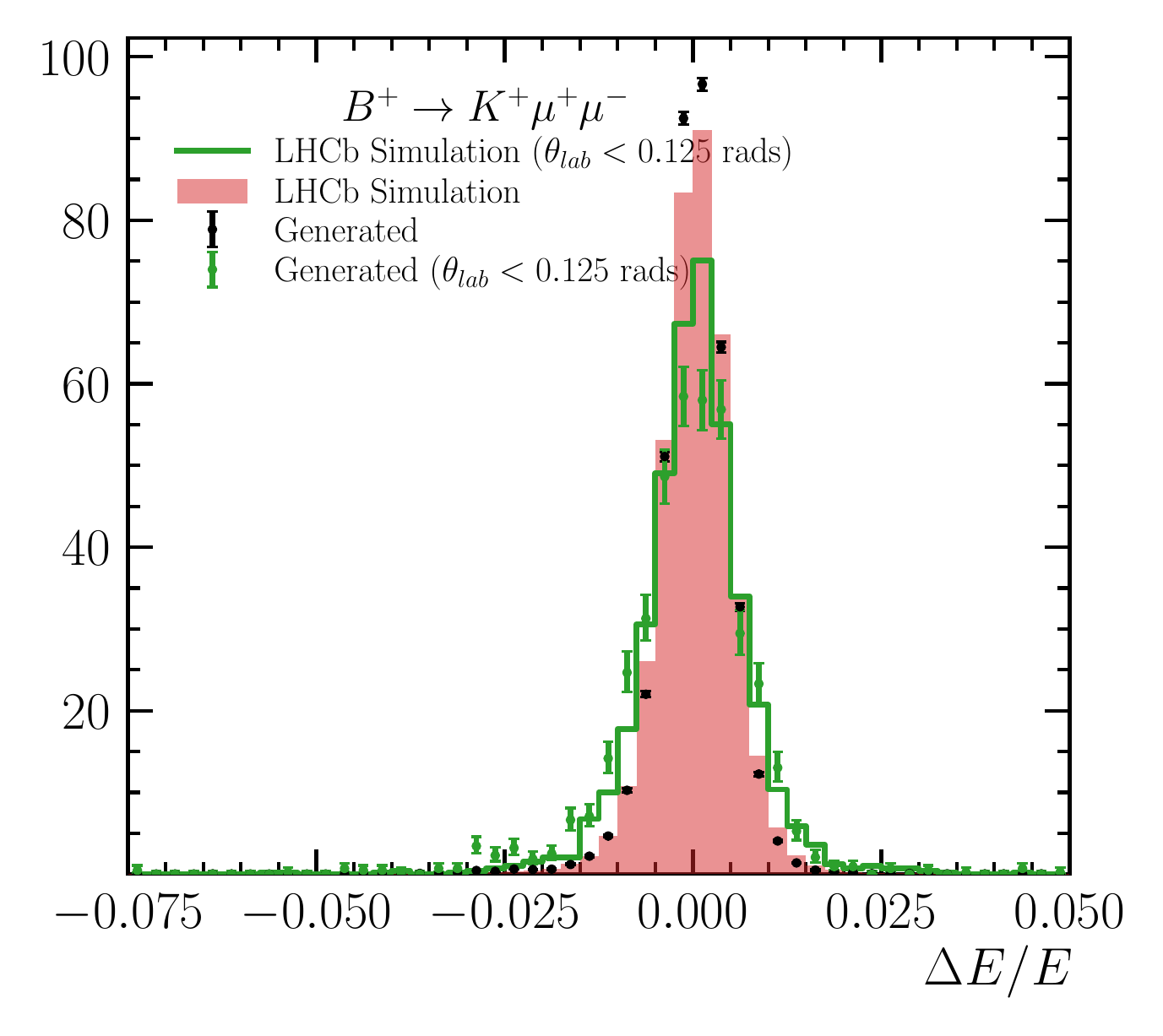}
            \vspace*{-0.5cm}
            \end{center}
            \caption{
            \small Distributions of $\Delta E/E$ for the $\ell^+$ in $B^+\to K^+\ell^+\ell^-$, where $\ell$ is $e$ (left) and $\mu$ (right). Both distributions are shown twice, firstly for all events, and then just for events where the angle between the leptons in the lab frame was smaller than $0.125$\;rads.}
            \label{fig:architecture:differential_smearing}
        \end{figure}
        
        The true momenta of each particle, as generated by \evtgen, are smeared to account for detector resolution by querying the smearing network with node types assigned according to true particle species. Figure~\ref{fig:performance:smearing} shows a comparison of distributions of $\Delta E/E$ between generated and simulated events, for different final state particles. Note the drastically different scale of the smearing for the electrons. Also provided in Fig.~\ref{fig:performance:smearing} are distributions of the reconstructed $B$-mass for the examples of $B^+\to K^+ e^+ e^-$ and $B^+\to K^+ \mu^+ \mu^-$. The individual distributions of $\Delta E/E$ appear well-modelled, and this translates into the full reconstructed masses, which show the behaviour is correctly captured out into the tails of the distribution. In each case, the network correctly picks up the required asymmetry in the smearing. 
    
        As described in Sec.~\ref{sec:architecture:smearing}, the advantage of smearing the momenta of particles on an event-by-event basis rather than independently is that the smearing can capture relationships between individual track smearings. Figure~\ref{fig:architecture:differential_smearing} shows the $\Delta E/E$ distributions for the $\ell^+$ in two examples of $B^+\to K^+\ell^+\ell^-$ decays, where $\ell=\;e,\;\mu$. Whilst the overall agreement is imperfect, for cases in which the angle between the leptons is small, the network correctly degrades the resolution. This is only possible because of the communication within the graph structure.
    
    \subsection{PID}\label{sec:performance:PID}

        \begin{figure}[h]
            \begin{center}
            \centering
            \includegraphics[page=54, width=0.48\textwidth,keepaspectratio]{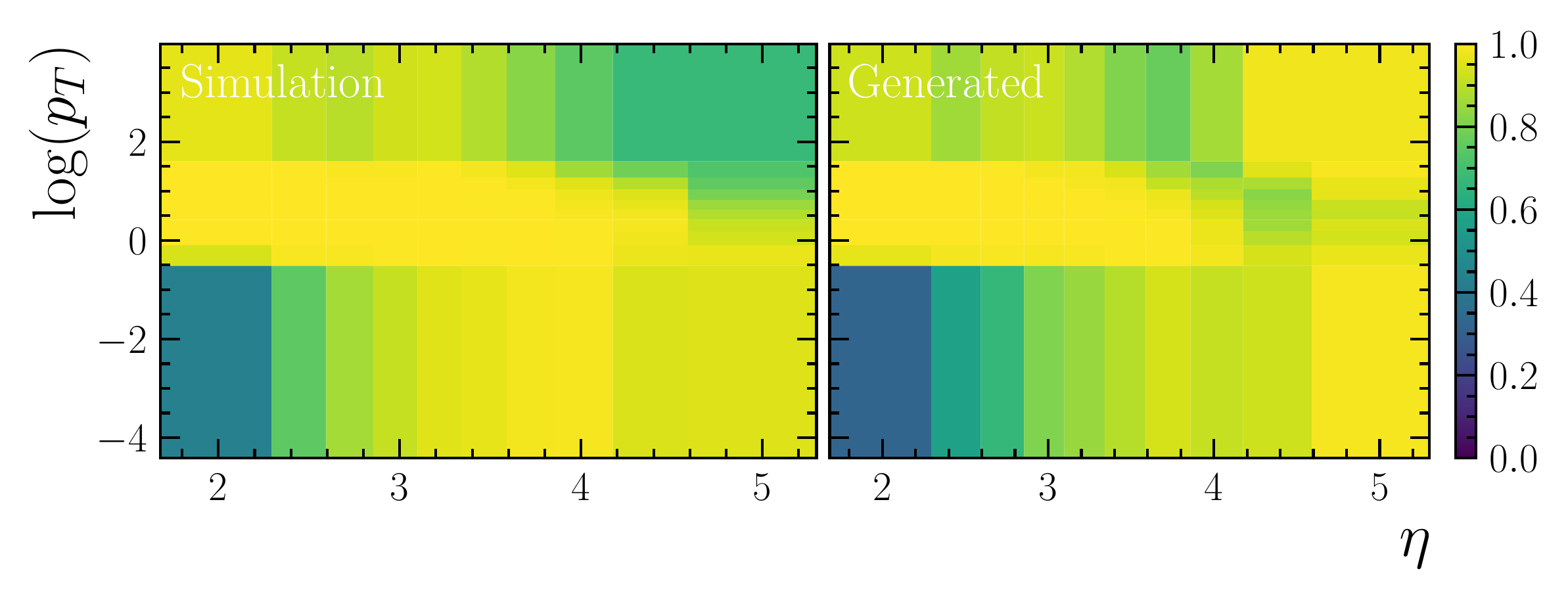}
            \includegraphics[page=59, width=0.48\textwidth,keepaspectratio]{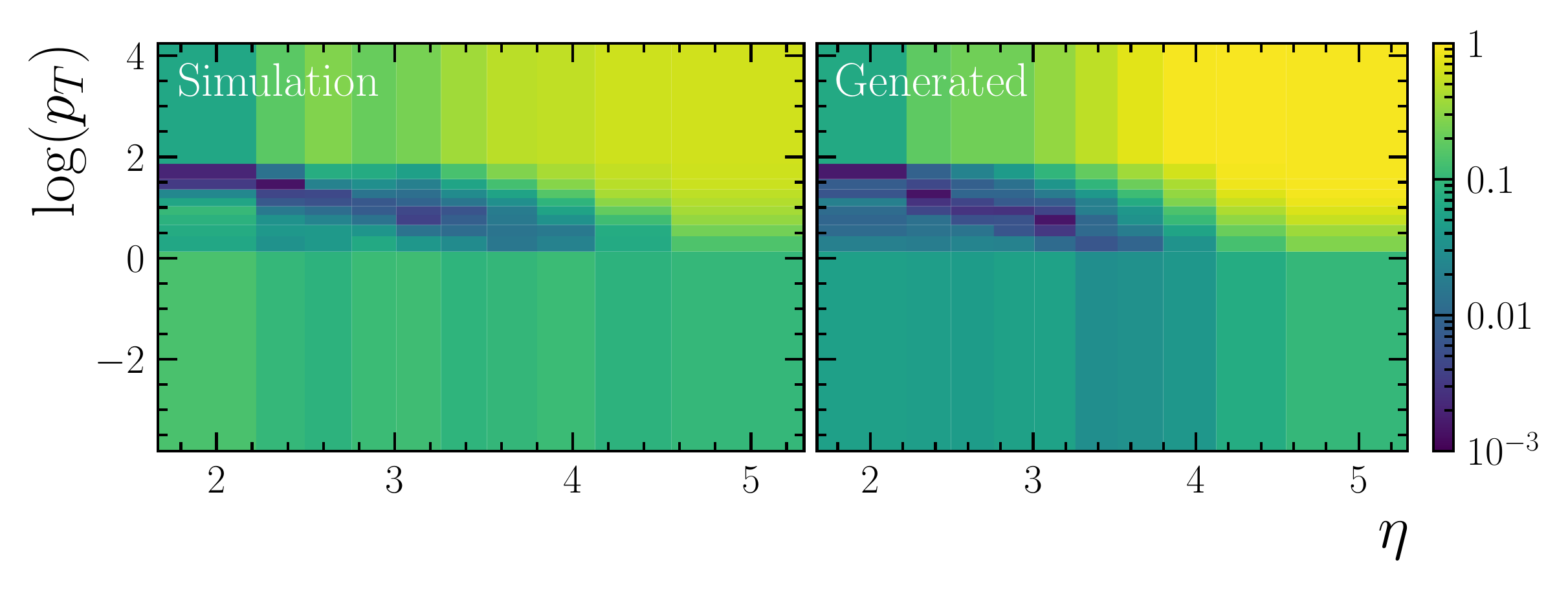}
            
            \includegraphics[page=65, width=0.48\textwidth,keepaspectratio]{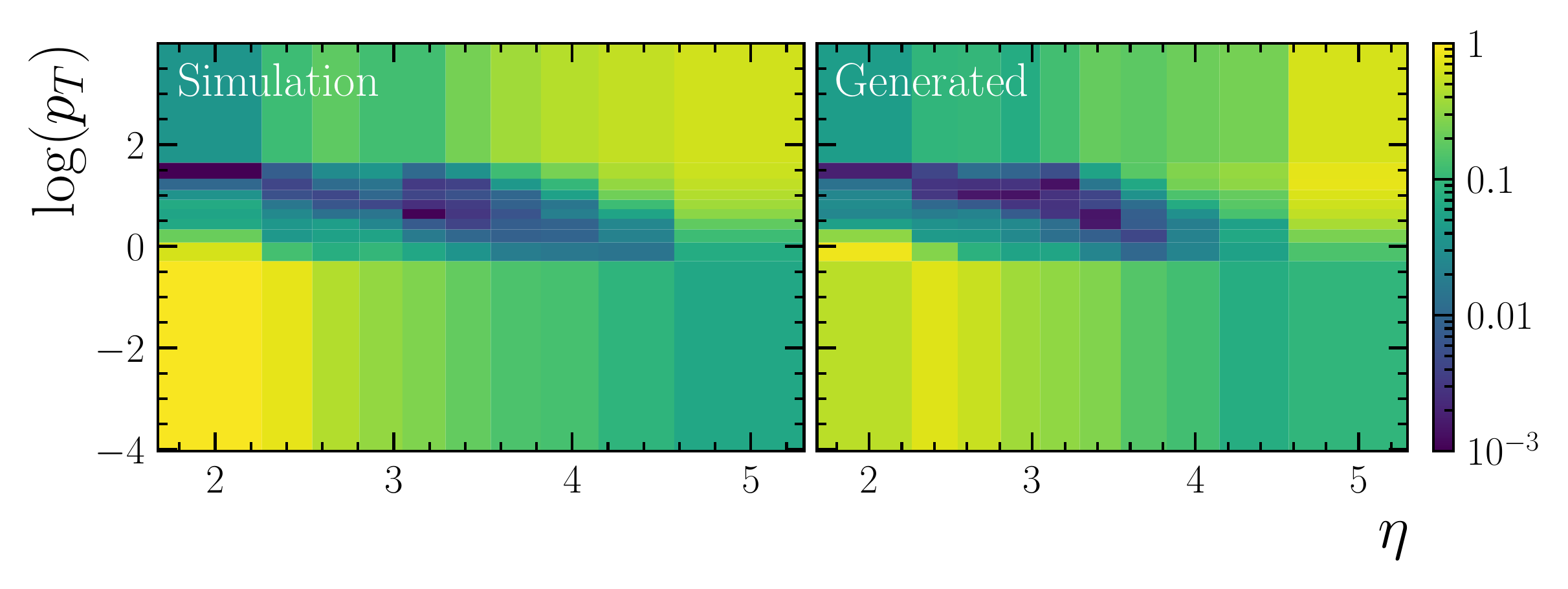}
            \includegraphics[page=72, width=0.48\textwidth,keepaspectratio]{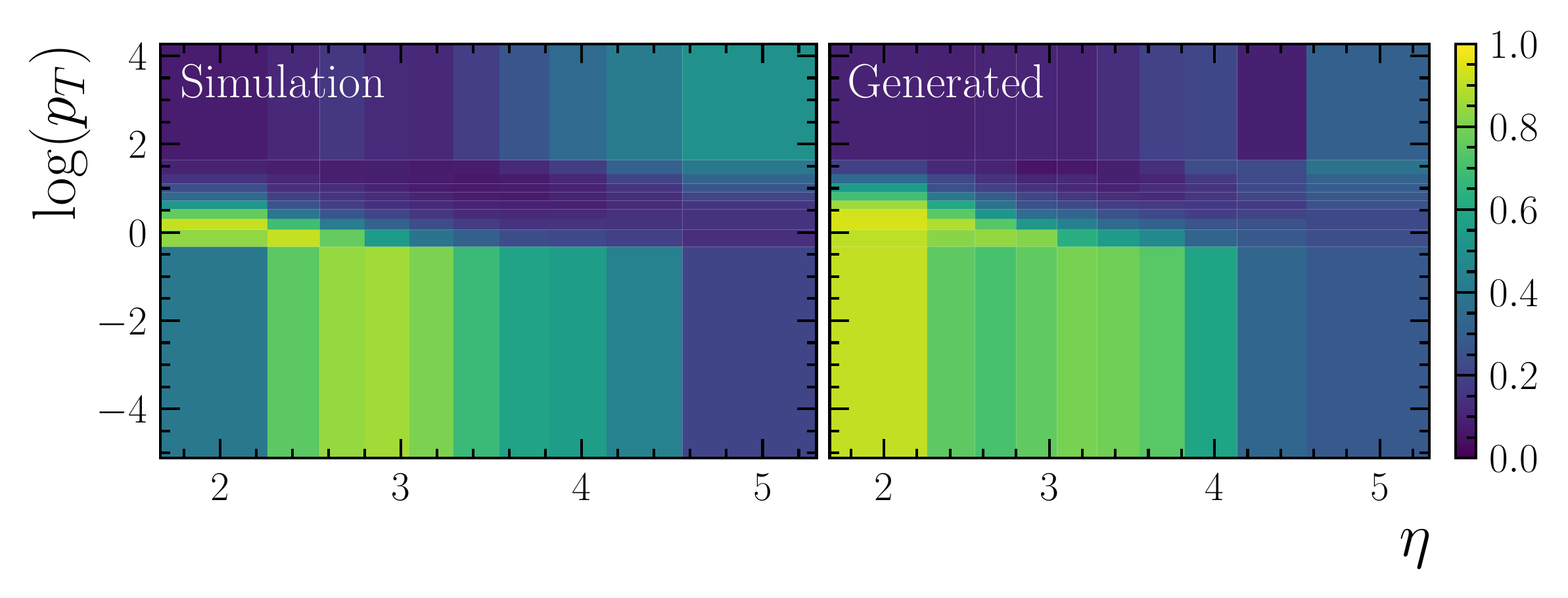}
            \vspace*{-0.5cm}
            \end{center}
            \caption{
            \small A comparison between simulated and generated maps of the efficiency of the cut "PIDK$>0$" across $\log(p_T)$ and $\eta$ for each true particle type. The binning scheme is adapted with quantiles to keep the number of events in each bin similar.}
            \label{fig:performance:PID2D}
        \end{figure}

        The efficiencies of an example cut of PIDK$>0$, as a function of $\log(p_T)$ and $\eta$ for supported particle types, are presented in Fig.~\ref{fig:performance:PID2D}. The network correctly models the shape and scale of these efficiencies. A full set of PID variables and similar efficiency maps for a variety of example cuts are presented in the appendix (see Figs.~\ref{fig:ack:PID_1D}~and~\ref{fig:ack:PID_2D}). In general, the model performs well, capturing the efficiency dependence across kinematic variables. Some variables, though, appear systematically mismodeled. In particular, the distribution of PIDe for electrons. We observe that this can be corrected by weighting events such that the distribution of $B$ kinematics more closely aligns with \pythia. Improving the generation of $B$ kinematics is listed as an area for future development in Sec.~\ref{sec:conclusion}. It is worth noting here that PID variables are one of the most poorly modelled parts of the LHCb full simulation, and it is therefore essential for analysts to weight even fully-simulated events to correct for this. Comprehensive data-driven tools have been developed within the collaboration to standardise this procedure~\cite{Anderlini:2016kco}. One is able to apply the same tools to weight samples generated with \rex, where any generated correlations with other variables will be included. 

    \subsection{Vertexing and reconstruction}\label{sec:performance:vtx}

\begin{figure}[h]
        \begin{subfigure}{\textwidth}
        \includegraphics[page=4, width=0.69\textwidth, keepaspectratio]{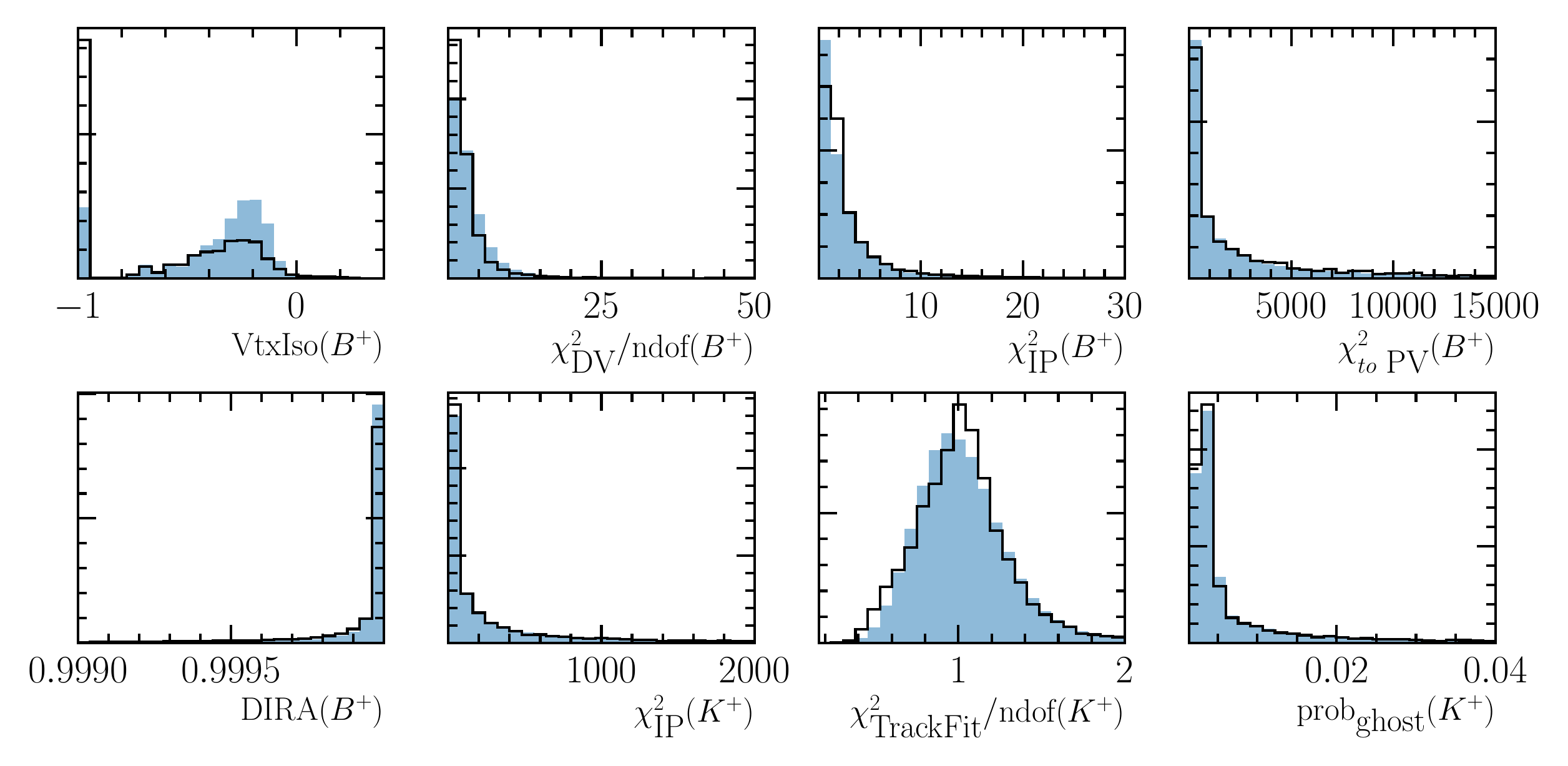}
        \caption{} \label{fig:1a}
        \end{subfigure}%
        
        \begin{subfigure}{\textwidth}
        \includegraphics[page=6, width=\textwidth, keepaspectratio]{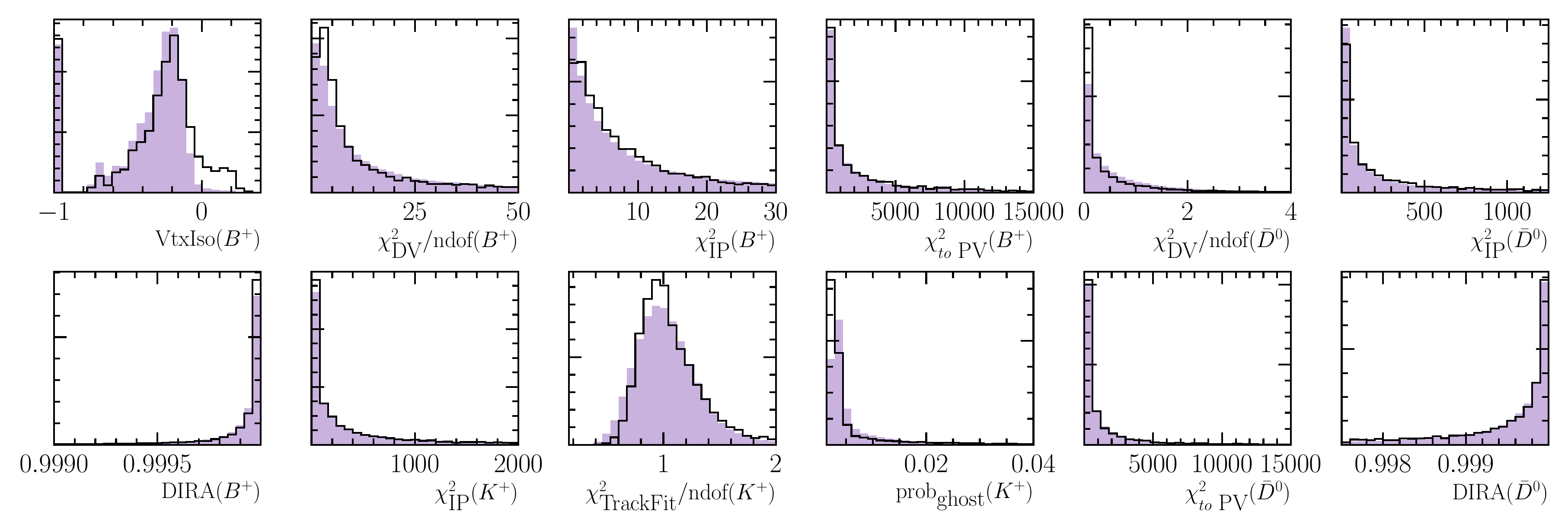}
        \caption{} \label{fig:1b}
        \end{subfigure}%
        \vspace*{-0.5cm}
        \caption{\small Distributions of generated vertexing and reconstruction quality variables for the channels $B^+\to K^+\mu^+\mu^-$ and $B^+\to \bar{D}^{0}(\to K^+e^-\bar{\nu}_e)\pi^+$. The full simulation is shown with the filled histograms, and the empty stepped histograms are derived from the output of \rex.
        }
        \label{fig:performance:vertex_distributions}
        \end{figure}

\begin{figure}[h]
            \begin{center}
            \centering
            \includegraphics[width=\textwidth,keepaspectratio]{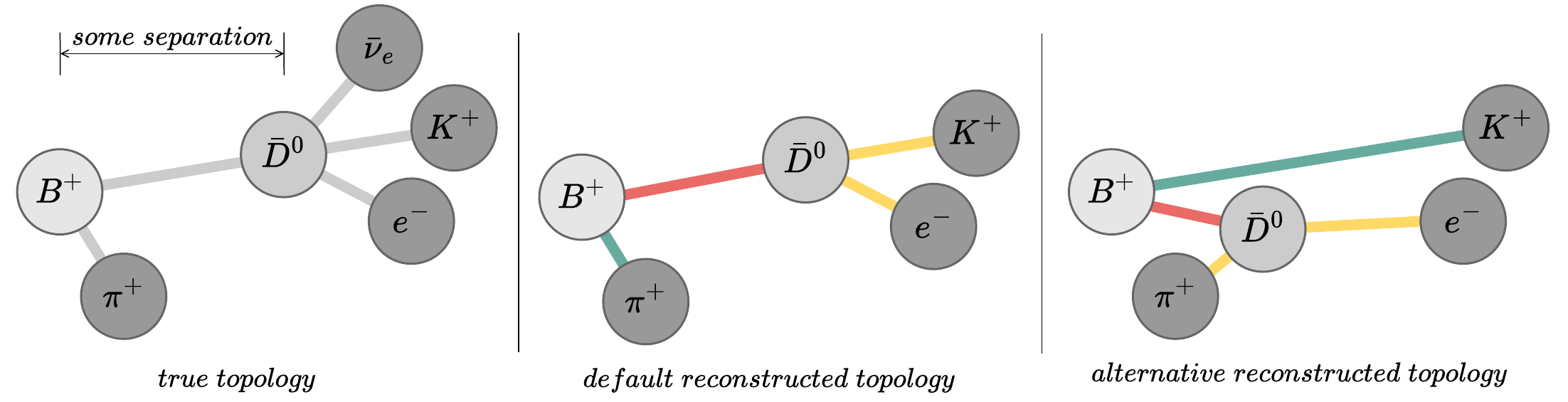}
            \vspace*{-0.5cm}
            \end{center}
            \caption{
            \small Illustrations of the generated and reconstructed topologies referred to in the text.}
            \label{fig:performance:alternativeD}
        \end{figure}

        Distributions of vertexing and reconstruction quality variables are presented in Fig.~\ref{fig:performance:vertex_distributions} for two example channels, $B^+\to K^+\mu^+\mu^-$ and $B^+\to \bar{D}^{0}(\to K^+e^-\bar{\nu}_e)\pi^+$. 
        Despite both having three final-state charged tracks, these channels have significantly different decay topologies, being fully- and partially-reconstructed decays, respectively. The final state particles in the $B^+\to K^+\mu^+\mu^-$ decays all originate from the $B^+$ decay vertex; this is not the case for the $B^+\to \bar{D}^{0}(\to K^+e^-\bar{\nu}_e)\pi^+$ decay in which the intermediate state $\bar{D}^{0}$ has a lifetime long enough to create some physical separation, meaning the $K^+$ and $e^-$ originate from a vertex displaced from that of the $B^+$.
        This is accounted for in the graph structure parsed within which the $\bar{D}^{0}$ intermediate is reconstructed. Consequently, the variables associated with this intermediate state are additionally presented for this channel in Fig.~\ref{fig:performance:vertex_distributions}.
        Illustrations of the generated and reconstructed topologies of $B^+\to \bar{D}^{0}(\to K^+e^-\bar{\nu}_e)\pi^+$ are shown in Fig.~\ref{fig:performance:alternativeD}, where, as the neutrino is not detected, it doesn't appear in the reconstructed topology.  
        Note that the variables presented in Fig.~\ref{fig:performance:vertex_distributions} only depend on the choice of reconstructed topology; they are unaffected by any potential track misidentifications and incorrect mass hypotheses. Several variables appear imperfectly modelled, which is likely due to instabilities during the convergence of the network. As mentioned in Sec.~\ref{sec:performance:uncertainties} and Sec.~\ref{sec:conclusion}, this is an area for future development.  

        An accurate description of these variables is achieved, with the model successfully capturing differences in kinematic and topological features between the two example decay modes. The clearest examples of differences are in the distributions of the vertex fit quality $\chi^2_{\textrm{DV}}/\textrm{ndof}$ and impact parameter significance $\chi^2_{\textrm{IP}}$ of the parent particle, where the distributions in the partially reconstructed case are significantly broader. This broadening is attributed to degraded vertex resolution caused by momentum lost in the undetected neutrino and the spatial displacement due to the flight distance of the intermediate $\bar{D}^{0}$.
        
\begin{figure}[h]
            \begin{center}
            \centering
            \begin{minipage}{0.48\textwidth}
            \includegraphics[page=4, width=\textwidth, keepaspectratio]{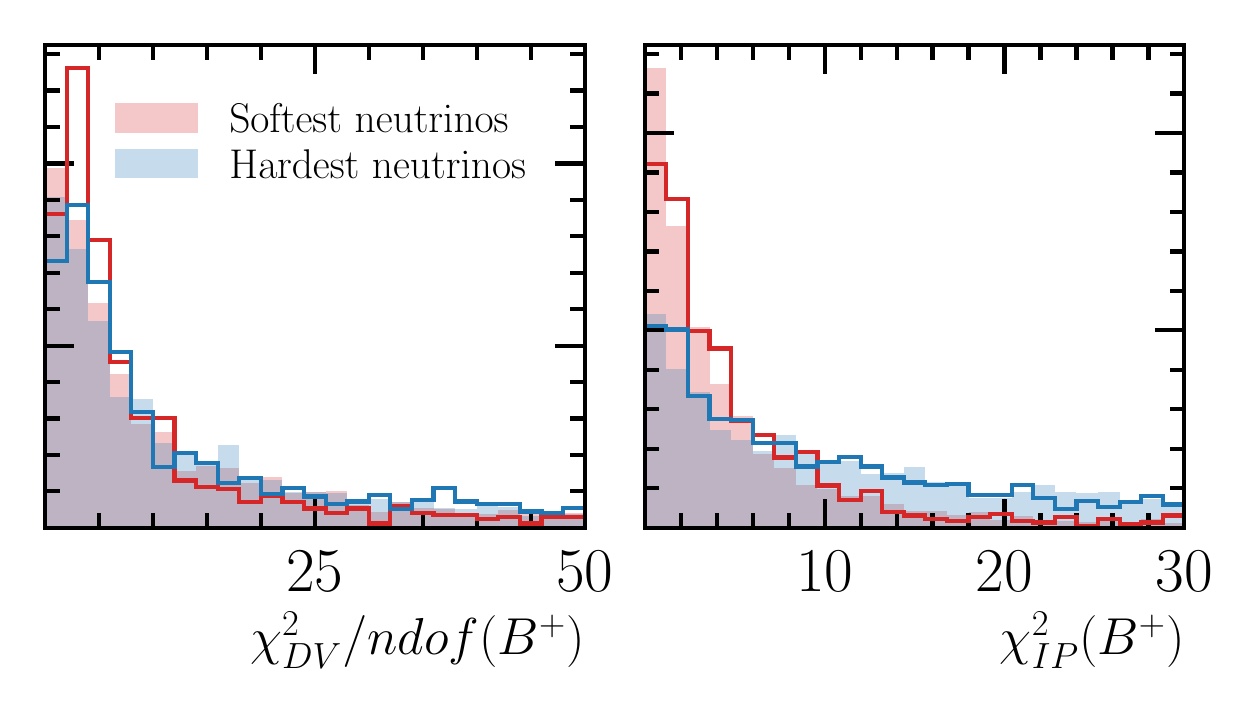}
            \end{minipage}
            \begin{minipage}{0.48\textwidth}
            \includegraphics[page=5, width=\textwidth, keepaspectratio]{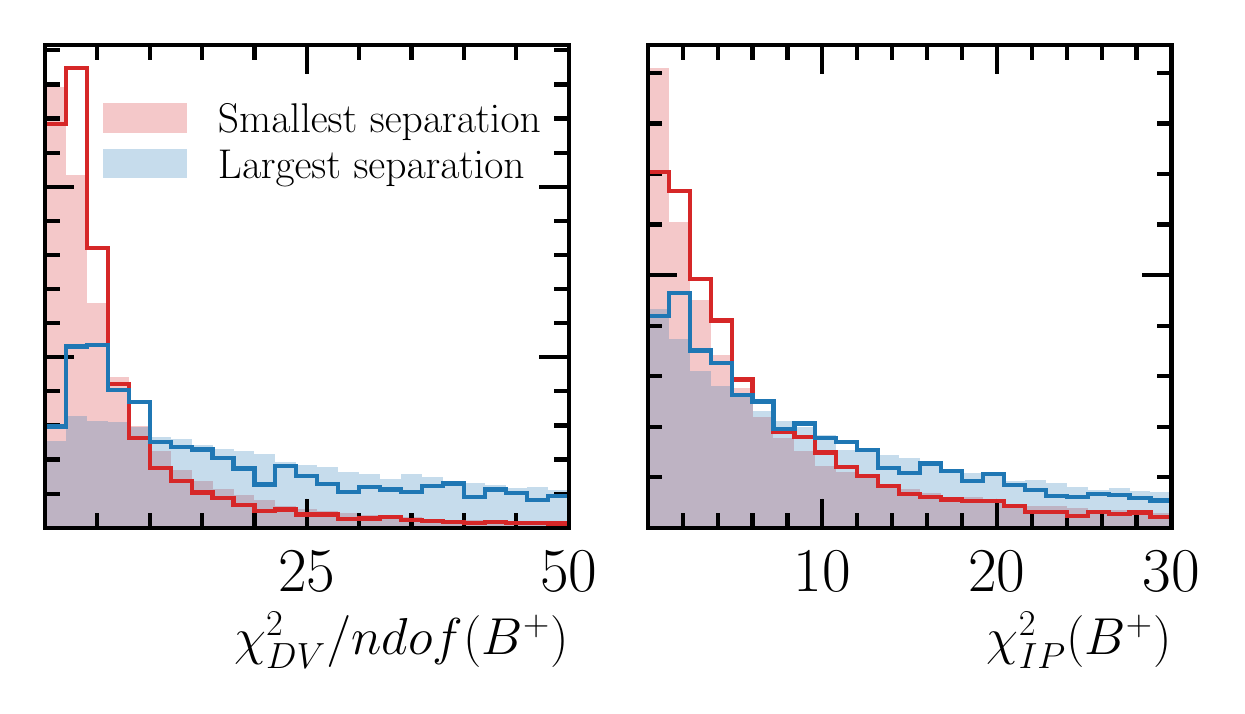}
            \end{minipage}
            \vspace*{-0.5cm}
            \end{center}
            \caption{\small Distributions of variables for the reconstructed $B^+$ in $B^+\to \bar{D}^{0}(\to K^+e^-\bar{\nu}_e)\pi^+$ events in extreme cases of $\bar{\nu}_e$ momentum and of $\bar{D}^{0}$ flight distance. The full simulation is shown with the filled histograms, and the empty stepped histograms are derived from the output of \rex.
            }
            \label{fig:performance:cut_nu_mom}
        \end{figure}

        To further explore the origin of the degradation in the $B^+\to \bar{D}^{0}(\to K^+e^-\bar{\nu}_e)\pi^+$ case, we partition the data into subsets containing the most extreme examples from the sample. In Fig.~\ref{fig:performance:cut_nu_mom}, distributions of the same $\chi^2_{\textrm{DV}}/\textrm{ndof}$ and  $\chi^2_{\textrm{IP}}$ variables are shown for events with the hardest (2.5th percentile) and softest neutrinos. Additionally, the sample is divided according to the largest and smallest $\bar{D}^{0}$ flight distances, which is also shown in Fig.~\ref{fig:performance:cut_nu_mom}. In each case, the network correctly models the impact on these variables.

\begin{figure}[h]
            \begin{center}
            \centering
            \includegraphics[page=2, width=\textwidth, keepaspectratio]{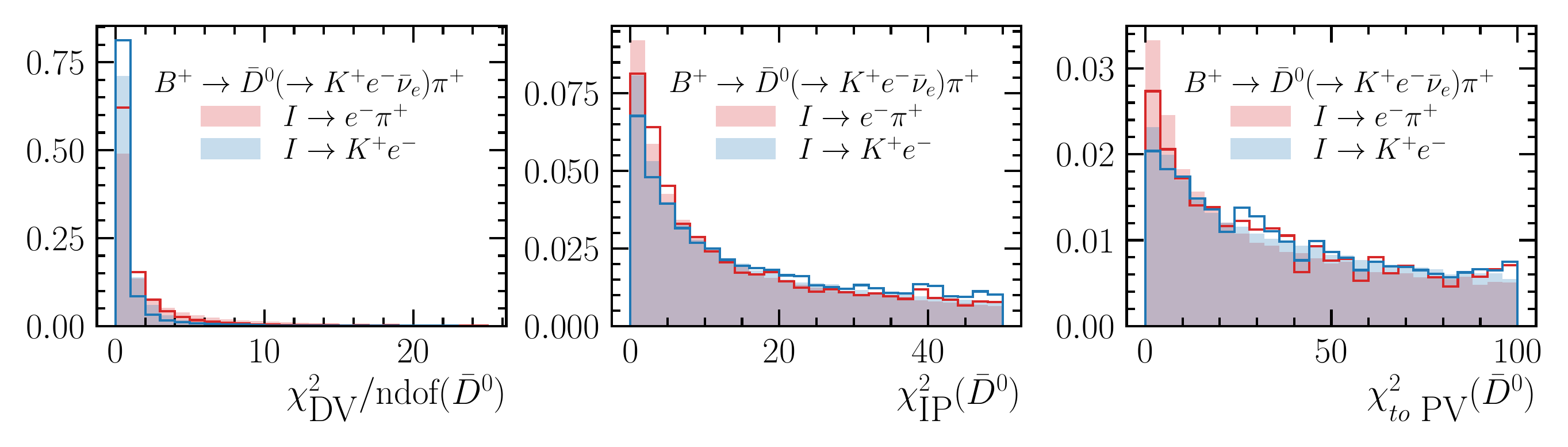}
            \vspace*{-0.5cm}
            \end{center}
            \caption{\small Distributions of generated vertexing quality variables assigned to the reconstructed intermediate of $B^+\to \bar{D}^{0}(\to K^+e^-\bar{\nu}_e)\pi^+$ in the correctly reconstructed case (blue) and swapped case (red). The full simulation is shown with the filled histograms, and the empty stepped histograms are derived from the output of \rex.}
            \label{fig:performance:intermediate_topology}
        \end{figure}

        In the examples shown up to this point, the $B^+\to \bar{D}^{0}(\to K^+e^-\bar{\nu}_e)\pi^+$ decays are correctly reconstructed with the $K^+$ and $e^-$ tracks forming the intermediate candidate. We can look at how the generated variables change in the case that one changes the reconstruction hypothesis and reconstructs the same events with the $e^-$ and $\pi^+$ tracks combined in the reconstruction to form the intermediate candidate, illustrated as the alternative reconstruction topology in Fig.~\ref{fig:performance:alternativeD}. In Fig.~\ref{fig:performance:intermediate_topology}, we see that the effect on the distributions of these variables is again modelled correctly. 
        
\begin{figure}[tb]
            \begin{center}
            \centering
            \includegraphics[page=3, width=0.32\textwidth,keepaspectratio]{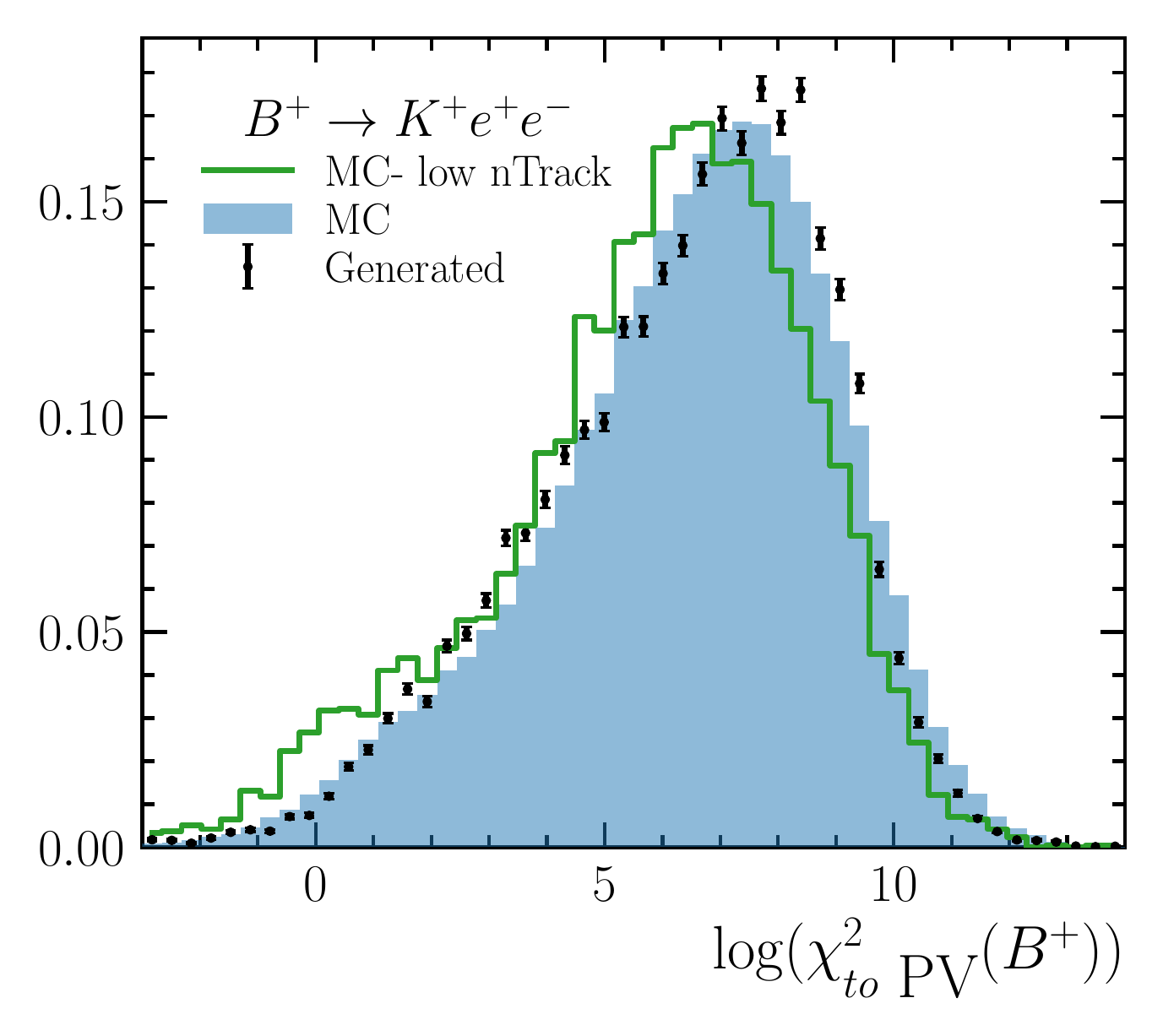}
            \includegraphics[page=8, width=0.32\textwidth,keepaspectratio]{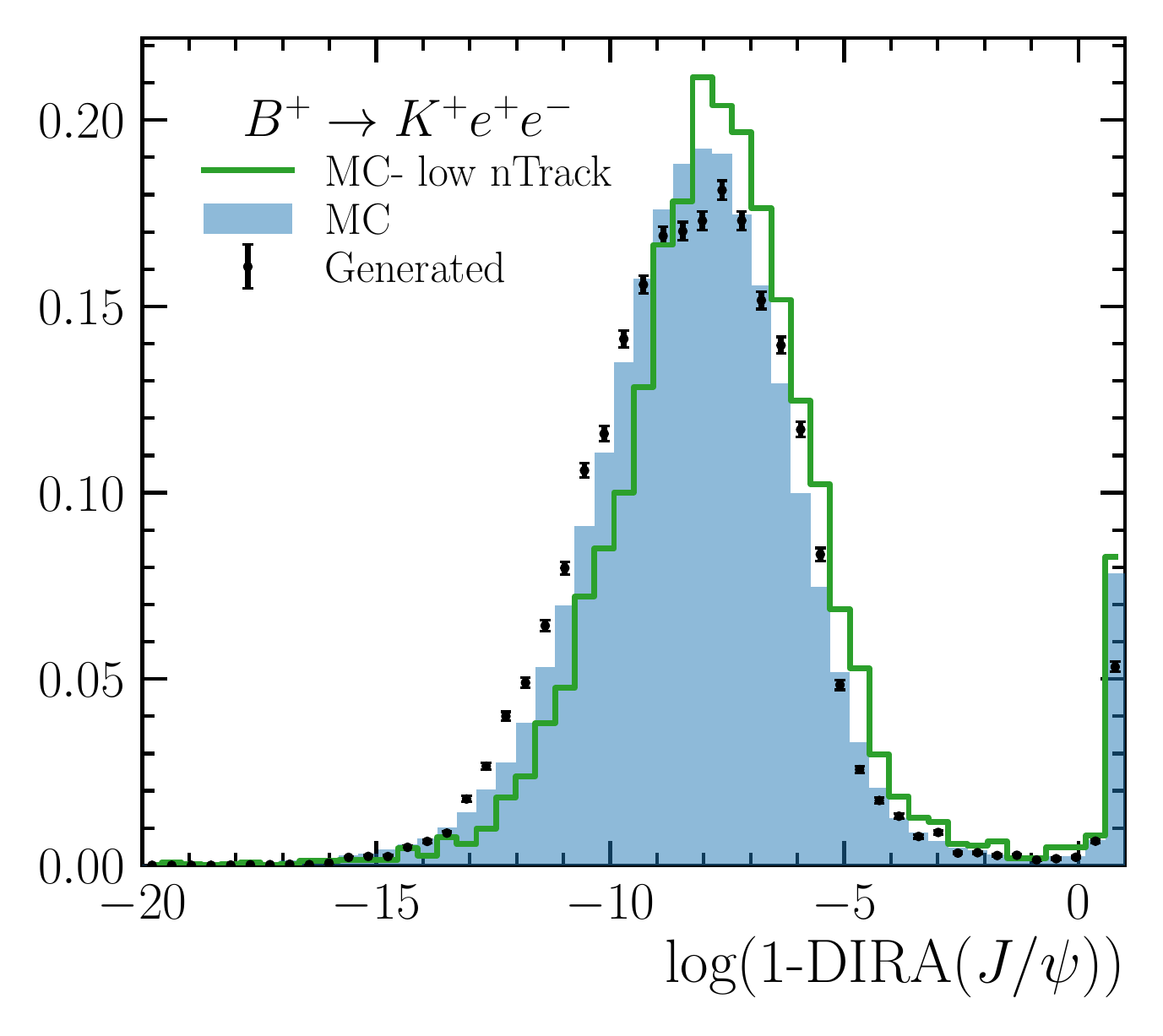}
            \includegraphics[page=9, width=0.32\textwidth,keepaspectratio]{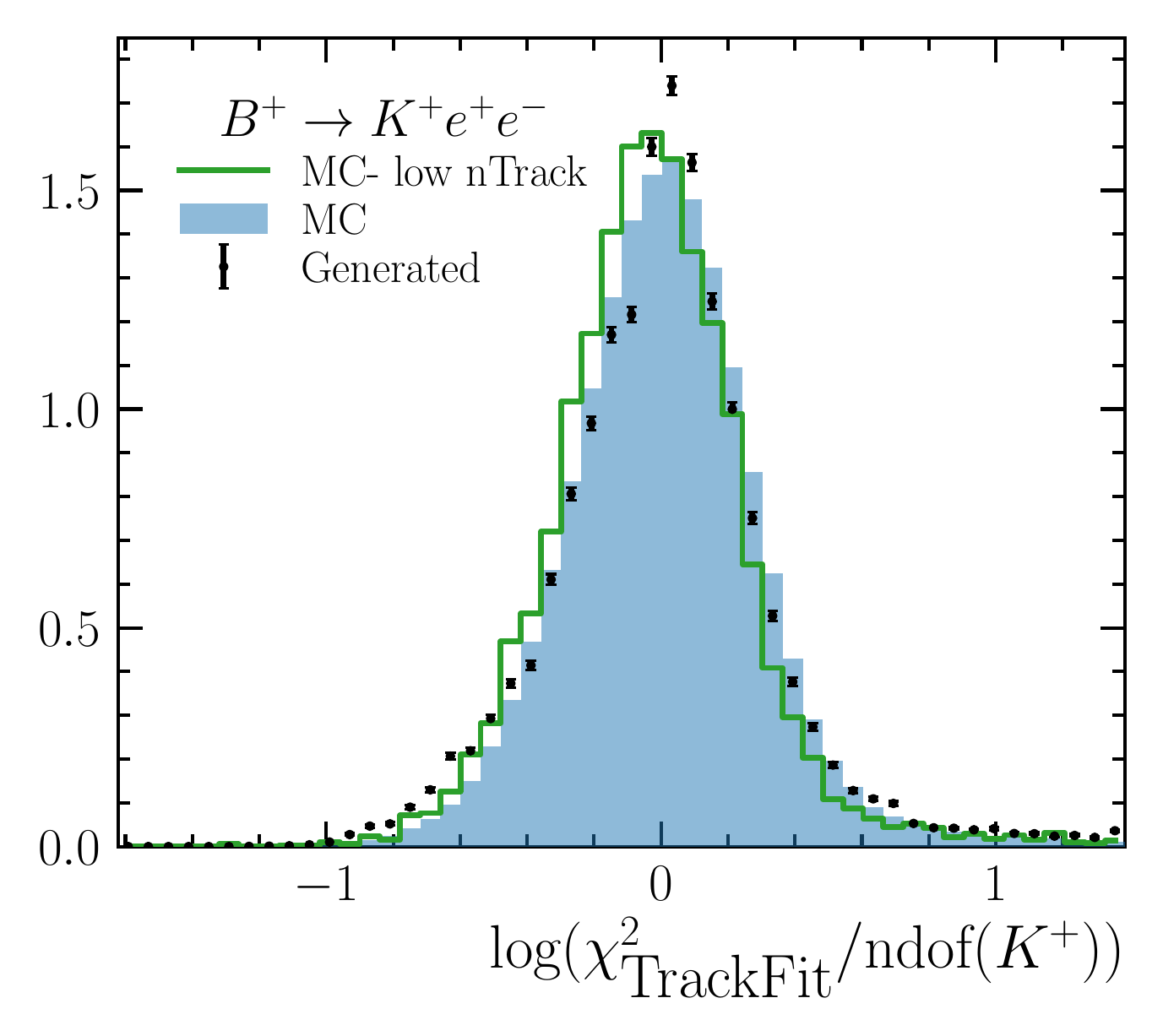}\\
            \includegraphics[page=30, width=0.32\textwidth,keepaspectratio]{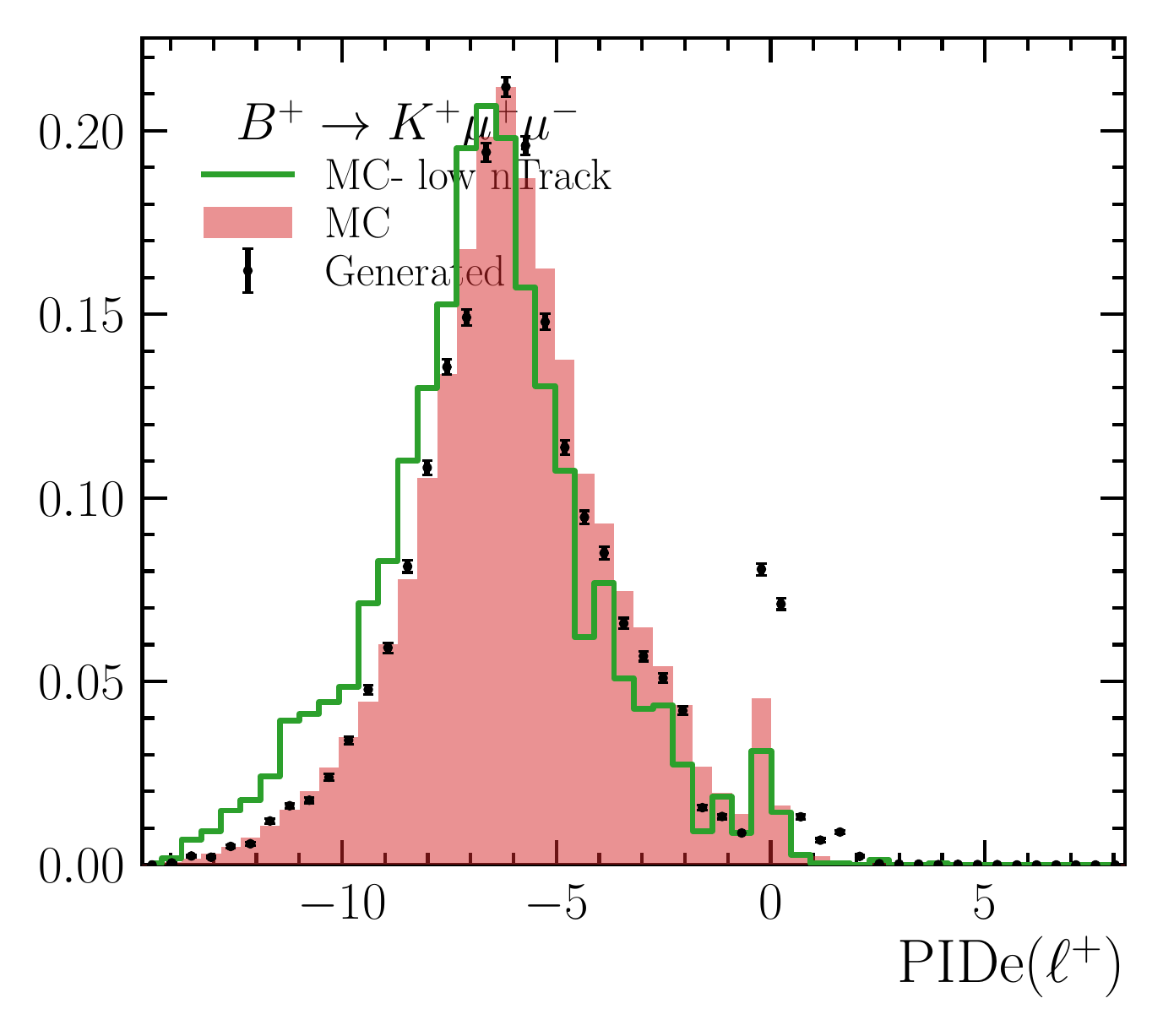}
            \includegraphics[page=14, width=0.32\textwidth,keepaspectratio]{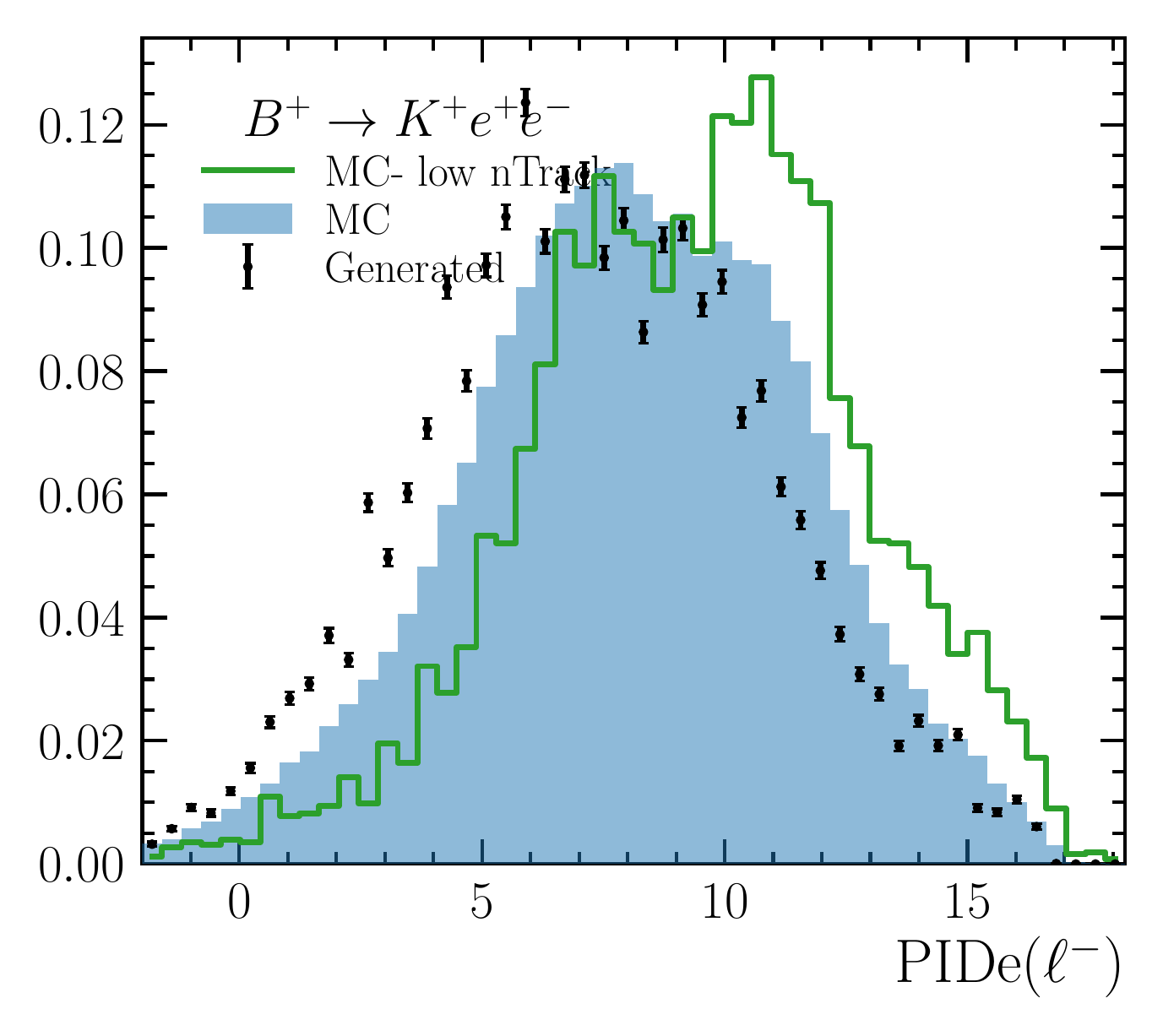}
            \includegraphics[page=34, width=0.32\textwidth,keepaspectratio]{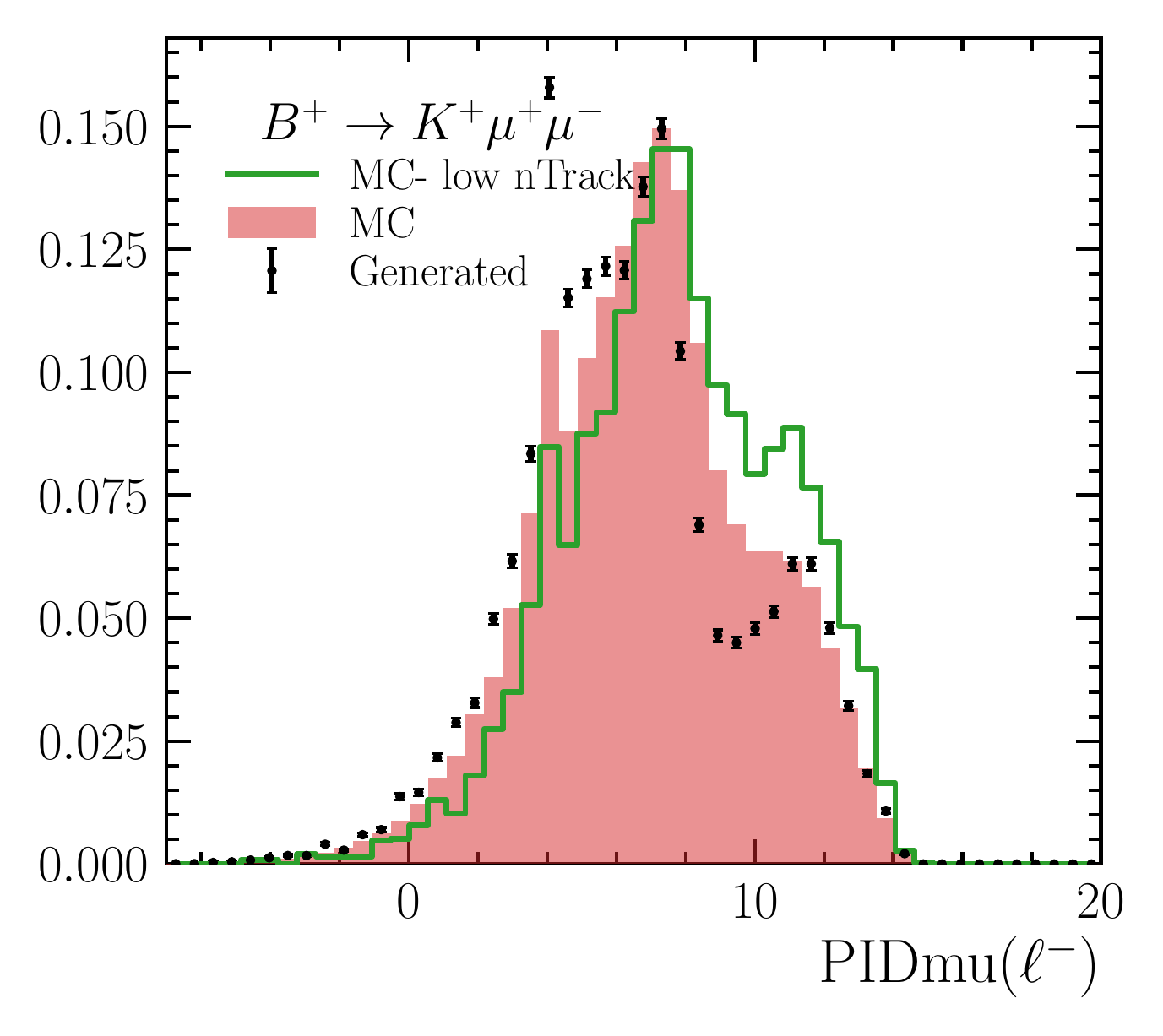}
            \vspace*{-0.5cm}
            \end{center}
            \caption{
            \small A comparison between distributions of various reconstruction variables in simulated and generated $B^+\to K^+ e^+ e^-$ and $B^+\to K^+ \mu^+ \mu^-$ samples reconstructed with the hypothesis of $B^+\to K^+J/\psi(\to \ell^+\ell^-)$. The simulated sample is additionally shown with a cut on the occupancy of the event.}
            \label{fig:performance:occupancy}
        \end{figure}

        As discussed in Sec.~\ref{sec:fast_framework}, we avoid the generation of the rest of the event in the interest of speed and instead include the impact from the surrounding event activity by training on the full simulation in which the effect is present. With this approach, the model effectively samples the size of the effect to apply such that it is consistent with the simulated conditions of the decay under study. There is, though, no explicit conditional control of the size of the effect, but it is nevertheless included in the output.
        While the impact from the rest of the event varies from event to event, it is generally correlated with event occupancy — a proxy for which is the number of reconstructed tracks, with vertexing quality improving in events with fewer reconstructed tracks. We demonstrate in Fig.\ref{fig:performance:occupancy} that the generated distributions follow the full distributions rather than the idealised low occupancy distributions. In future versions, the network should include occupancy variables in the output so that any correlations with other variables are modelled and in the case that an analyst wishes to include cuts on occupancy in a selection, the events that pass have the correct properties. 
    
        \subsection{Example analysis selection chain}\label{sec:performance:analysis}

            To assess the performance of \rex in a realistic analysis context, we emulate the full sequence of an analysis selection criteria on both signal and a range of background channels. For this study, we look at the analysis of the $B^+\to K^+ e^+ e^-$ decay. This channel was selected as its associated backgrounds represent a broad range of topologies commonly encountered within other LHCb analyses; these backgrounds also include tracks from all particle species supported within this version of \rex. Channels with electrons are particularly useful test cases, as electrons are both underrepresented in the training sample and exhibit the most distinctive reconstruction characteristics.
            
            All signal and background candidates are reconstructed under the same decay hypothesis of $B^+\to K^+ I(\to e^+ e^-)$, where $I$ denotes a fictitious intermediate state representing the di-lepton system. The squared invariant mass of the intermediate state $I$ defines the variable $q^2$. The following background processes are considered:
            \begin{itemize}
                \item $B^+\to K^+J/\psi(\to e^+e^-)$: This channel closely resembles the signal but is confined to a narrow $q^2$ region corresponding to the invariant mass of the $J/\psi$.
                \item $B^0\to K^{*0}(\to K^+\pi^-)e^+e^-$: This channel contributes as a partially reconstructed background in cases where the $\pi^-$ is not detected. 
                \item $B^+\to \bar{D}^{0}(\to K^+e^-\bar{\nu}_e)\pi^+_{\to e^+}$: This is another partially reconstructed channel, due to the undetected neutrino; however, this channel presents several additional challenges for this tool. As discussed in Sec.~\ref{sec:performance:vtx}, the finite lifetime of the $\bar{D}^0$ meson results in a spatial separation between the decay vertices of the final-state particles, which can distort vertex-related observables. Furthermore, the true intermediate particle is the $\bar{D}^0$, formed from the $K^+$ and $e^-$, whereas the reconstruction assumes the intermediate state $I$ is composed of the $e^-$ and $\pi^+$, introducing additional effects which must be accounted for in order to correctly reproduce the characteristics of fully-simulated events. Finally, this background includes a particle misidentification, where the $\pi^+$ is incorrectly reconstructed as an $e^+$.
                \item $B^+\to K^+\mu^+_{\to e^+}\mu^-_{\to e^-}$: This background shares the same topology as the signal but involves two misidentified muons reconstructed as electrons. This background is well controlled by PID requirements; however, the performance of the tool can still be studied here before these selection requirements are made.  
            \end{itemize}
            
\begin{figure}[h]
                \begin{center}
                \centering
                \begin{minipage}{0.5\textwidth}
                \includegraphics[page=6, width=\textwidth, keepaspectratio]{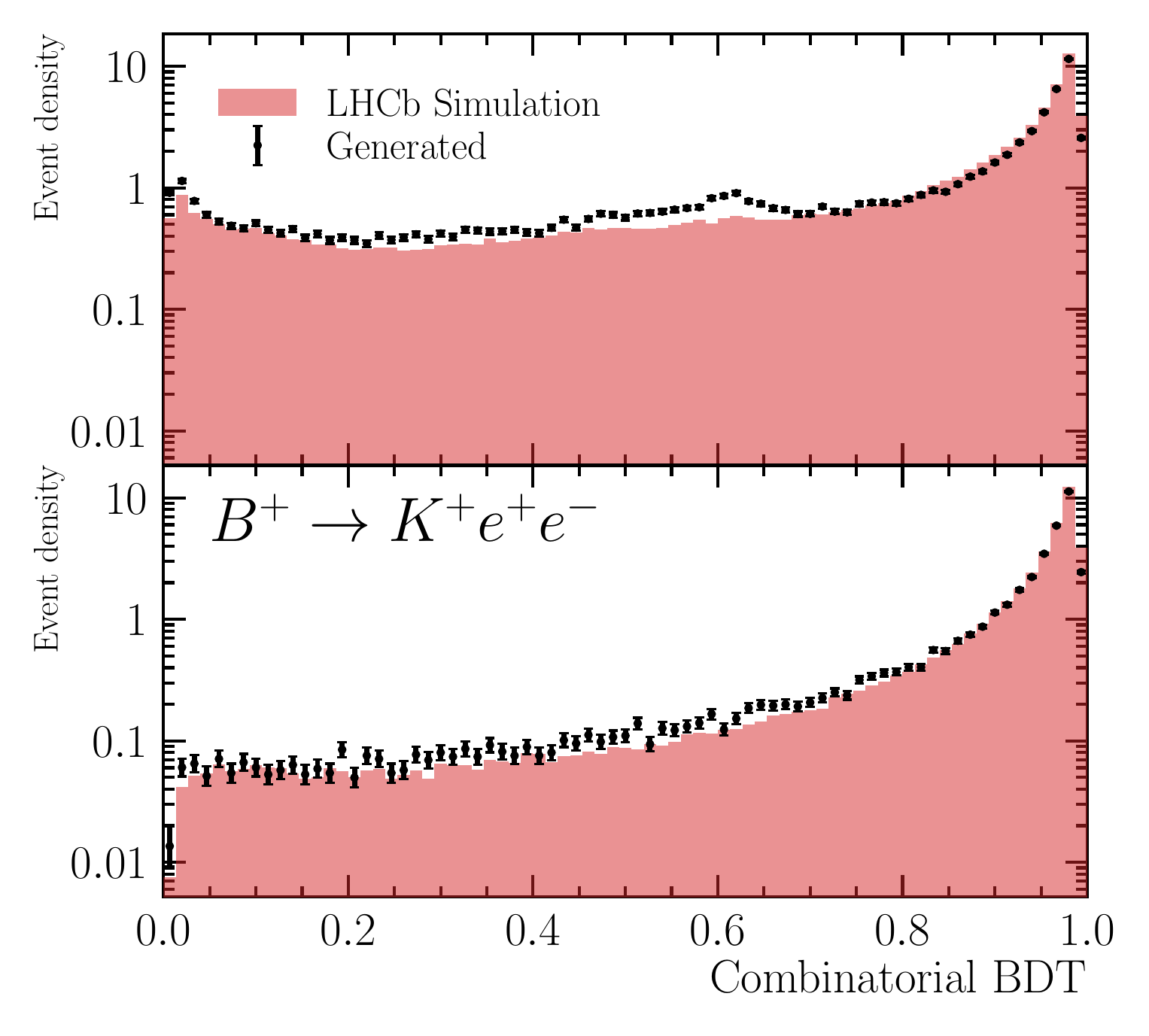}
                \end{minipage}
                \begin{minipage}{0.24\textwidth}
                \includegraphics[page=13, height=0.85\textwidth, keepaspectratio]{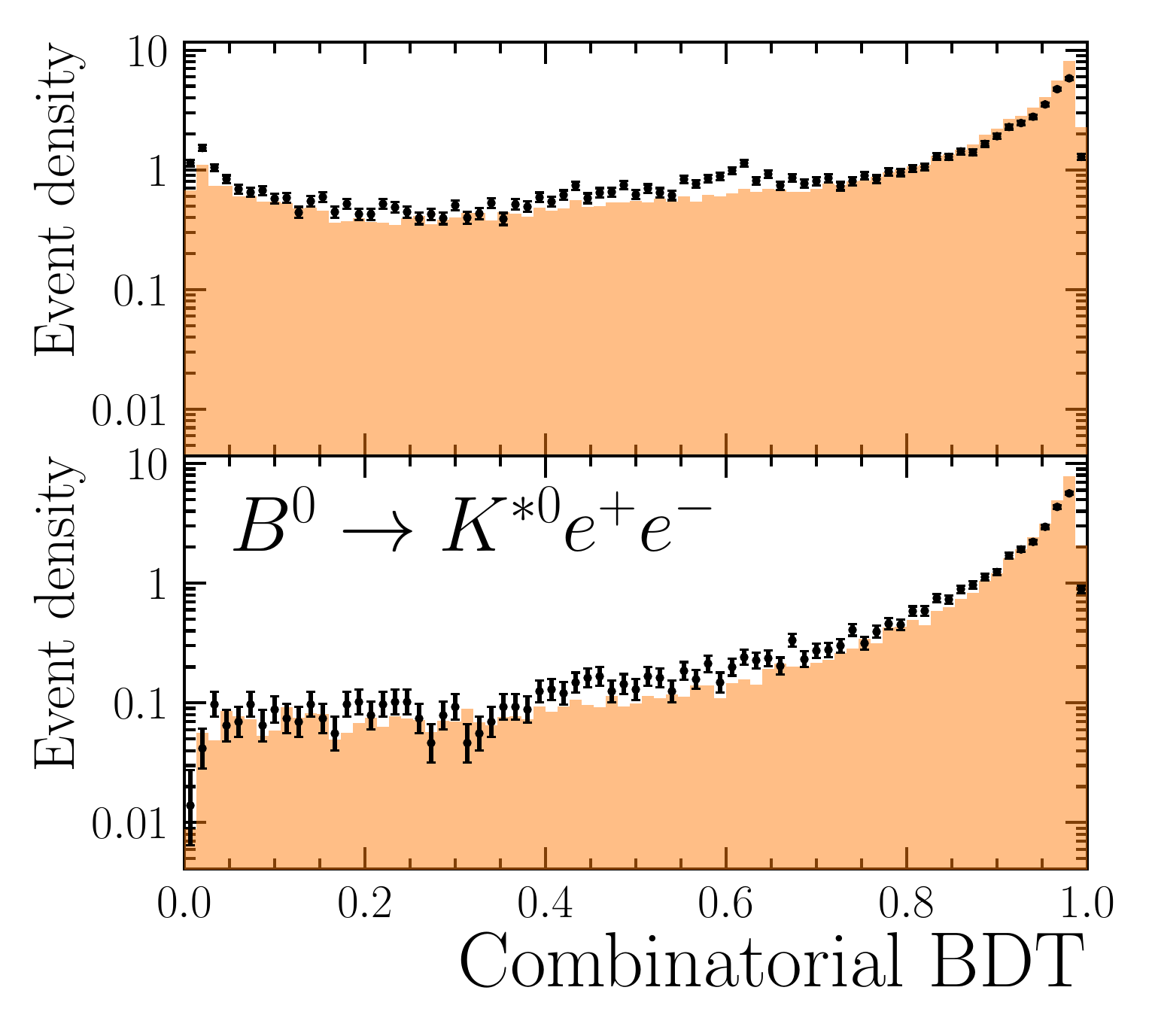}
                \includegraphics[page=20, height=0.85\textwidth, keepaspectratio]{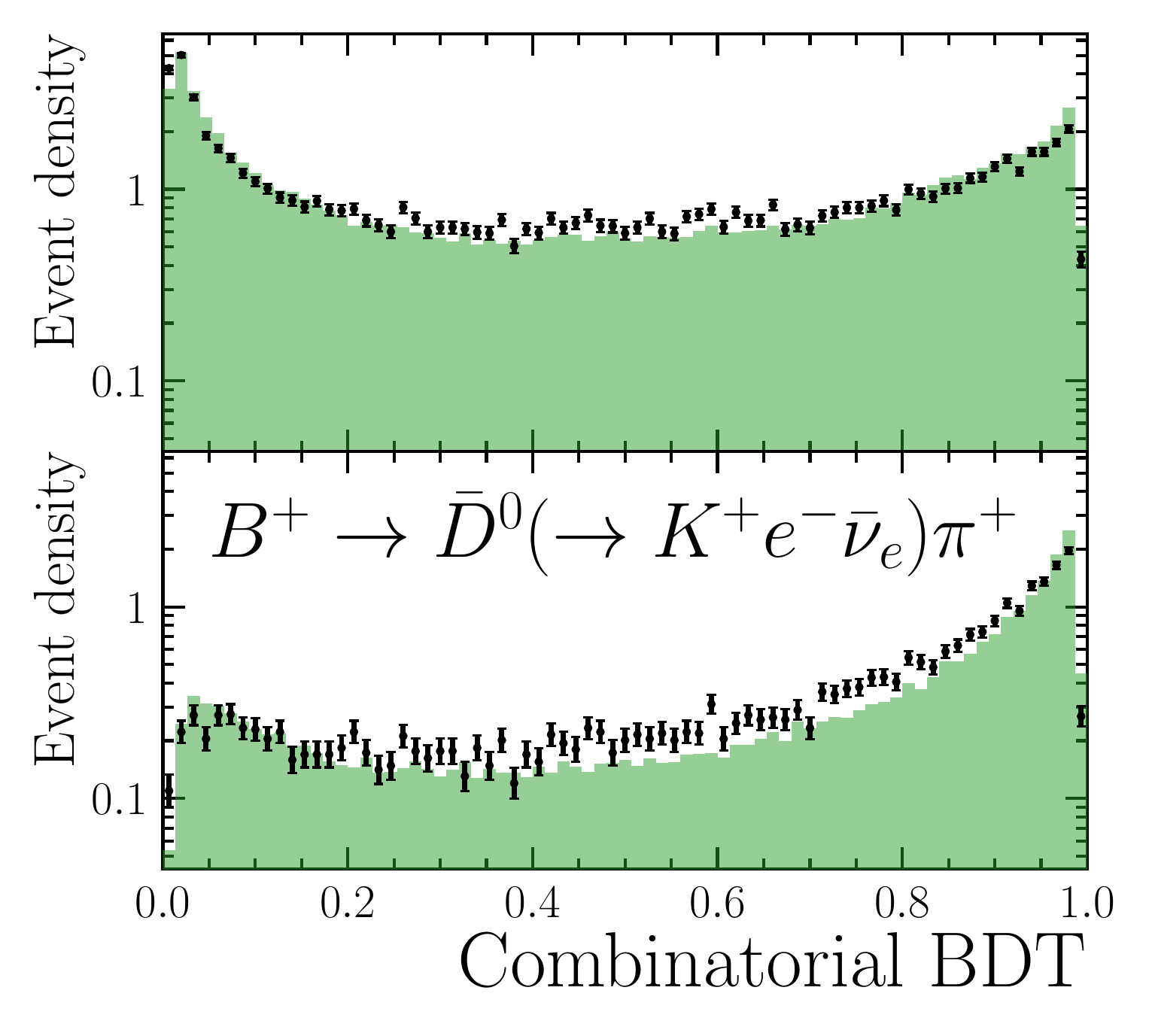}
                \end{minipage}
                \begin{minipage}{0.24\textwidth}
                \includegraphics[page=27, height=0.85\textwidth, keepaspectratio]{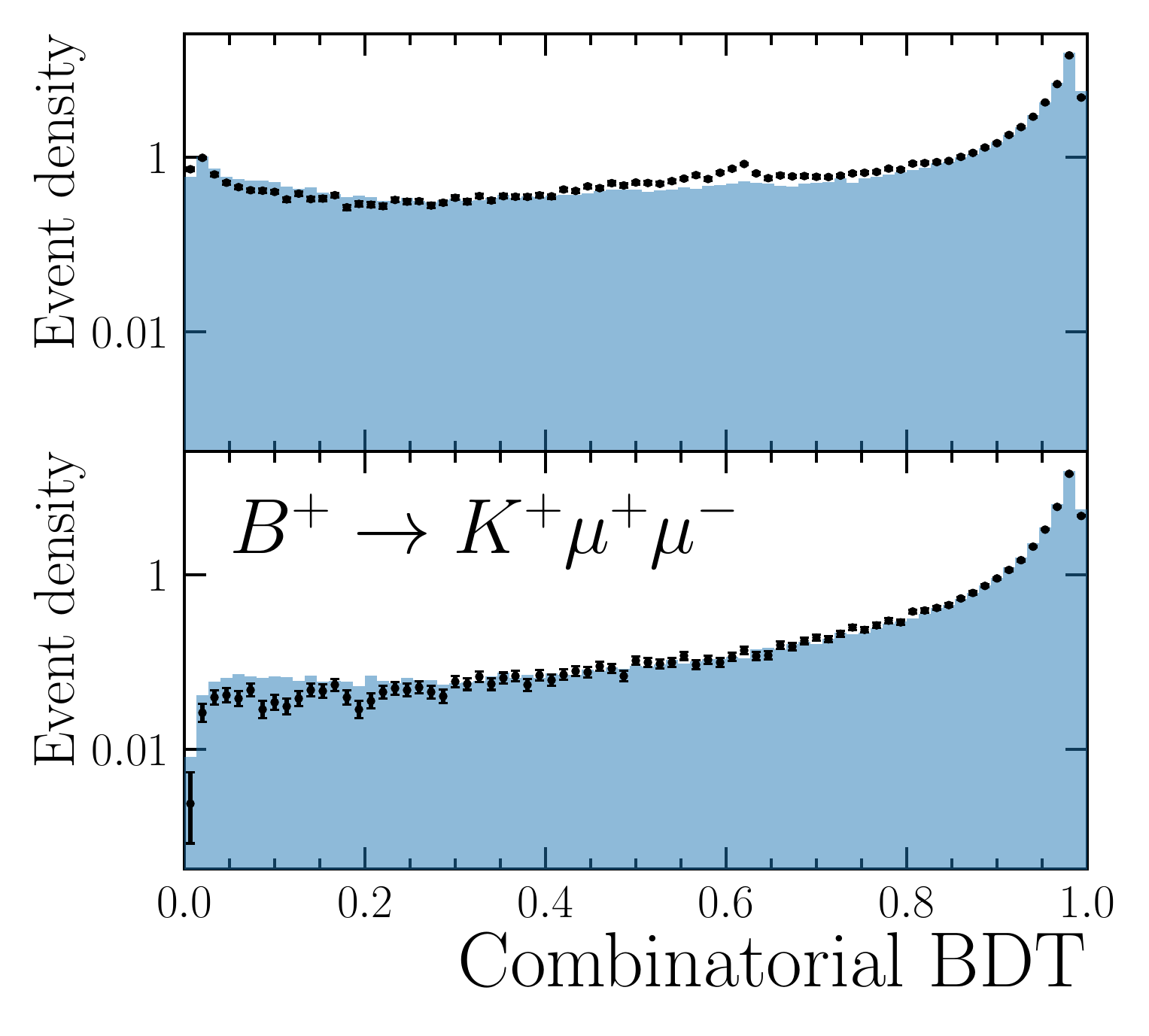}
                \includegraphics[page=34, height=0.85\textwidth, keepaspectratio]{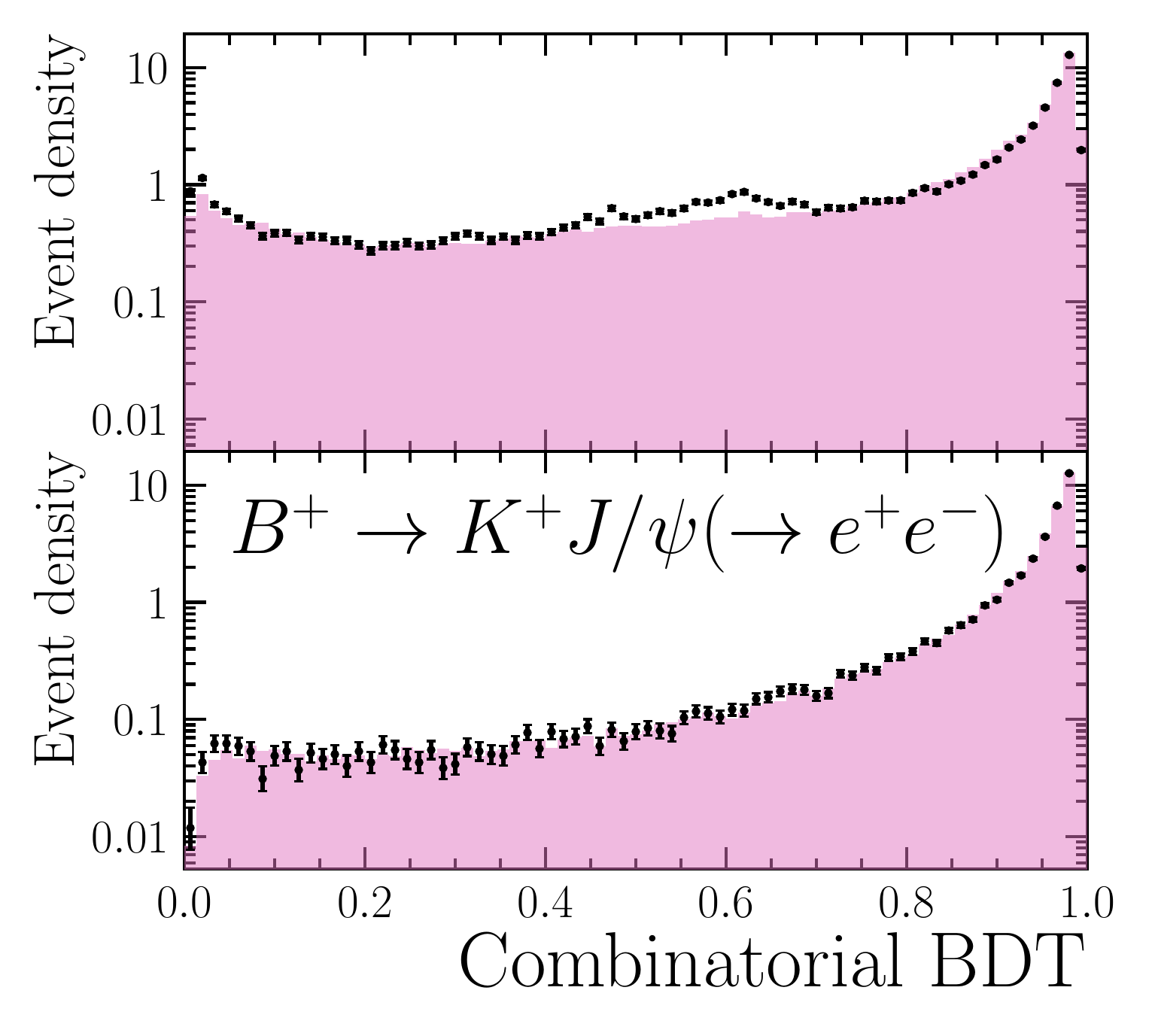}
                \end{minipage}
                \vspace*{-0.5cm}
                \end{center}
                \caption{\small Combinatorial BDT output of all events (top panels), and events passing cut-based selection (bottom panels) for each example in the case study.}
                \label{fig:performance:combinatorialBDT}
            \end{figure}

\begin{figure}[h]
                \begin{center}
                \centering
                \begin{minipage}{0.48\textwidth}
                \includegraphics[page=1, width=\textwidth, keepaspectratio]{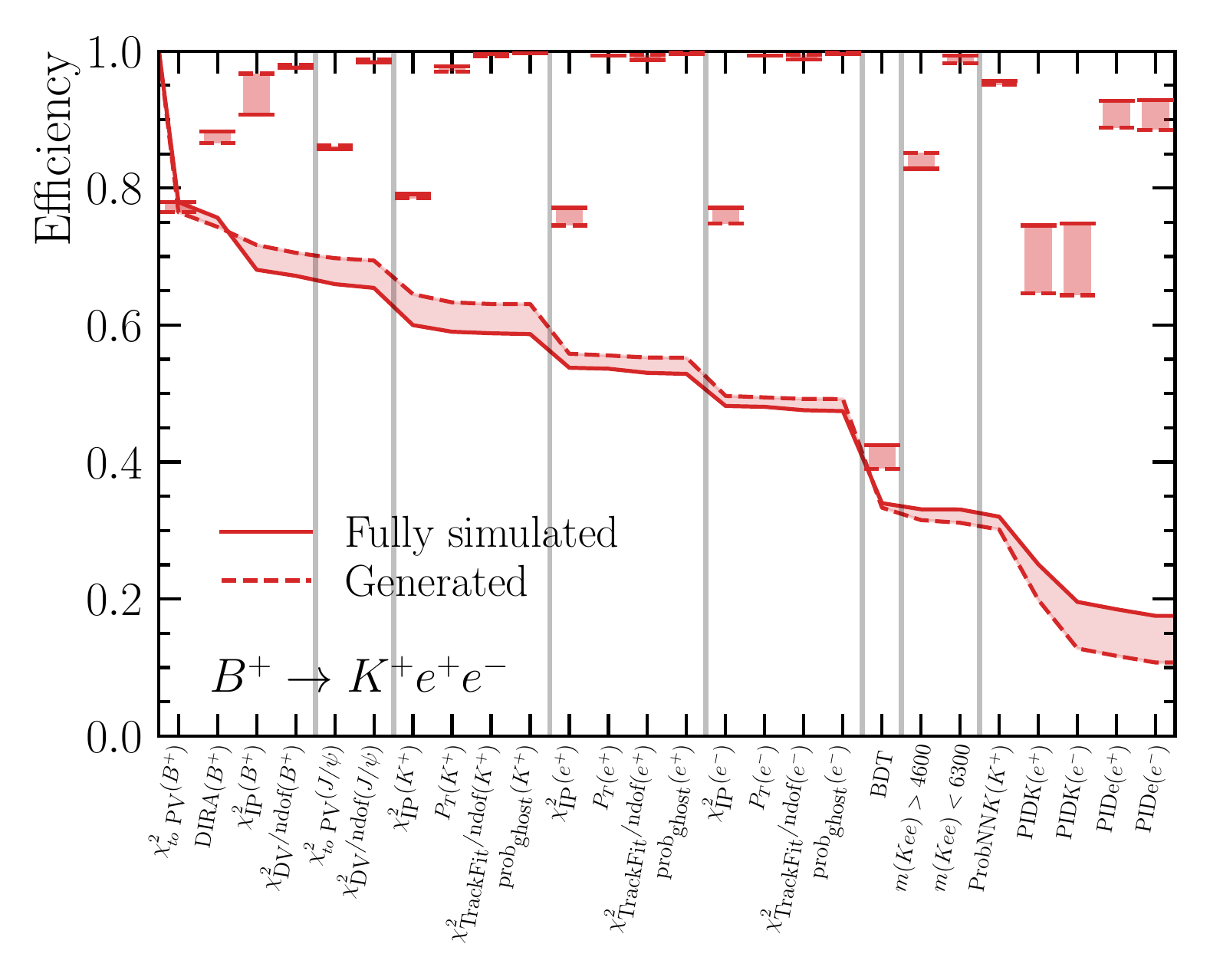}
                \end{minipage}
                \begin{minipage}{0.48\textwidth}
                \includegraphics[page=2, width=\textwidth, keepaspectratio]{figures/overlays_everythingbreakdown.pdf}
                \end{minipage}
                
                \begin{minipage}{0.48\textwidth}
                \includegraphics[page=3, width=\textwidth, keepaspectratio]{figures/overlays_everythingbreakdown.pdf}
                \end{minipage}
                \begin{minipage}{0.48\textwidth}
                \includegraphics[page=4, width=\textwidth, keepaspectratio]{figures/overlays_everythingbreakdown.pdf}
                \end{minipage}
                
                \begin{minipage}{0.48\textwidth}
                \includegraphics[page=5, width=\textwidth, keepaspectratio]{figures/overlays_everythingbreakdown.pdf}
                \end{minipage}
                \vspace*{-0.5cm}
                \end{center}
                \caption{\small A comparison of the efficiencies of the applied selection criteria, as computed using full simulation (solid) and \rex (dashed). Individual cut efficiencies are presented alongside the cumulative (running) efficiency.
                }
                \label{fig:performance:runningeff}
            \end{figure}

            \begin{table}[!htbp]
                \caption{List of selection requirements and variables which are included in the combinatorial BDT.}
                \label{tab:stripping}
                \begin{center}
                \begin{minipage}{0.45\textwidth}
                    \centering
                    \begin{tabular}{>{\centering\arraybackslash}p{1.5cm}| >{\centering\arraybackslash}p{1.3cm} >{\centering\arraybackslash}p{1.0cm} >{\centering\arraybackslash}p{0.7cm}}
                        & \multicolumn{3}{c}{Selection cuts} \\
                        \hline
                        \multirow{4}{*}{$\Bp$}
                        & $\chi^2_{to~PV}$ & $>$                   & $100$                 \\
                        & $\cos(\dira)$    & $>$                   & $0.995$               \\
                        & \ipChiSq         & $<$                   & $25$                  \\
                        & \vtxChiSqNdf     & $<$                   & $9$                   \\
                        \hline
                        \multirow{4}{*}{$I$}
                        & $\chi^2_{to~PV}$      & $>$                   & $16$                  \\
                        & $\cos(\dira)$                  \\
                        & \ipChiSq                          \\
                        & $\vtxChiSqNdf$   & $<$                   & $9$                   \\
                        \hline
                        \multirow{3}{*}{all tracks}
                        & \ghostProb       & $<$ & $0.3$ \\
                        & \chiSqTrack      & $<$ & $3$ \\
                        & \ipChiSq         & $>$                   & $9$                   \\
                    \end{tabular}
                \end{minipage}
                \begin{minipage}{0.45\textwidth}
                    \centering
                    \begin{tabular}{>{\centering\arraybackslash}p{1.5cm}| >{\centering\arraybackslash}p{1.3cm} >{\centering\arraybackslash}p{1.0cm} >{\centering\arraybackslash}p{0.7cm}}
                        & \multicolumn{3}{c}{Included in BDT} \\
                        \hline
                        \multirow{4}{*}{$\Bp$}
                        & $\chi^2_{to~PV}$ & & \checkmark                                      \\
                        & $\cos(\dira)$    & & \checkmark                                   \\
                        & \ipChiSq         & & \checkmark                                       \\
                        & \vtxChiSqNdf     & & \checkmark                                       \\
                        \hline
                        \multirow{4}{*}{$I$}
                        & $\chi^2_{to~PV}$ & & \checkmark                                      \\
                        & $\cos(\dira)$    & &                                    \\
                        & $\ipChiSq$       & & \checkmark                                       \\
                        & $\vtxChiSqNdf$   & &                                        \\
                        \hline
                        \multirow{3}{*}{all tracks}
                        & \ghostProb       & &   \\
                        & \chiSqTrack      & & \checkmark   \\
                        & \ipChiSq         & & \checkmark                                    \\
                    \end{tabular}
                \end{minipage}
                \end{center}
            \end{table}
            A typical analysis selection includes cuts on reconstruction quality variables, PID requirements, a boosted decision tree (BDT) to suppress combinatorial background, and finally, cuts on reconstructed invariant mass variables. The cuts used for this example are taken from Ref.~\cite{Aaij:2931511} and are detailed in Tab.~\ref{tab:stripping}. These cuts serve to ensure that basic reconstruction quality requirements are met. Following this, the BDT of Ref.~\cite{Aaij:2931511} is employed to reduce contamination from combinatorial background candidates formed in random combinations of unrelated particles from the same event. The BDT is trained to separate fully simulated signal candidates, labelled as 1, from candidates in the data selected from a region known to be dominated by combinatorial background, labelled as 0. At no point is the BDT trained on candidates generated using \rex. This BDT uses the variables listed in Tab.~\ref{tab:stripping}, this list does not include PID variables nor vertex isolation variables, which are the least well-modelled variables in \rex. However, this is a typical list of the most important features used across LHCb to suppress combinatorial backgrounds. Accurately reproducing the efficiency of this BDT requires the model to capture both the individual variable distributions and the correlations between them correctly. Figure~\ref{fig:performance:combinatorialBDT} shows the output of the BDT for all events (top panels) and for those that pass the cut-based selection (bottom panels), for each channel. The distributions are well-reproduced in all cases, both before and after the cut-based selection is applied. Note that the distributions are normalised before selection, meaning that both the shape and scale must be correctly modelled for good agreement. For the rest of the examples in this paper, events are selected with a threshold of $>0.9$ on the BDT output. The efficiencies of each individual cut, the BDT, some PID cuts (also from Ref.~\cite{Aaij:2931511}) and the kinematic $B$-mass cuts as used in the analysis, along with the running efficiency, are shown in Fig.~\ref{fig:performance:runningeff}. 

\begin{figure}[h]
                \begin{center}
                \centering
                \begin{minipage}{0.5\textwidth}
                \includegraphics[page=43, width=\textwidth, keepaspectratio]{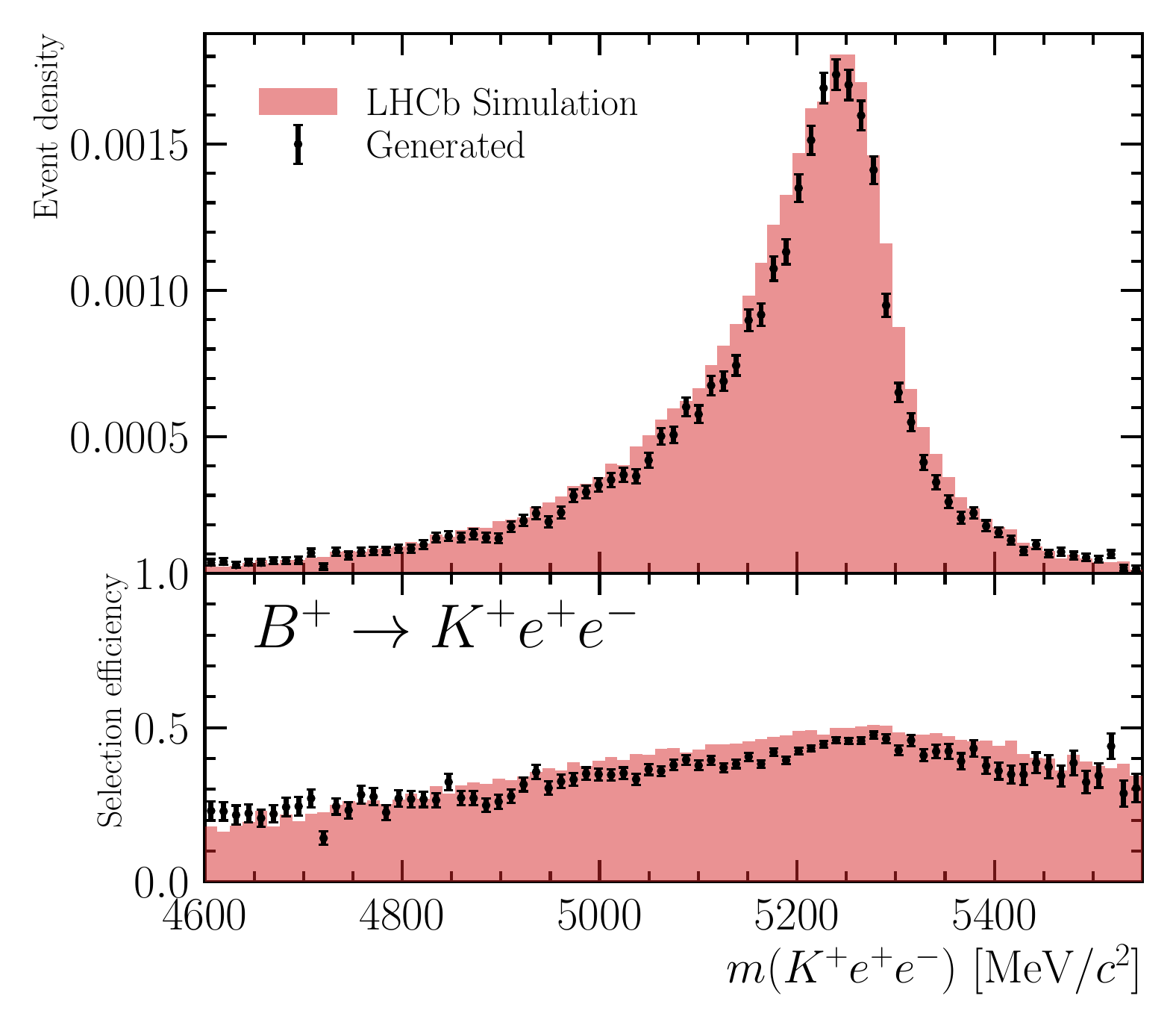}
                \end{minipage}
                \begin{minipage}{0.24\textwidth}
                \includegraphics[page=50, height=0.85\textwidth, keepaspectratio]{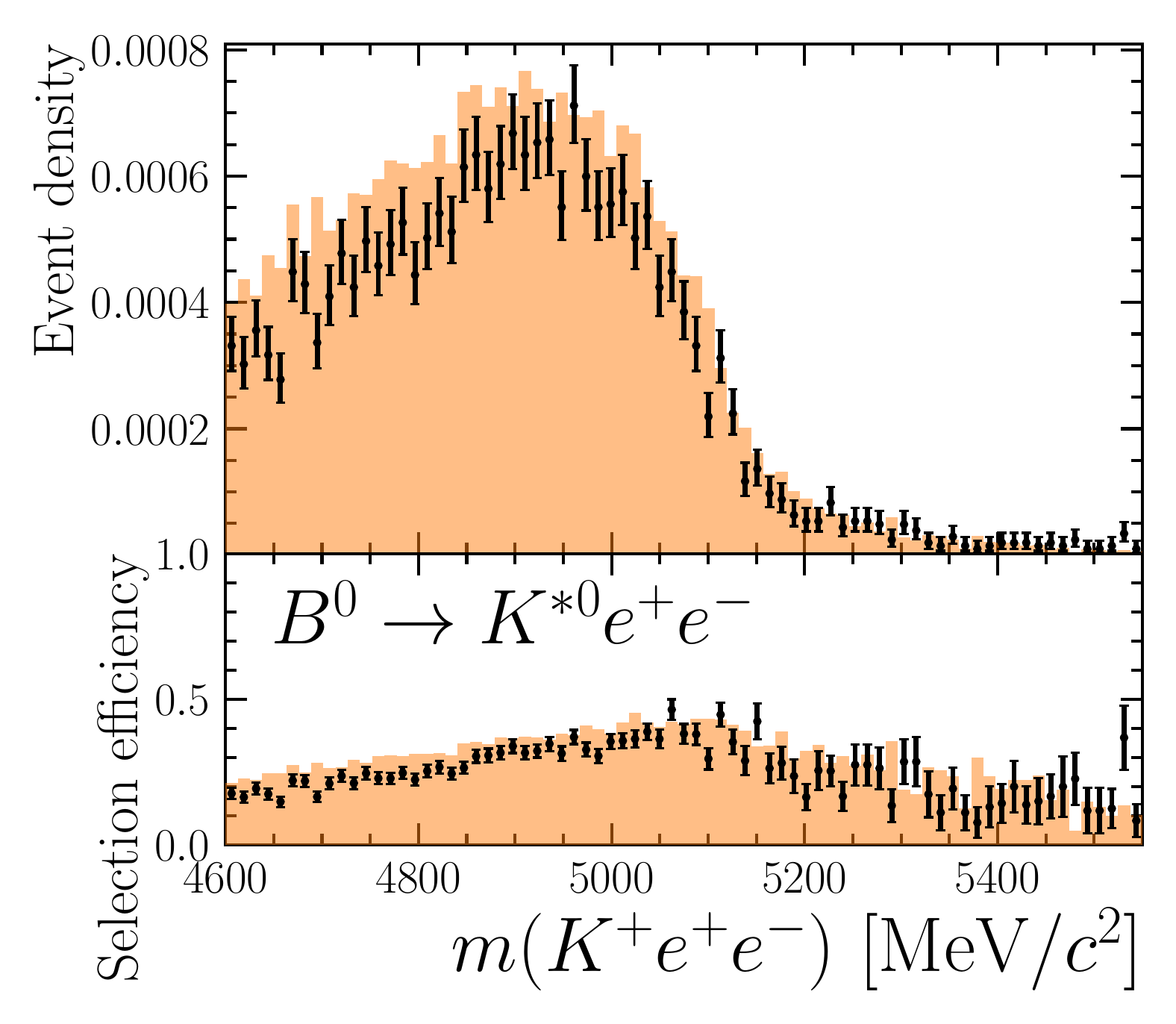}
                \includegraphics[page=57, height=0.85\textwidth, keepaspectratio]{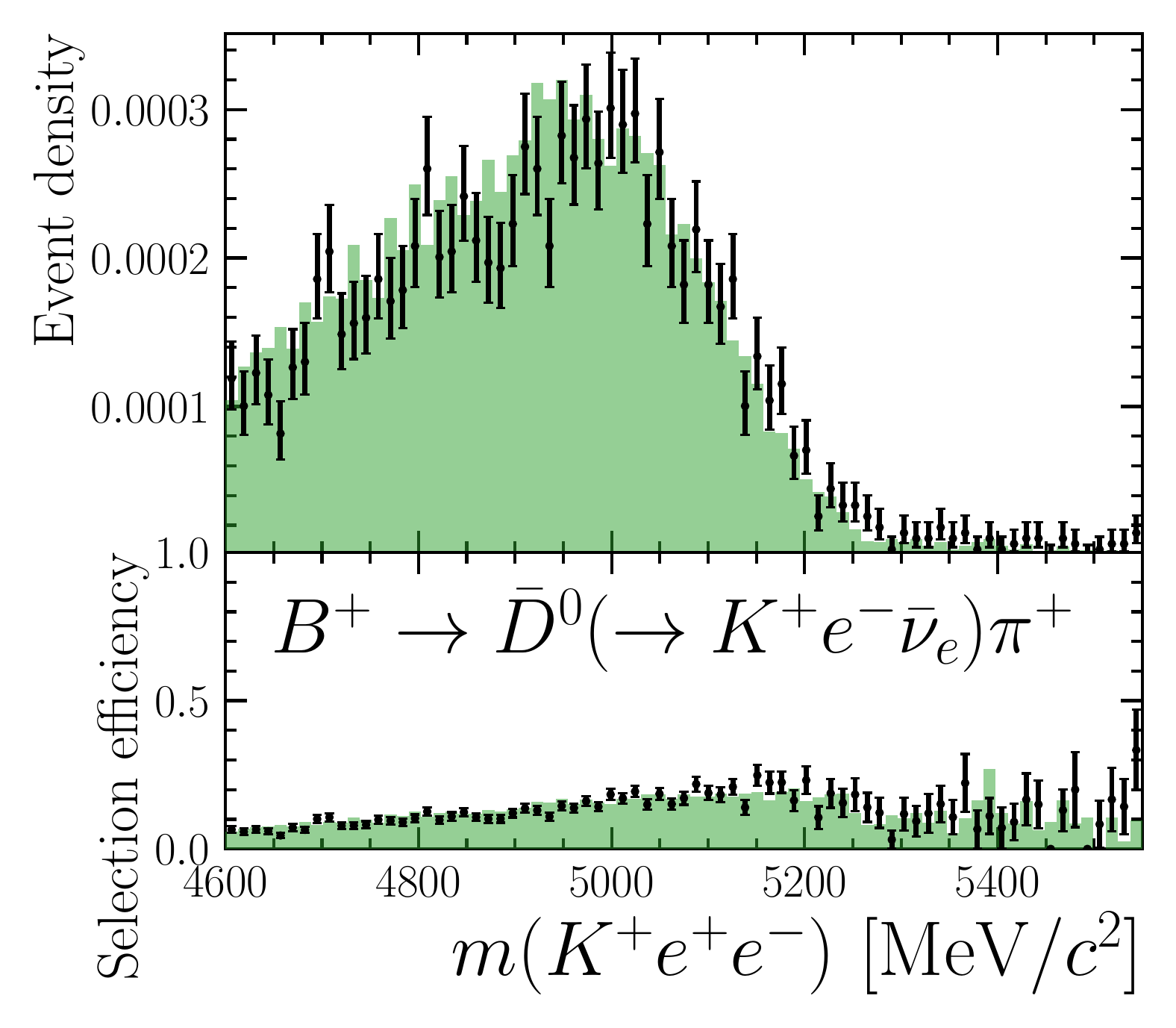}
                \end{minipage}
                \begin{minipage}{0.24\textwidth}
                \includegraphics[page=64, height=0.85\textwidth, keepaspectratio]{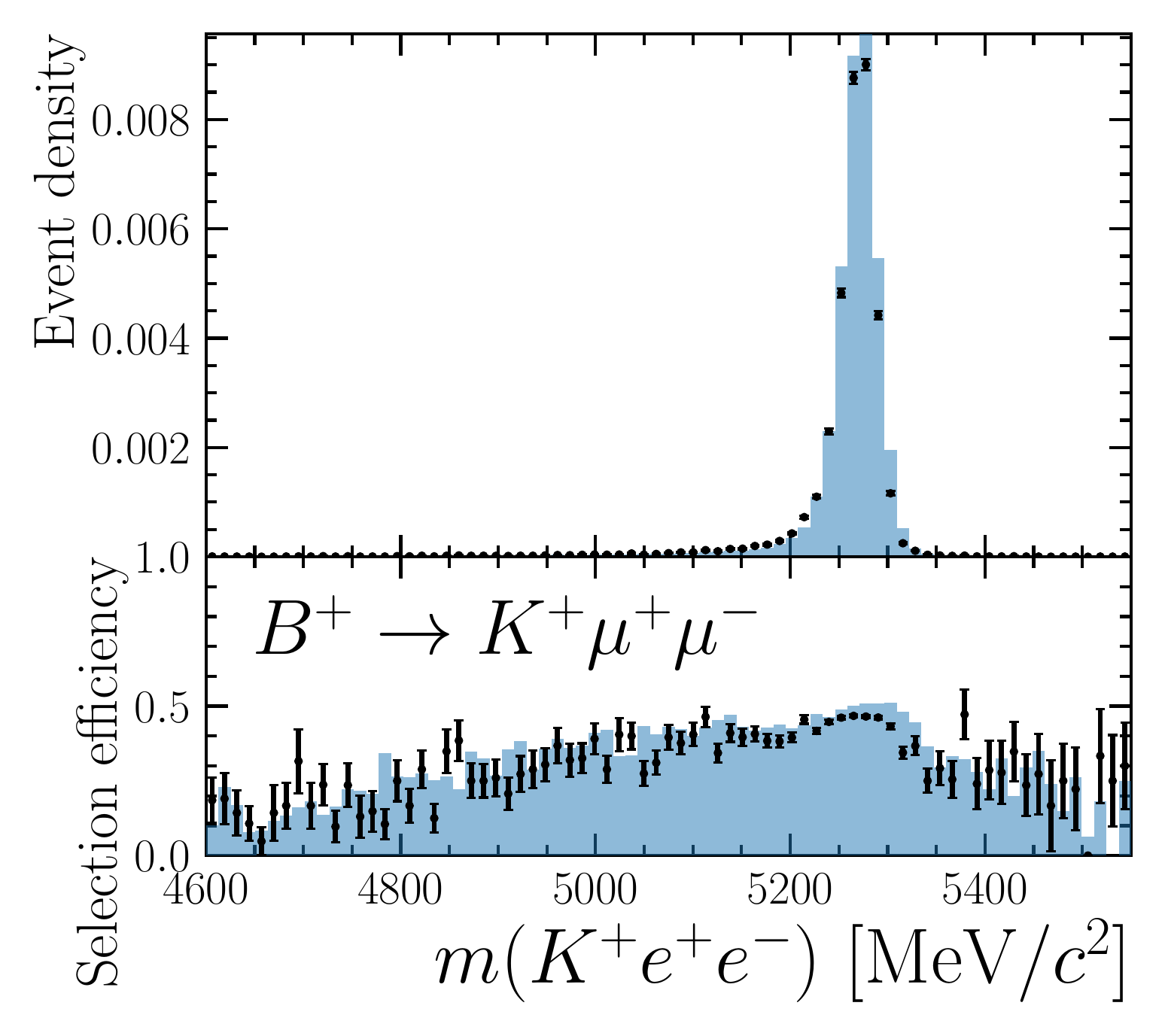}
                \includegraphics[page=71, height=0.85\textwidth, keepaspectratio]{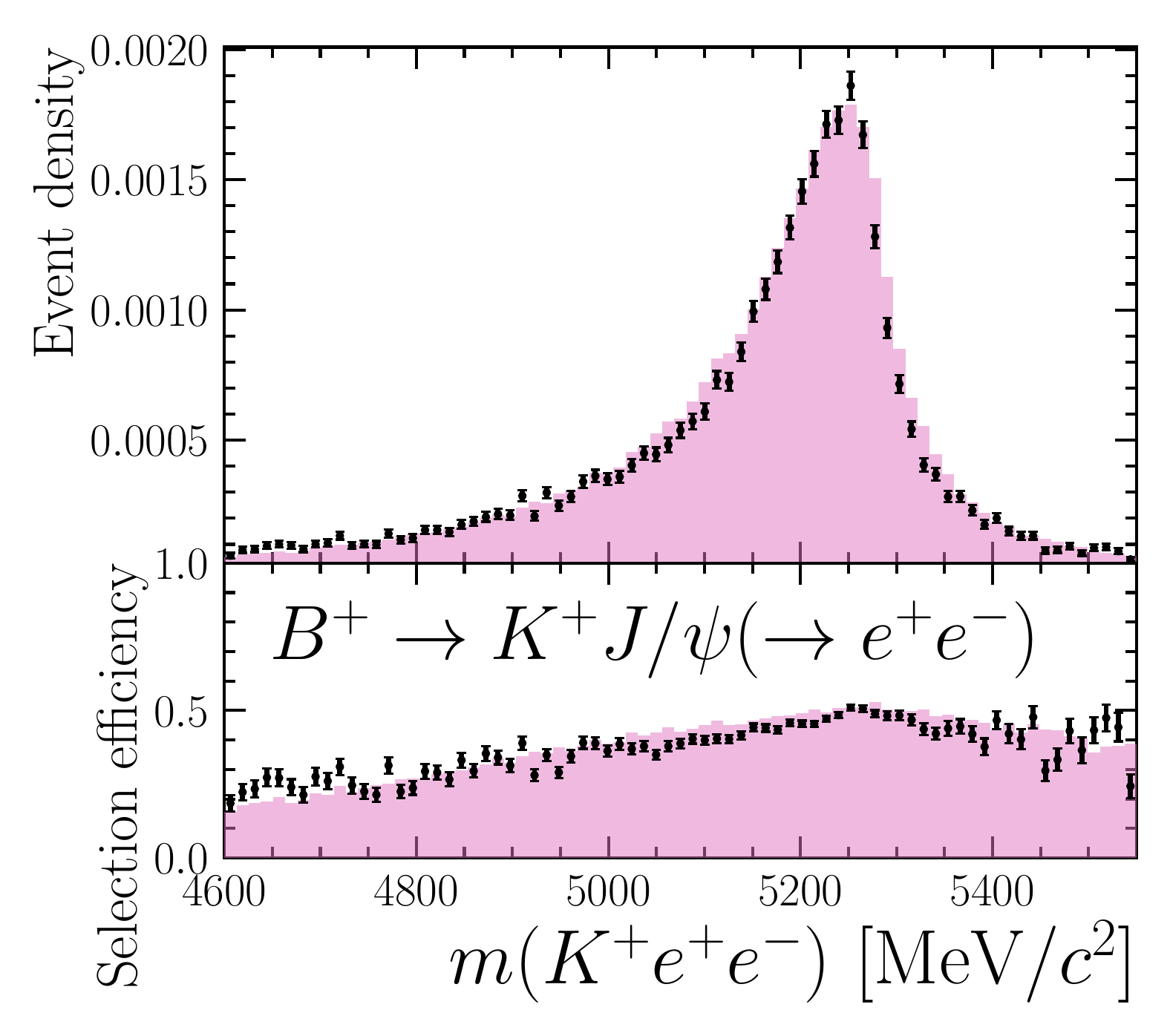}
                \end{minipage}
                \vspace*{-0.5cm}
                \end{center}
                \caption{\small Reconstructed $B$-mass distributions (top panels) using the signal hypothesis of $B^+\to K^+ e^+ e^-$ for the signal channel and the various backgrounds, and efficiency of combinatorial BDT and cut-based selection excluding PID cuts (bottom panels) as a function of reconstructed mass.}
                \label{fig:performance:masses}
            \end{figure}

            The top panels of Fig.~\ref{fig:performance:masses} show the distribution of reconstructed $B$-mass in the hypothesis of $B^+\to K^+ e^+ e^-$ after both the cut-based and the BDT selection are made. The bottom panels show the efficiency of the selection as a function of reconstructed $B$-mass. Similar distributions are produced for other kinematic variables and are available in the appendix (see Fig.~\ref{fig:ack:kinematic_variables}). To reiterate, these invariant kinematic variables are not given to the network in the conditions, yet the model correctly captures the dependence of the efficiency as a function of them. Resolution effects also appear well modelled in each case. 

            Following the procedure outlined in Ref.~\cite{LHCb:2022qnv} describing the measurement of $R_K$ in $B^+\to K^+\ell^+\ell^-$ transitions, we find that the replacement of the fully simulated samples with ones produced through \rex would lead to a systematic shift of $0.5\%$ in $R_K$. This shift would constitute a subleading source of systematic uncertainty for this measurement. Further details regarding this study are given in Appendix~\ref{measuring_RK}. A more stringent validation of the performance of \rex is provided by the measurement of $r_{J/\psi}$, also described in Ref.~\cite{LHCb:2022qnv}, which probes lepton universality in $J/\psi$ decays from $B^+\to K^+J/\psi$ transitions. Unlike $R_K$, which makes use of a double ratio to cancel systematic uncertainties related to efficiency determination and to reduce the impact of mismodelling differences between electrons and muons, $r_{J/\psi}$ is based on a single efficiency ratio and is therefore directly sensitive to such mismodelling effects. In this case, using \rex to produce the simulated samples, the shift to $r_{J/\psi}$ is $2\%$, which is of similar order to existing systematic effects. Updates to come in future releases of \rex, as listed in the conclusions of this paper, will improve on this.  

    \subsection{Uncertainties}\label{sec:performance:uncertainties}

        Inaccuracies in the output of \rex may be ascribed to four principal sources:
        
        \paragraph{Convergence-related uncertainties}  
            The network’s performance is sensitive to instabilities in the training process. These may arise from premature or late stopping — a consequence of the difficulty in defining a convergence criterion that generalises well. The model may also become saturated, lacking the capacity to capture the required behaviour across phase space; this may manifest in continual trade-offs between different regions. Instabilities may also emerge in localised parts of condition space, latent space, or output space due to the stochastic nature of batch ordering during training. The GAN loss itself is infamously unstable, often requiring significant tuning of hyperparameters and of training schedules to maintain the balance between the discriminator and generator required to achieve convergence. These effects are more likely to manifest in sparsely populated or high-variance regions of phase space. Repeating training or resampling with different random seeds can help in assessing these uncertainties.
        
        \paragraph{Model-related limitations}  
            Systematic biases may arise from limitations in the model architecture. For example, it may be the case that the network or graph structures do not sufficiently allow or unknowingly prohibit certain relationships between objects. In complex topologies, skip connections or other architectural components may be required to accurately model correlations between variables. A more likely cause of biases within this category is insufficient conditioning: we are inherently restricted in the set of variables we can use as conditions. For any effects that depend on something not parsed, any behaviour will only be captured on average. This could result in biases when the network is queried for specific channels. Studying the performance across a range of decay topologies may help mitigate these issues, though edge cases may remain vulnerable.
        
        \paragraph{Inaccuracies in conditions}  
            Even with a perfectly trained network, any discrepancies between the training conditions (as derived from simulation) and those generated for inference will cause systematic bias. Known mismodellings include differences in the kinematics of hadrons originating from the primary vertex and a simplified geometric acceptance. These effects appear to be small, and future versions of \rex will aim to address them. Physically motivated perturbations to the conditions can be used to probe the sensitivity of the outputs to such differences. With the simplified geometrical acceptance, the tool is currently unable to account for particles in acceptance that may not have been reconstructed. This will be accounted for in later versions by employing an improved probabilistic geometrical acceptance map.
            
        \paragraph{Simulation inaccuracies}  
            The full simulation is known to imperfectly describe data. Analyses at LHCb routinely apply significant data-driven weighting procedures to improve the fidelity of simulated distributions. This tool inherits any inaccuracies present in the underlying simulation, forming a source of irreducible uncertainty.
        \newline
        
        \noindent Beyond identifying the sources of uncertainty, we also must consider how best to present them to end users. Given the generalised nature \rex, with outputs in the form of raw variables rather than derived quantities, the most appropriate approach is likely to provide multiple alternative sets of output variables. Users can then propagate each set through their analysis workflows (e.g., selection criteria) and study the resulting changes in efficiency or yield. While this tool could incorporate some built-in functionality to support that workflow, the diversity of use cases suggests that the final implementation details are best left to the user.

\section{Conclusion}\label{sec:conclusion}

    This paper presents a set of novel heterogeneous graph-based GAN network architectures designed to emulate the LHCb simulation pipeline in a manner that is as general as possible. The tool is trained in such a way as to be agnostic to any specific final states, to support any physical decay topology and any arbitrary analysis selection criteria.
    Where possible, the architectures employed have been designed to embed the logic of each interdependent stage of modelling into the structure of the networks themselves. 

    Using a representative case study, we have demonstrated that the tool is capable of modelling selection efficiencies and invariant mass distributions of a broad spectrum of topologies, including fully reconstructed decays, partially reconstructed decays, final states involving long-lived intermediates, and various types of misidentification with excellent fidelity. Furthermore, we have demonstrated that second-order effects, such as interactions between tracks and interference from the rest of the event, are also correctly included.

    This work represents the first iteration of \rex and is available as a user-friendly Python package capable of modelling the sculpting of a decay phase space by arbitrary selection criteria. The tool reads simple configuration files and outputs in the same format as the full LHCb simulation, allowing for seamless integration with existing analysis workflows. Estimates of absolute efficiencies of any selection chain, including tracking, vertexing, impact parameter, and flight distance variables, can be produced across a wide range of decay topologies. Developments to come in future versions of \rex will reduce these discrepancies. 

    On the inference side, the next steps are to address the limitations mentioned in Sec.~\ref{sec:fast_framework} in the modelling of the detector’s geometrical acceptance and the generation of kinematics of heavy-flavour hadrons originating from the PV, as mentioned in Sec ~\ref{sec:performance:PID} this is likely to improve the modelling of the PID variables. Additionally, there are a number of commonly used variables which are not currently supported by the model, most notably trigger decisions and calorimetric variables. Note that extending the list of variables is possible without a change to the architecture. Finally, updated versions of \rex should also include extensions to support protons and higher multiplicity final states. Then, regarding the performance, the next steps are to begin some trials within ongoing analyses and to begin quantifying the agreement with the full simulation and the size of the various sources of uncertainty. Here, improvements in the training stability and convergence criteria may be required; this could be driven by a move away from GANs to diffusion models, to avoid the complexities of adversarial training. Beyond this, it should be possible to implement inference mechanisms to estimate new background types, such as combinatorial background or backgrounds involving particles from the primary vertex, using the existing network architecture. Finally, looking further ahead, it may even be possible to develop methods that enable training \rex directly on data, potentially avoiding some of the accuracy limitations in the full simulation.

    The speed-ups achieved by the methods presented in this paper have the potential to fundamentally reshape the simulation workflow at LHCb. Trained models can be stored, one set for each configuration of data-taking conditions, and used repeatedly for on-demand generation of analysis-ready samples, reducing the existing need for long-term storage of large samples of full detector readout. This workflow would offer significant savings in both compute and storage, and opens the door to a more decentralised workload that could offer more flexibility to analysts.
    
    The architectures and methods presented in this paper can easily be employed to emulate reconstruction processes at other particle physics experiments, many of which are seeing similar pressures relating to growing demands for simulation.
\clearpage
\section*{Acknowledgments}

We are extremely grateful to colleagues from the LHCb Simulation Project, the quality of this tool and the ability to train it at all rely on many years of development of the LHCb software. Additionally, we would like to thank colleagues from across the Data Processing and Analysis (DPA) project who have provided feedback during the development of this tool. Finally, this paper was significantly improved based on comments from Mark Whitehead and Mark Smith. 

The University of Bristol group acknowledges support from the STFC under Grant No. ST/W000490/1. The UZH group acknowledges support from the Swiss National Science Foundation (SNSF) under Grants No. 209263 and 204238, and from Forschubngskredit under the UZH Candoc Grant No. FK-25-086. The Imperial College group acknowledges support from the STFC (United Kingdom) and The Leverhulme Trust (United Kingdom).
\appendix

\section{Remeasuring $R_K$ with \rex}\label{measuring_RK}

    \begin{figure}[h]
        \begin{center}
        \centering
        \begin{minipage}{0.48\textwidth}
        \includegraphics[width=\textwidth, keepaspectratio]{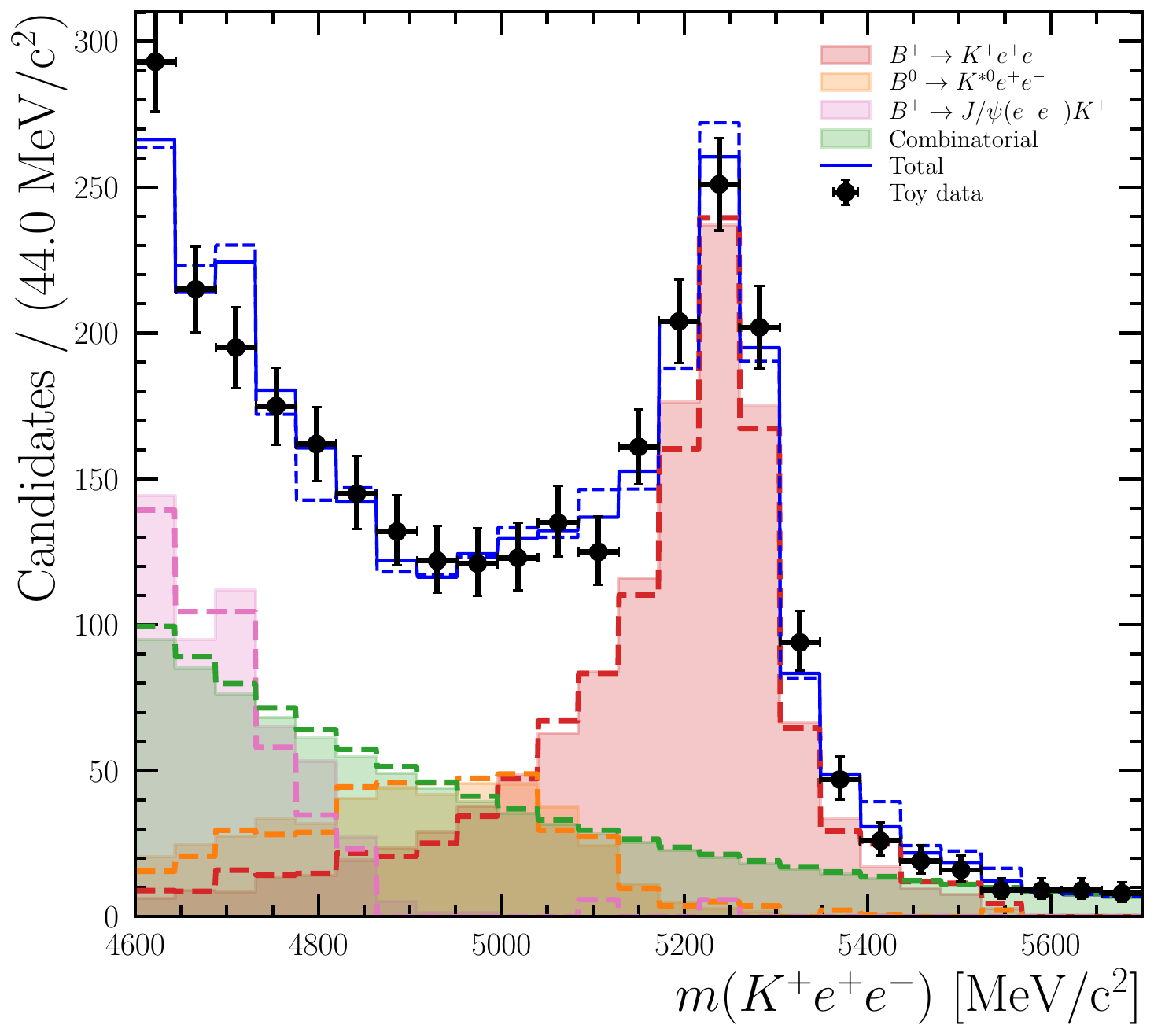}
        \end{minipage}
        \begin{minipage}{0.48\textwidth}
        \includegraphics[width=\textwidth, keepaspectratio]{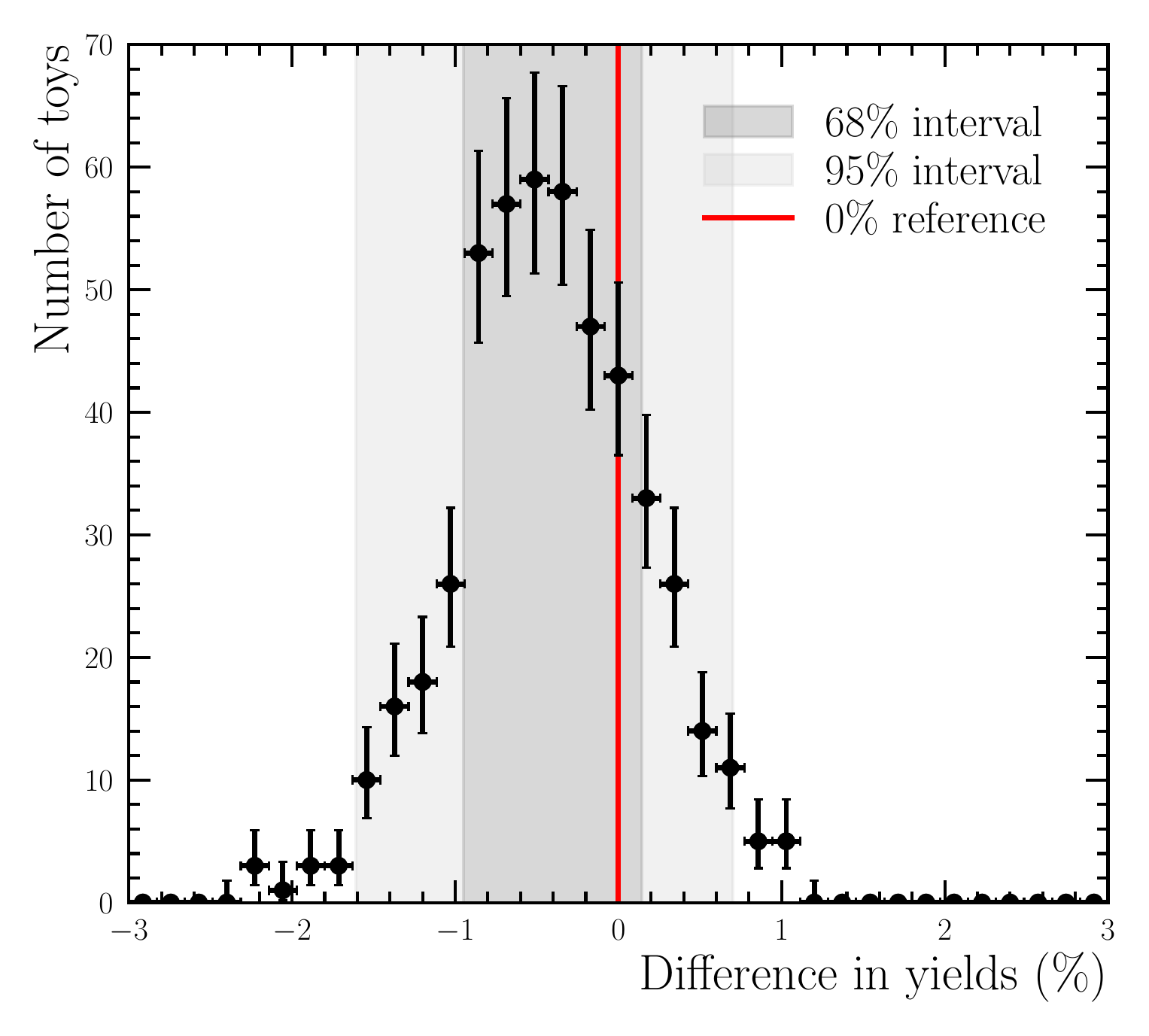}
        \end{minipage}
        \end{center}
        \caption{\small An example pseudo-dataset for the measurement of $B^+\to K^+e^+e^-$ and fitted components, with those from \rex being dashed (left), and the observed differences obtained from fitting pseudo-datasets with \rex relative to the full simulation (right).}
        \label{fig:performance:Kee_toys}
    \end{figure}

    We can employ fully simulated tuples and tuples generated by \rex to get a handle on how different $R_K$ would be if one were to perform the measurement using only \rex.
    
    First, we look at the fit to the reconstructed $B$-mass, $m(K^+e^+e^-)$, in the electron mode. We generate toy event sets using distributions from selected fully simulated tuples and using yields in accordance with the fits of Ref.~\cite{LHCb:2022qnv}. An exponential component is used to represent the combinatorial background. We fit each event set with the same set of fit components as obtained from the full simulation and additionally a set obtained using \rex, then compare the extracted signal yield. In each case, the signal yield is obtained via a binned likelihood template fit in which the only parameters not fixed are the yields of the components. An example toy, and the fitted components are presented in the left panel of Fig.~\ref{fig:performance:Kee_toys}. In the right panel is a distribution of the difference in the extracted signal yield expressed as a percentage. For the vast majority of cases, this difference is $<1\%$, and the most likely difference is a $\sim0.5\%$ shift.

    In a measurement of a branching fraction or of a ratio of branching fractions, such as $R_K$, as well as the measured yield, one also requires integrated efficiencies. These appear as the $\epsilon$ quantities, and the yields appear as the $N$ quantities, in the following expression for $R_K$, 
    \begin{equation}
        R_K = \frac{N_{K\mu\mu}}{N_{Kee}} \cdot \frac{\epsilon_{Kee}}{\epsilon_{K\mu\mu}} \cdot \frac{N_{KJ/\psi(ee)}}{N_{KJ/\psi(\mu\mu)}} \cdot \frac{\epsilon_{KJ/\psi(\mu\mu)}}{\epsilon_{KJ/\psi(ee)}}.
    \end{equation}
    \begin{table}
        \centering
        \resizebox{\textwidth}{!}{
        \begin{tabular}{r|cccc}
             & $B^+\to K^+e^+e^-$ & $B^+\to J/\psi(e^+e^-)K^+$ & $B^+\to K^+\mu^+\mu^-$ & $B^+\to J/\psi(\mu^+\mu^-)K^+$ \\
             \hline
            $\epsilon_{\textsc{Rex}} / \epsilon_{\text{LHCb}}$ & 0.93 & 0.92 & 0.95 & 0.94\\
        \end{tabular}
        }
        \caption{Ratios of efficiencies for channels involved in the extraction of $R_K$. }
        \label{tab:RK_effs}
    \end{table}
    The ratios of the efficiencies as produced by \rex and by the full simulation for the channels involved in $R_K$ are presented in Tab.~\ref{tab:RK_effs}. Here, we can see that efficiencies are mismodeled by \rex, in its current form. For example, \rex underestimates the efficiency of $B^+\to K^+e^+e^-$ by $\sim7\%$. However, it is important to stress that measurements at LHCb are almost always conducted with respect to control channels, which, by design, have final states and decay topologies as similar as possible to the channels of interest, such that in the measurement, mismodelings cancel. Indeed, it is the case too that in \rex the mismodellings of control channels, in this case the $J/\psi$ channels, are similar to their corresponding signal channels. Therefore, in the measurement scenario of $R_K$, effects from mismodellings in \rex are suppressed to a $<0.1\%$ effect. Leaving the full shift expected if one were to measure $R_K$ with \rex to be $\sim0.5\%$, with this being dominated by the difference in the extraction of the signal yield as observed in Fig.~\ref{fig:performance:Kee_toys}. Another related quantity is the single ratio $r_{J/\psi}$, which is employed as a validation channel in $R_K$ measurements, it is defined as,
    \begin{equation}
        r_{J/\psi} = \frac{N_{KJ/\psi(\mu\mu)}}{N_{KJ/\psi(ee)}} \cdot \frac{\epsilon_{KJ/\psi(ee)}}{\epsilon_{KJ/\psi(\mu\mu)}}.
    \end{equation}
    Studying any systematic shift here is a much more stringent test of the performance of \rex as cancellations within the quantity $r_{J/\psi}$ are not as powerful as in $R_K$, and the ratio remains vulnerable to differences in the extent to which the detector response to electrons and to muons is mismodeled. In this case, $r_{J/\psi}$ shifts by $2\%$ when using simulated samples from \rex, which is commensurate with the systematic uncertainty on $r_{J/\psi}$ reported in Ref.~\cite{LHCb:2021trn}. The shifts presented here are computed using tuples before any data-simulation correction weights that are applied to the full simulation tuples in any analysis at LHCb. Applying such weights would more closely align \rex and the full simulation and reduce systematic shifts reported here.

\FloatBarrier
\clearpage

\section{Additional PID plots}

    \begin{figure}[h]
        \begin{center}
        \centering
        \includegraphics[page=1, width=0.24\textwidth,keepaspectratio]{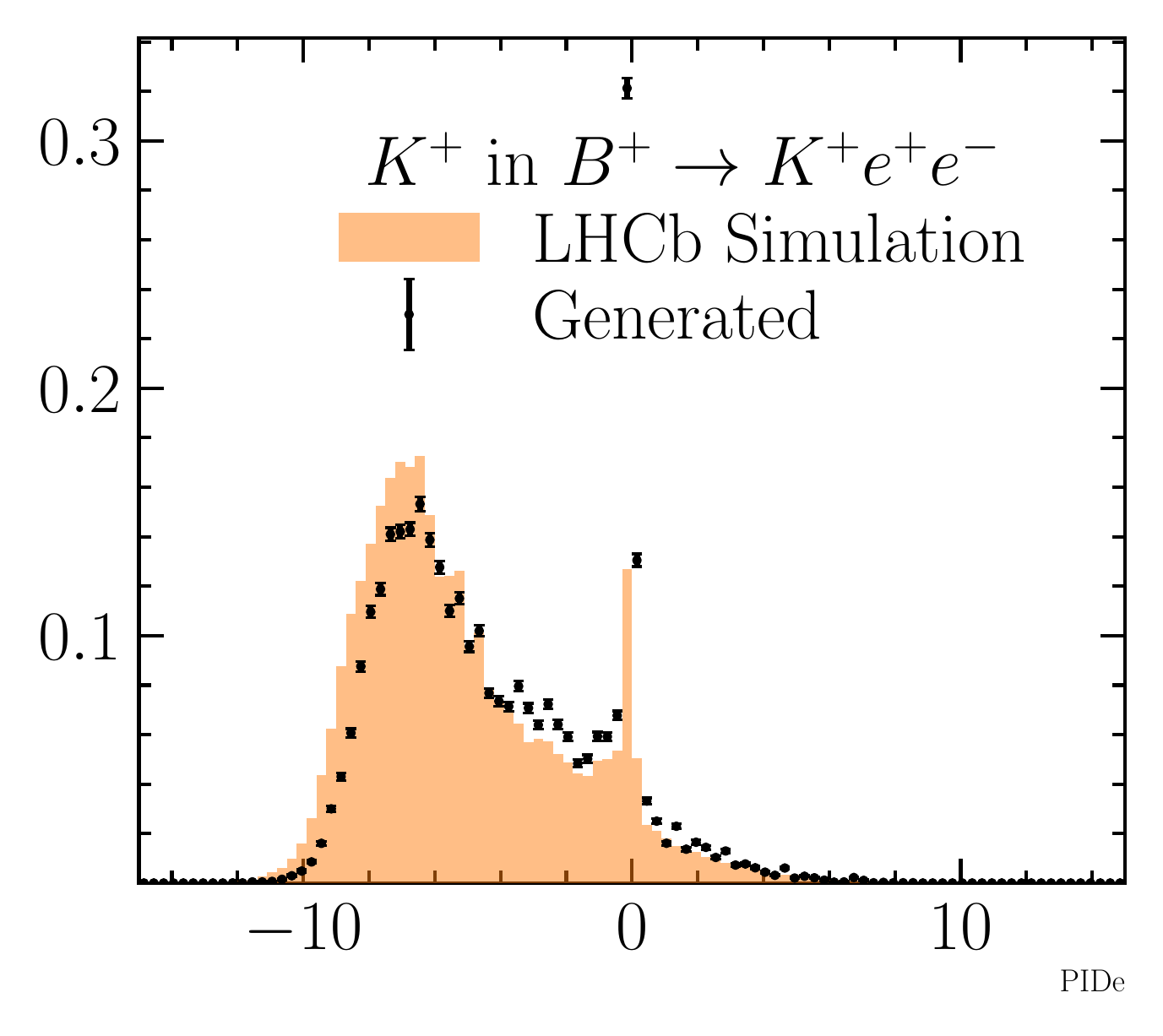}
        \includegraphics[page=2, width=0.24\textwidth,keepaspectratio]{figures/PID_plots.pdf}
        \includegraphics[page=3, width=0.24\textwidth,keepaspectratio]{figures/PID_plots.pdf}
        \includegraphics[page=4, width=0.24\textwidth,keepaspectratio]{figures/PID_plots.pdf}
        
        \includegraphics[page=5, width=0.24\textwidth,keepaspectratio]{figures/PID_plots.pdf}
        \includegraphics[page=6, width=0.24\textwidth,keepaspectratio]{figures/PID_plots.pdf}
        \includegraphics[page=7, width=0.24\textwidth,keepaspectratio]{figures/PID_plots.pdf}
        \includegraphics[page=8, width=0.24\textwidth,keepaspectratio]{figures/PID_plots.pdf}
        
        \includegraphics[page=9, width=0.24\textwidth,keepaspectratio]{figures/PID_plots.pdf}
        \includegraphics[page=10, width=0.24\textwidth,keepaspectratio]{figures/PID_plots.pdf}
        \includegraphics[page=11, width=0.24\textwidth,keepaspectratio]{figures/PID_plots.pdf}
        \includegraphics[page=12, width=0.24\textwidth,keepaspectratio]{figures/PID_plots.pdf}
        
        \includegraphics[page=13, width=0.24\textwidth,keepaspectratio]{figures/PID_plots.pdf}
        \includegraphics[page=14, width=0.24\textwidth,keepaspectratio]{figures/PID_plots.pdf}
        \includegraphics[page=15, width=0.24\textwidth,keepaspectratio]{figures/PID_plots.pdf}
        \includegraphics[page=16, width=0.24\textwidth,keepaspectratio]{figures/PID_plots.pdf}
        \vspace*{-0.5cm}
        \end{center}
        \caption{
        \small Example distributions of generated PID variables.}
        \label{fig:ack:PID_1D}
    \end{figure}

    \begin{figure}[h]
        \begin{center}
        \centering
        \includegraphics[page=5, width=0.48\textwidth,keepaspectratio]{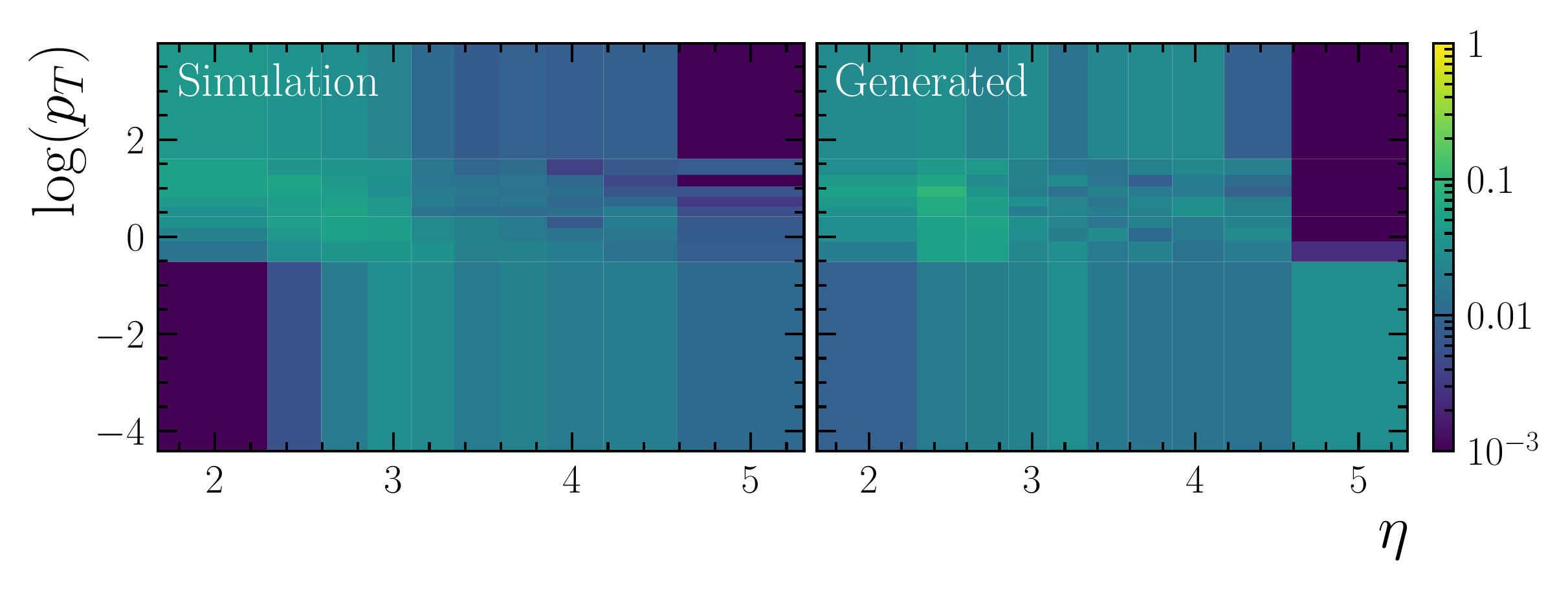} 
        \includegraphics[page=11, width=0.48\textwidth,keepaspectratio]{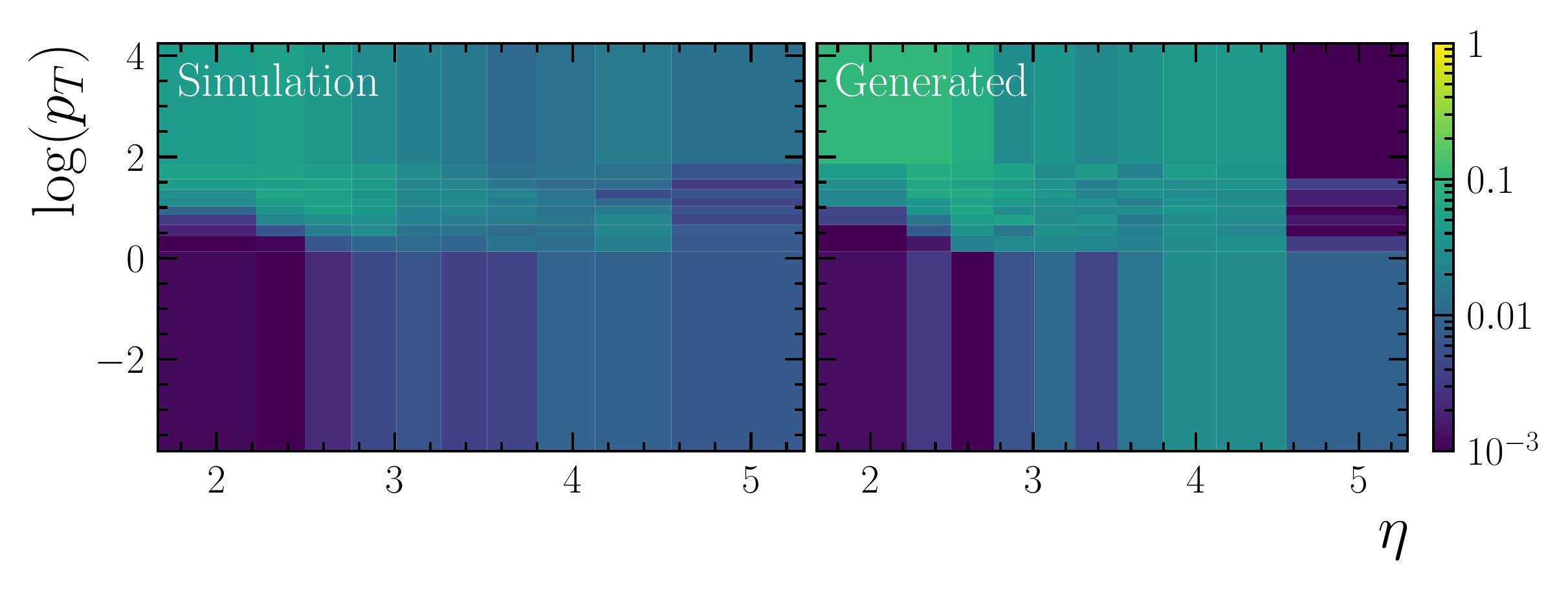}\\ 
        \includegraphics[page=17, width=0.48\textwidth,keepaspectratio]{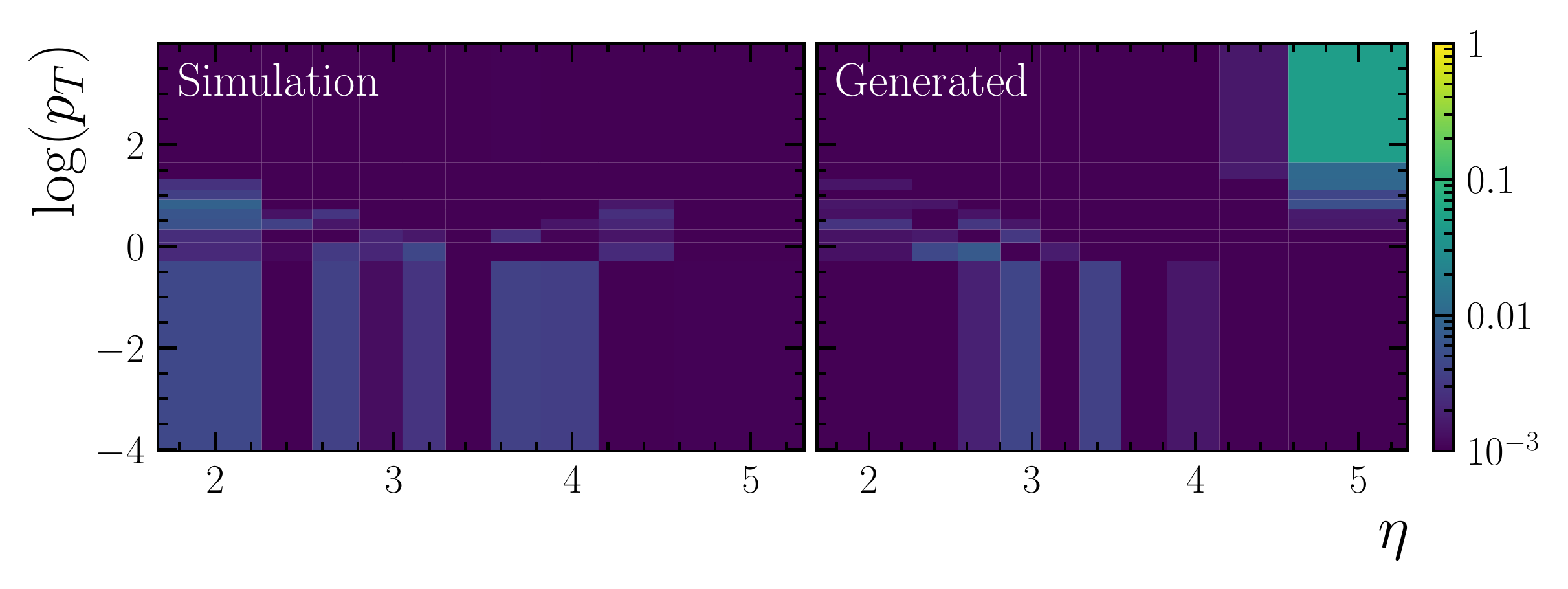} 
        \includegraphics[page=24, width=0.48\textwidth,keepaspectratio]{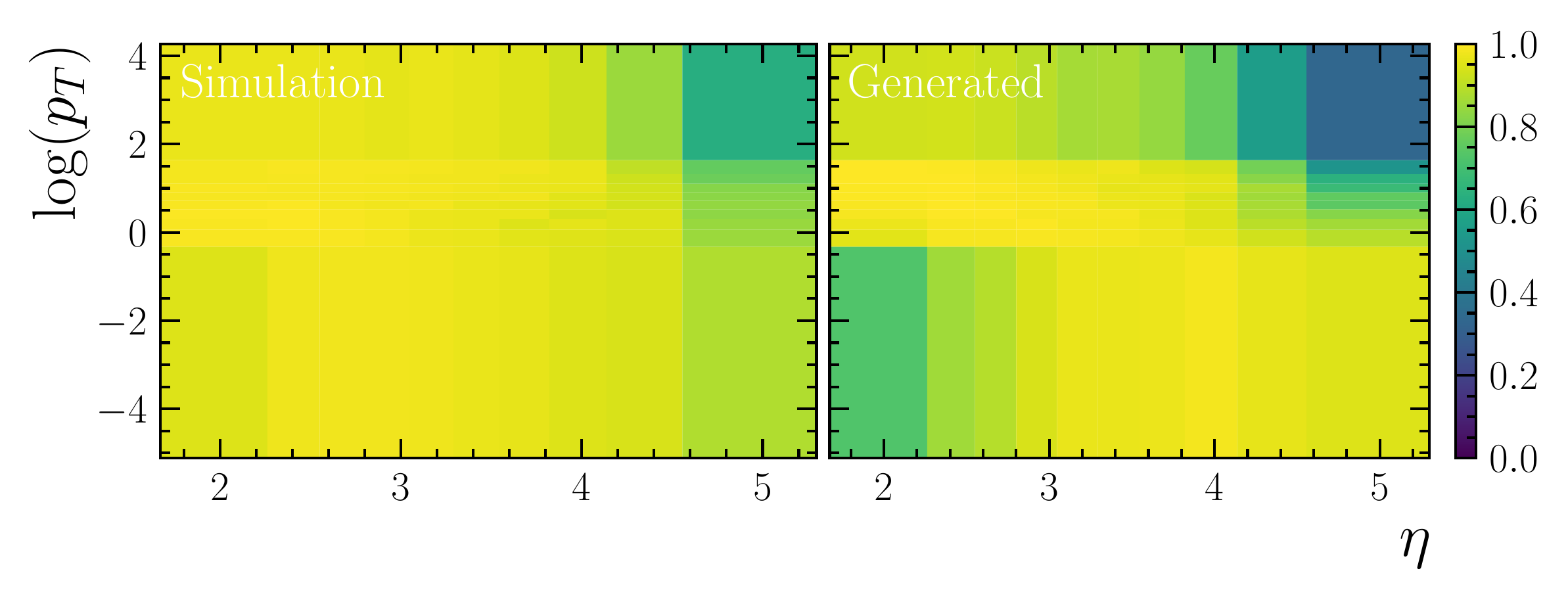} 

        \includegraphics[page=29, width=0.48\textwidth,keepaspectratio]{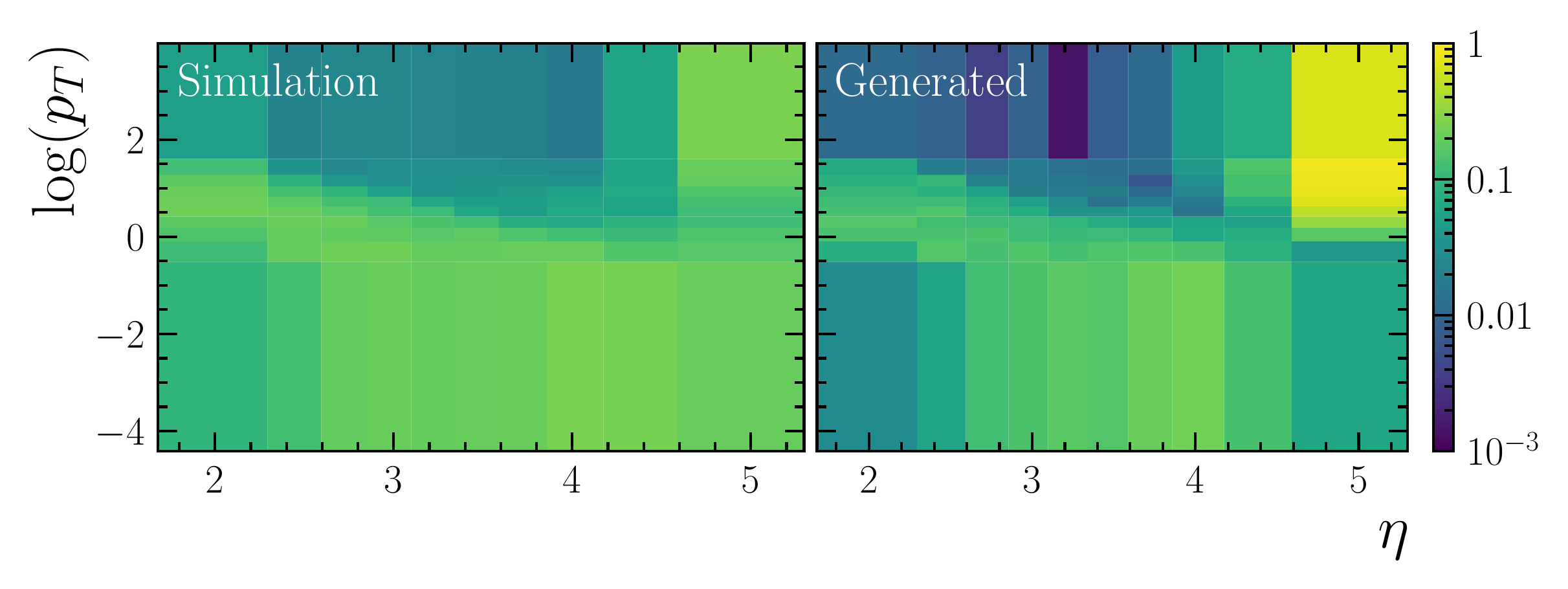}
        \includegraphics[page=35, width=0.48\textwidth,keepaspectratio]{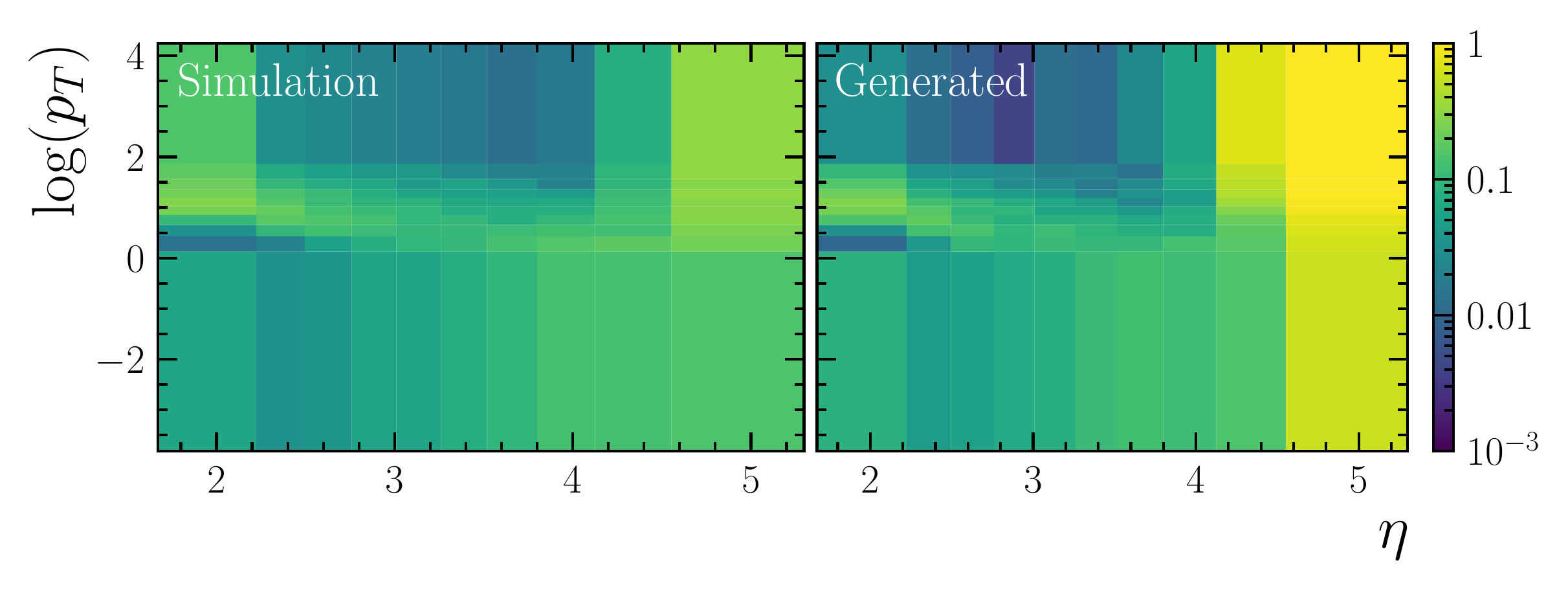}\\
        \includegraphics[page=42, width=0.48\textwidth,keepaspectratio]{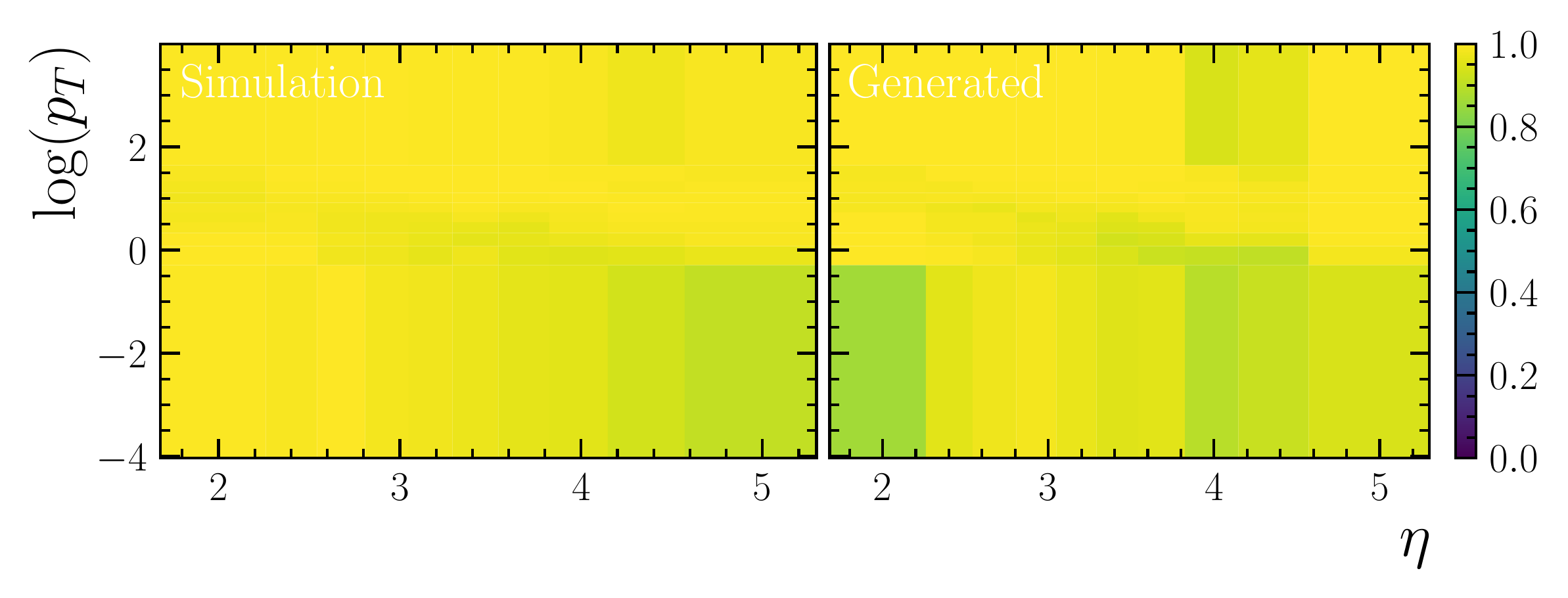}
        \includegraphics[page=48, width=0.48\textwidth,keepaspectratio]{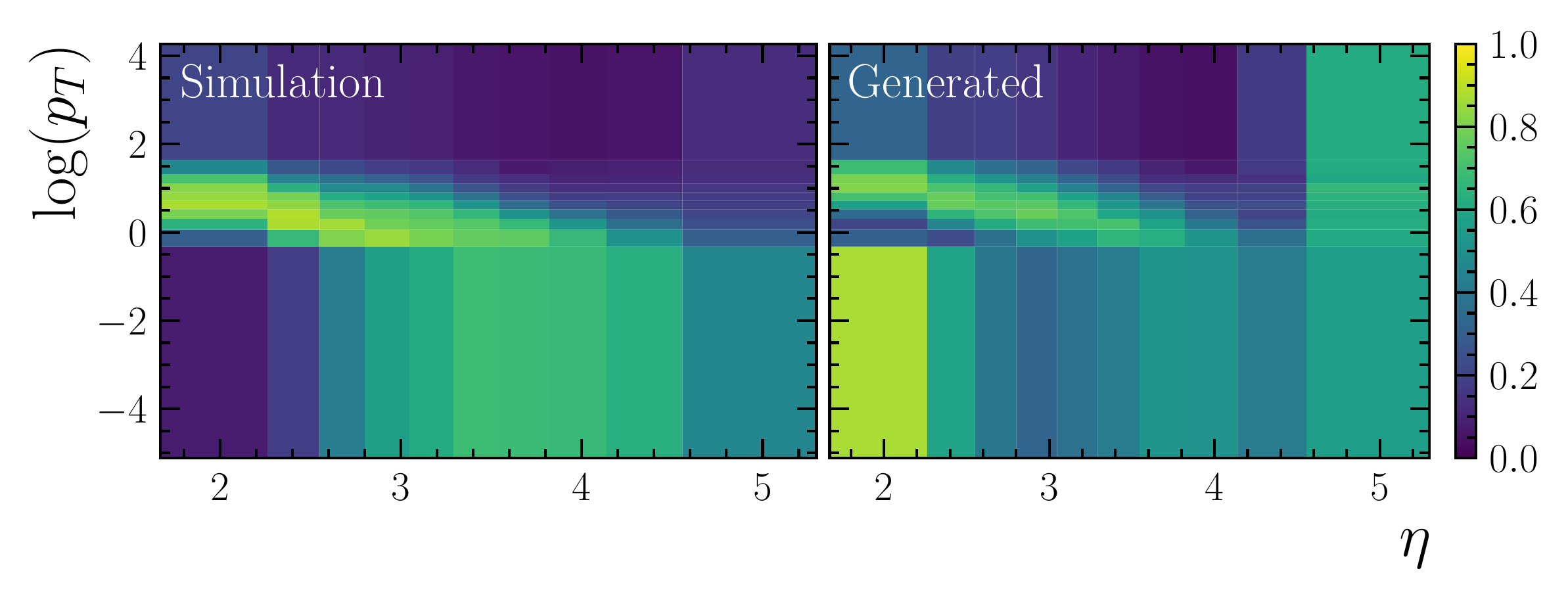}
        
        \includegraphics[page=78, width=0.48\textwidth,keepaspectratio]{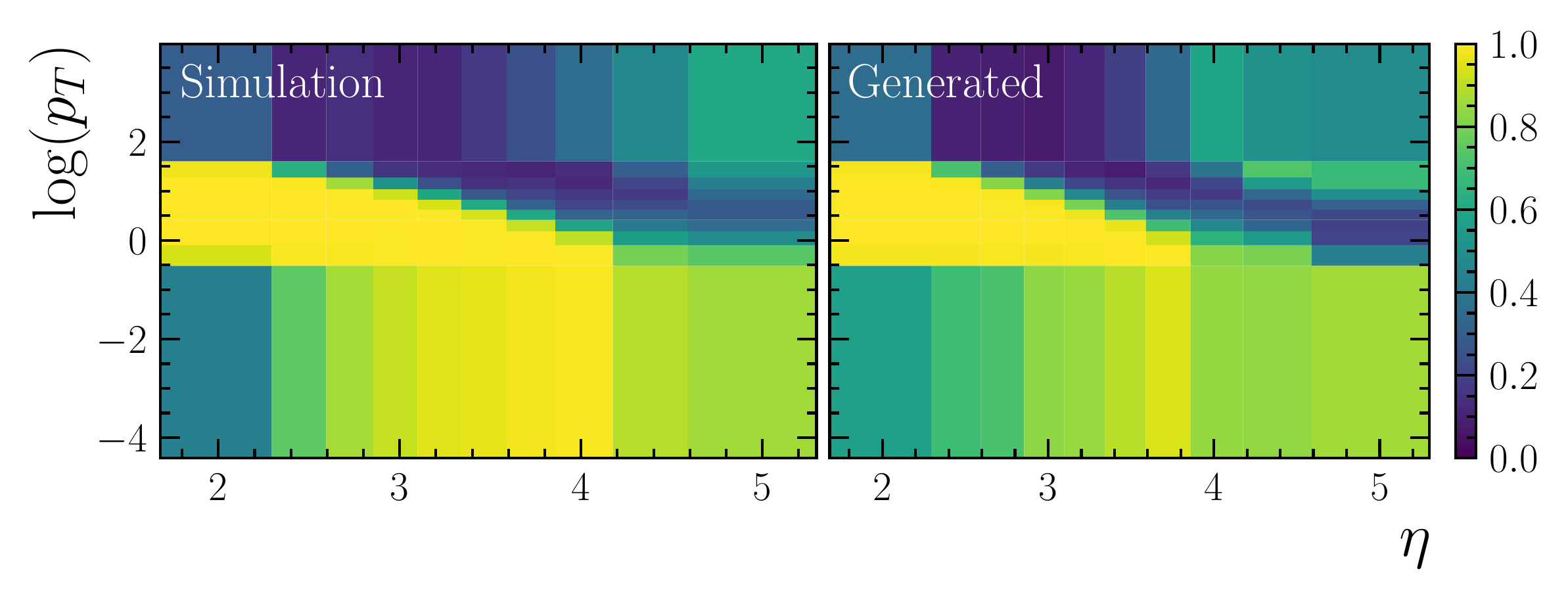}
        \includegraphics[page=83, width=0.48\textwidth,keepaspectratio]{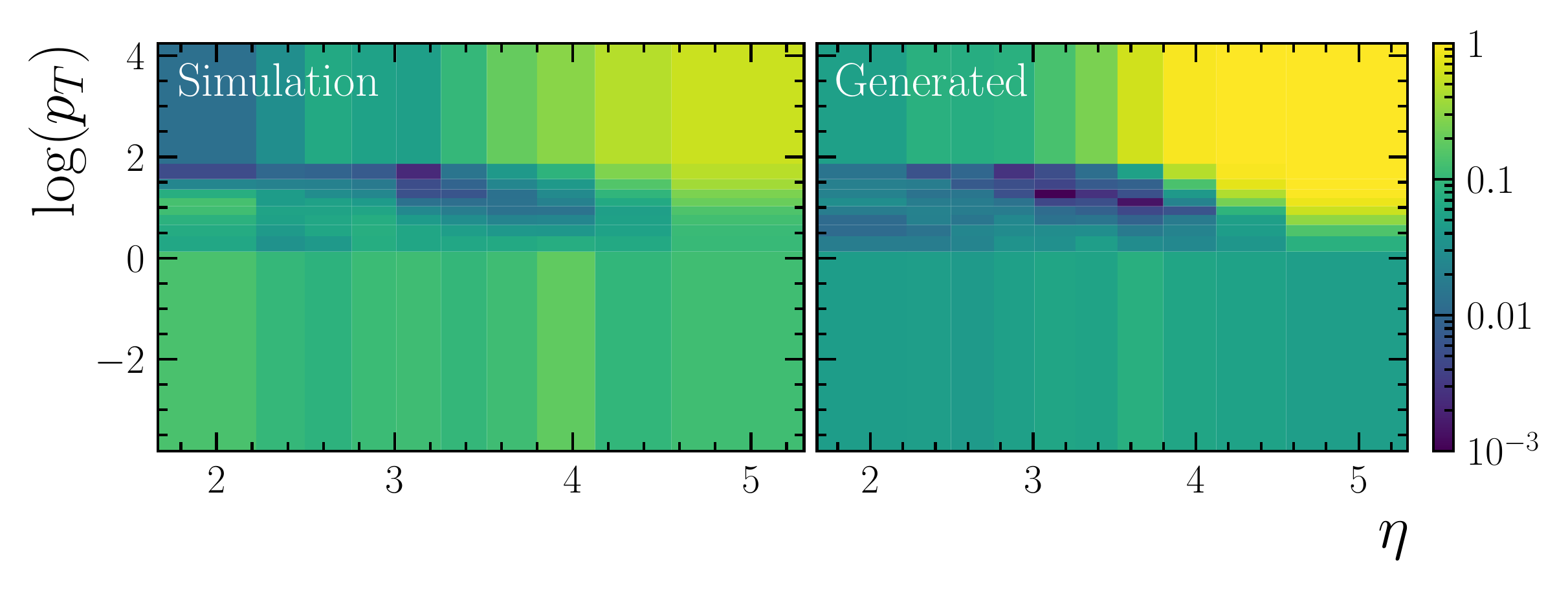}\\
        \includegraphics[page=89, width=0.48\textwidth,keepaspectratio]{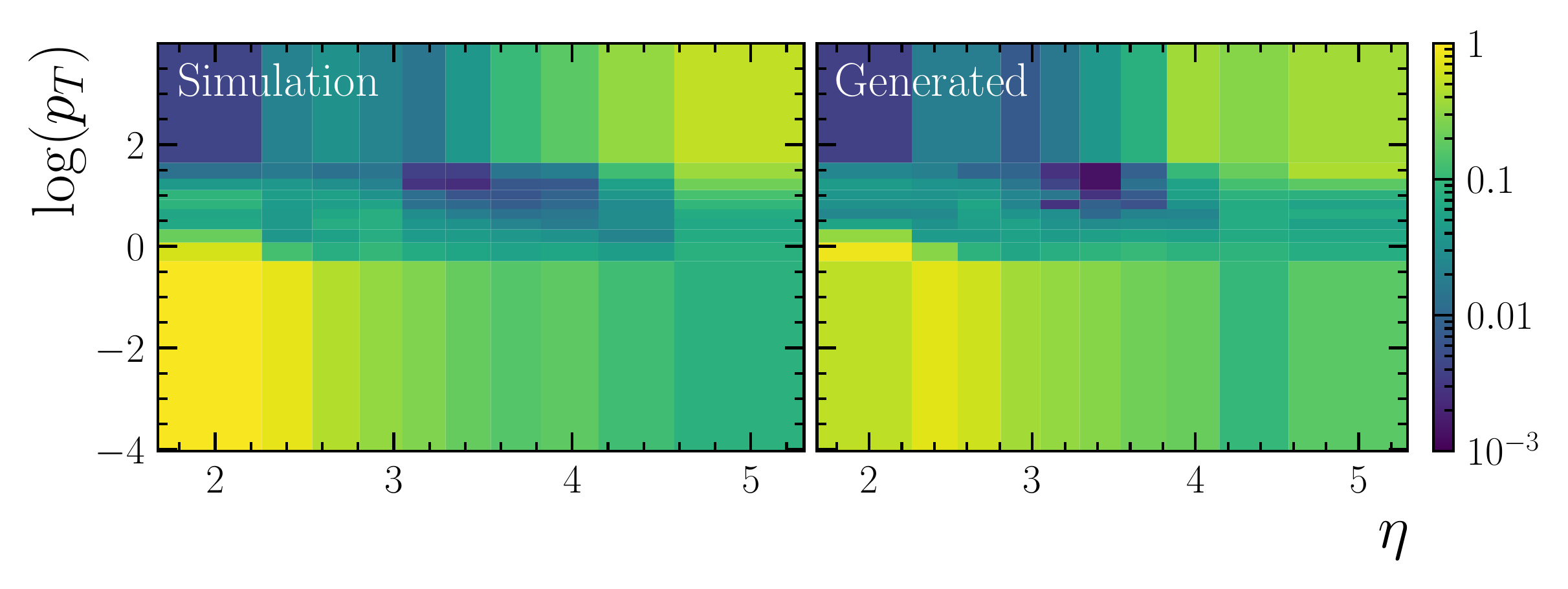}
        \includegraphics[page=96, width=0.48\textwidth,keepaspectratio]{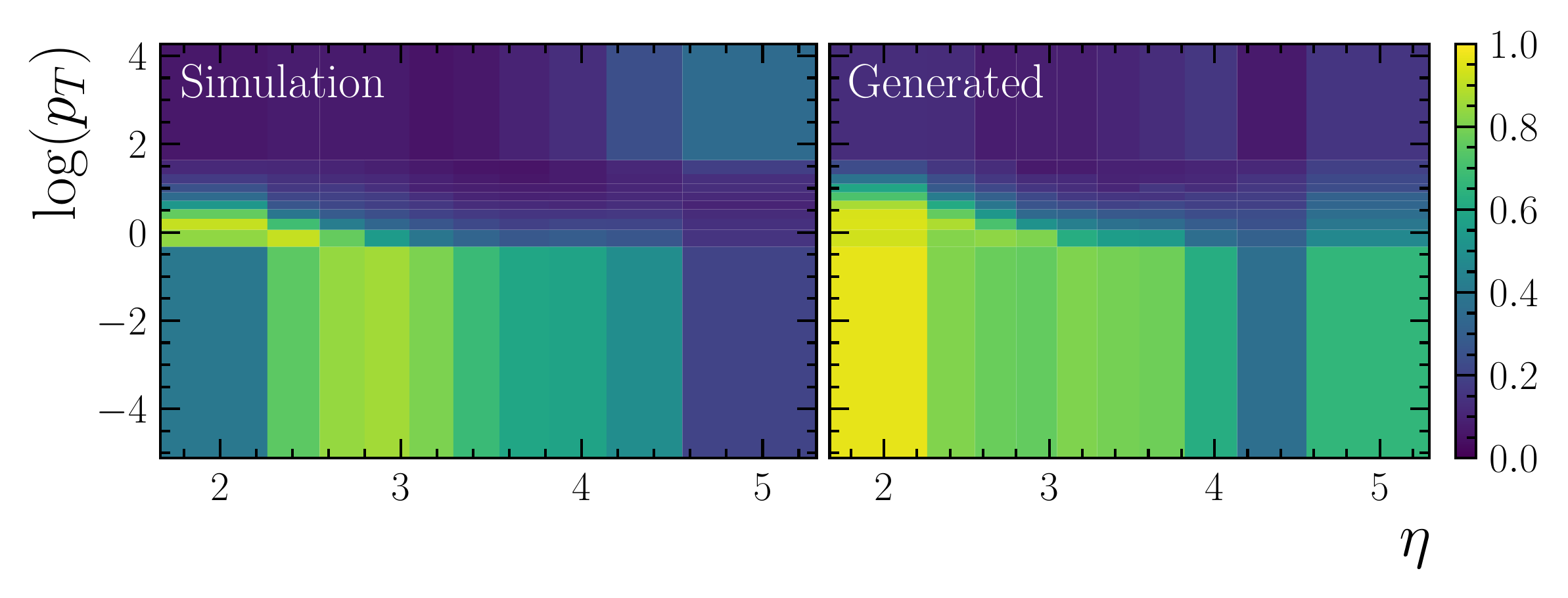}
        \vspace*{-0.5cm}
        \end{center}
        \caption{
        \small Efficiency maps (in $\log(p_T)$ and $\eta$) for various PID cuts.}
        \label{fig:ack:PID_2D}
    \end{figure}
    
\FloatBarrier
\clearpage
\section{Additional vertexing plots}

    \begin{figure}[h]
        \begin{center}
        \centering
        \begin{minipage}{0.47\textwidth}
        \includegraphics[page=79, width=\textwidth, keepaspectratio]{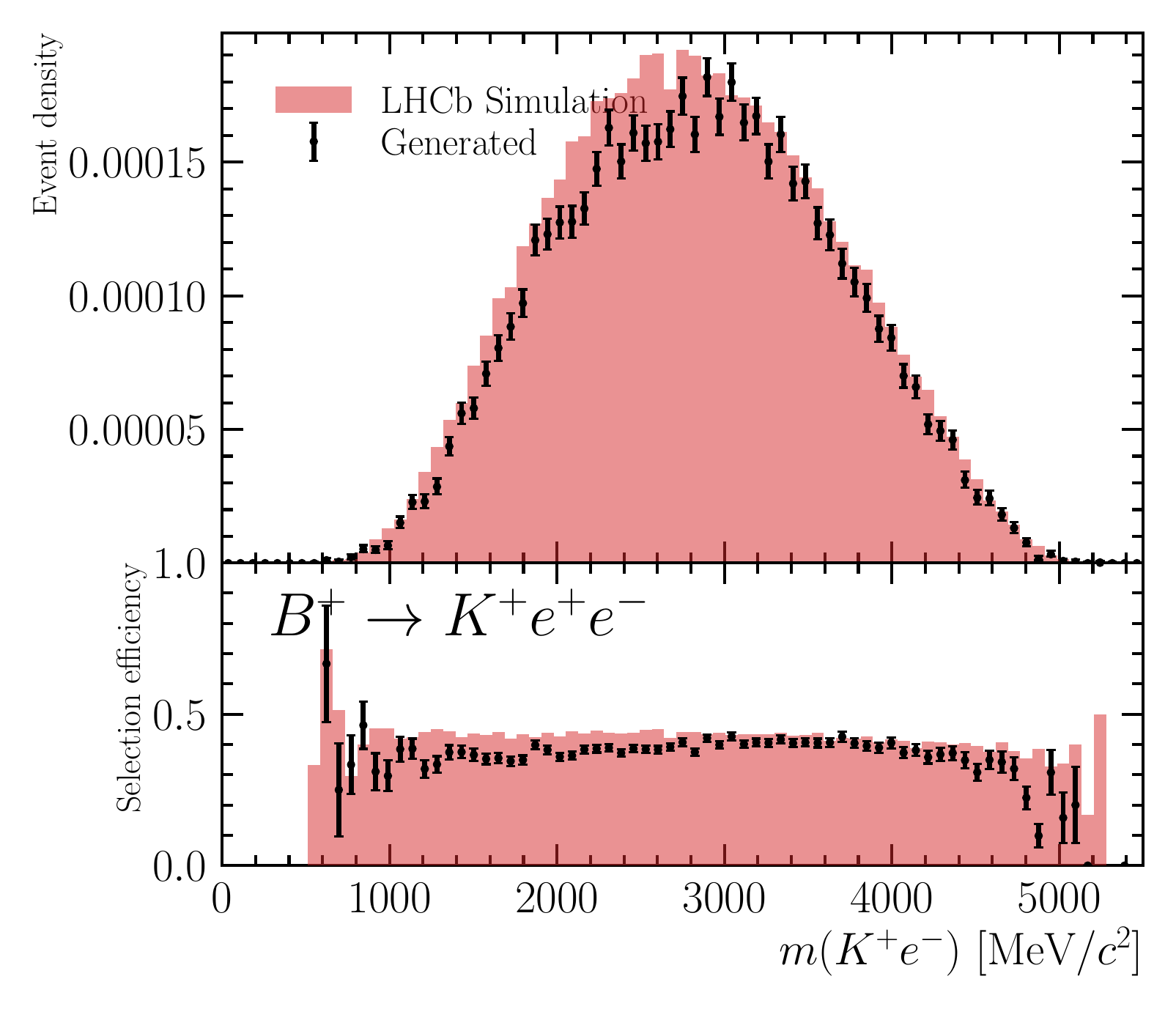}
        \end{minipage}
        \begin{minipage}{0.23\textwidth}
        \includegraphics[page=86, height=0.85\textwidth, keepaspectratio]{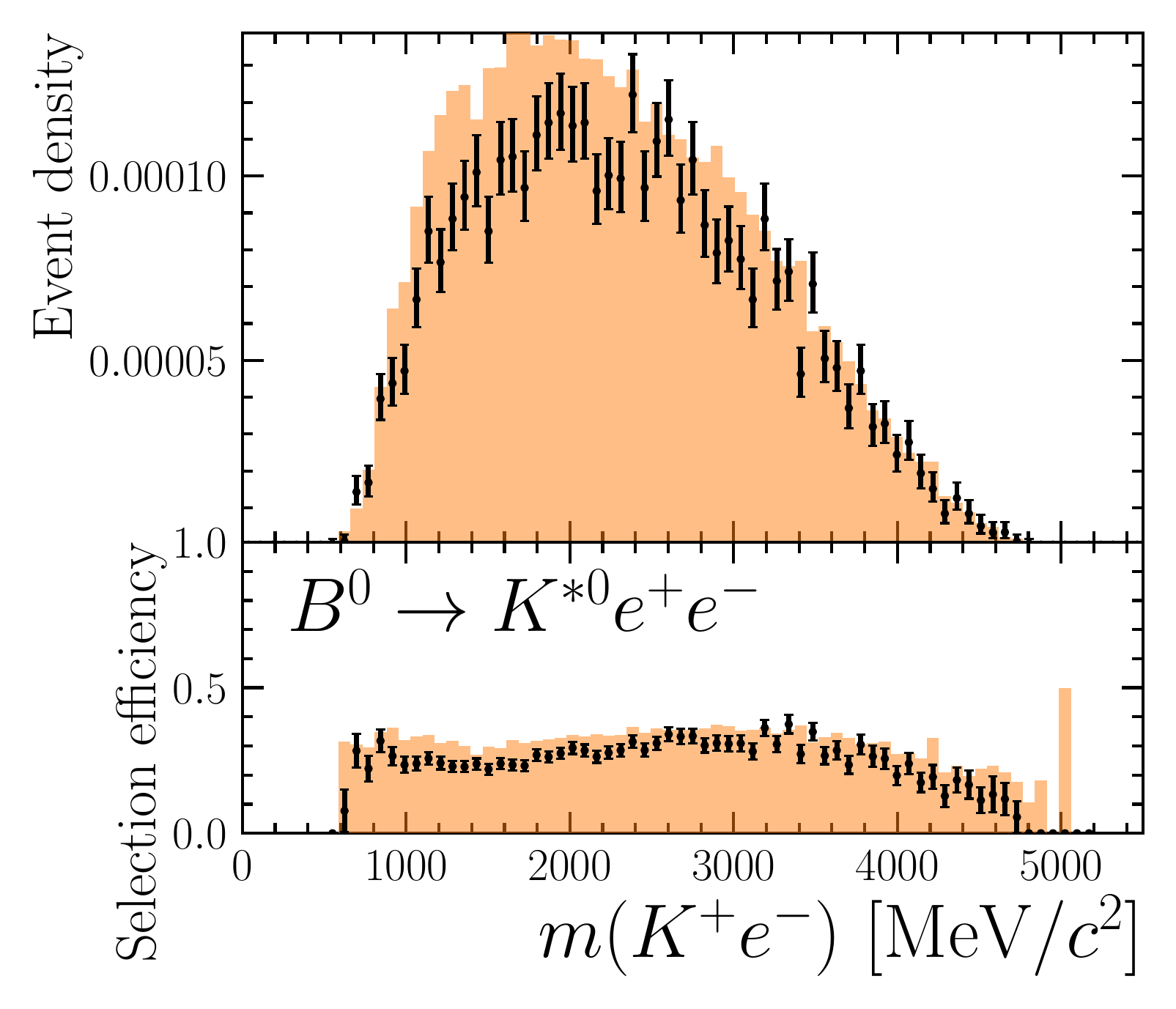}
        \includegraphics[page=93, height=0.85\textwidth, keepaspectratio]{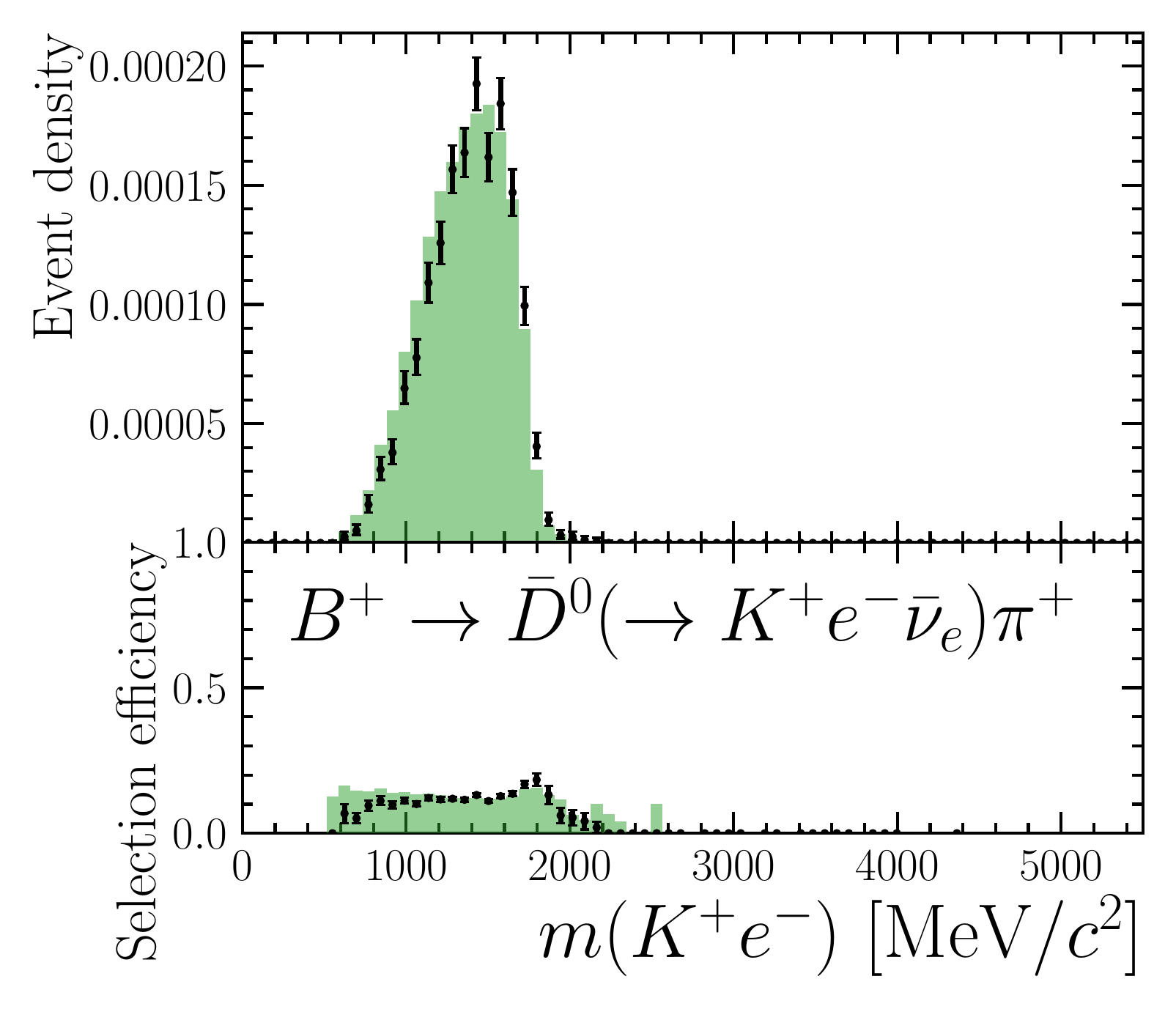}
        \end{minipage}
        \begin{minipage}{0.23\textwidth}
        \includegraphics[page=100, height=0.85\textwidth, keepaspectratio]{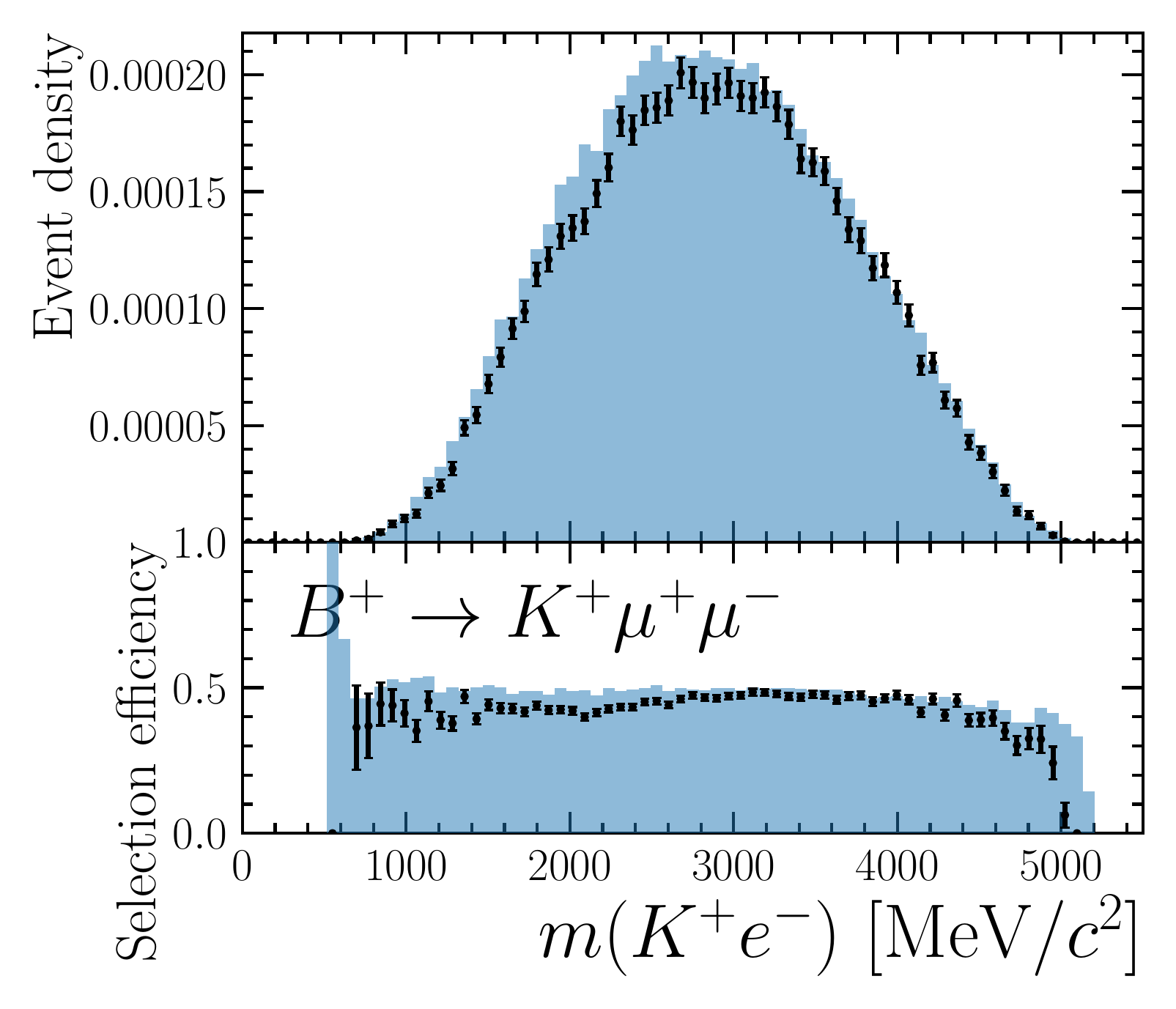}
        \includegraphics[page=107, height=0.85\textwidth, keepaspectratio]{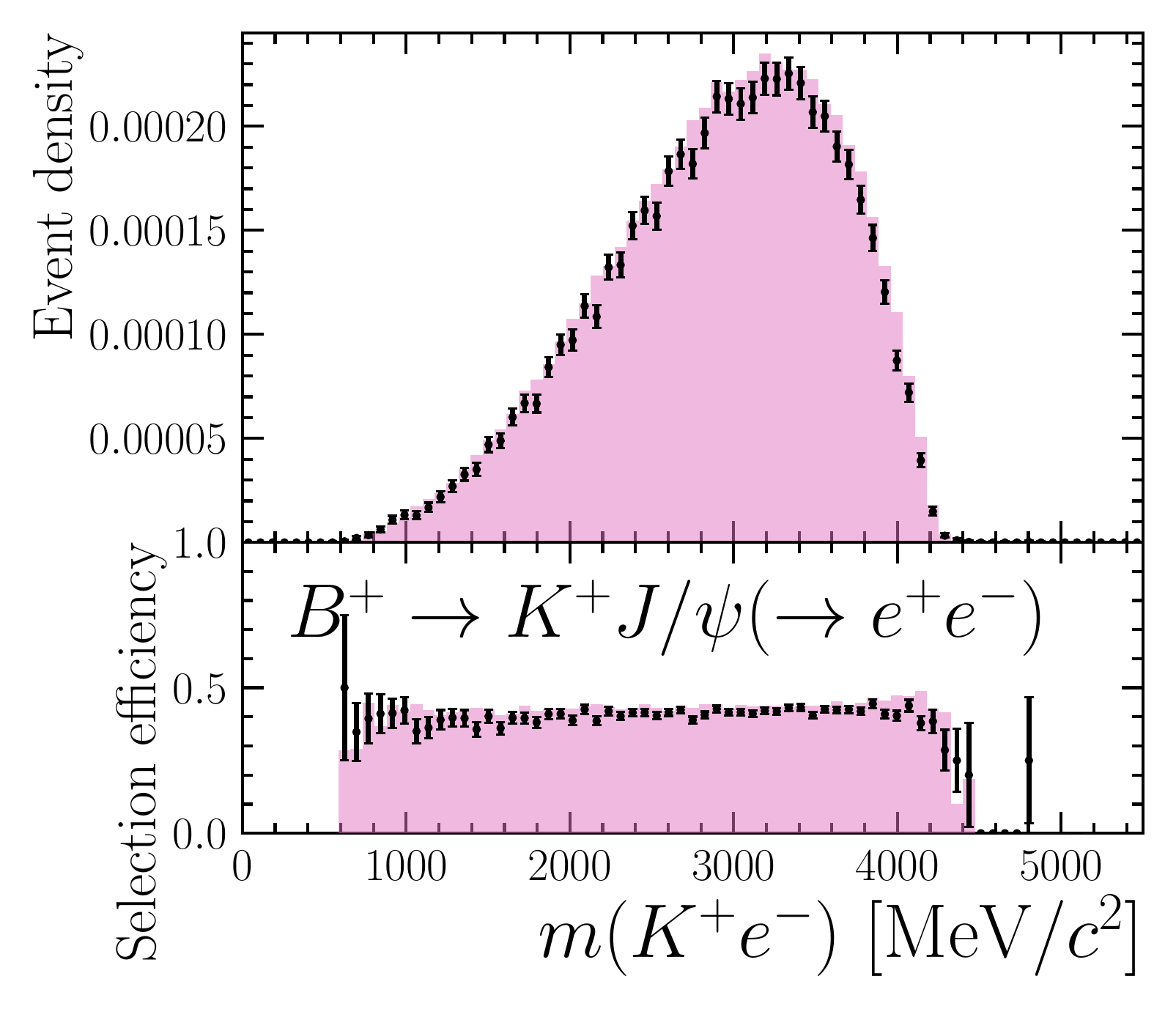}
        \end{minipage}
        
        \begin{minipage}{0.47\textwidth}
        \includegraphics[page=115, width=\textwidth, keepaspectratio]{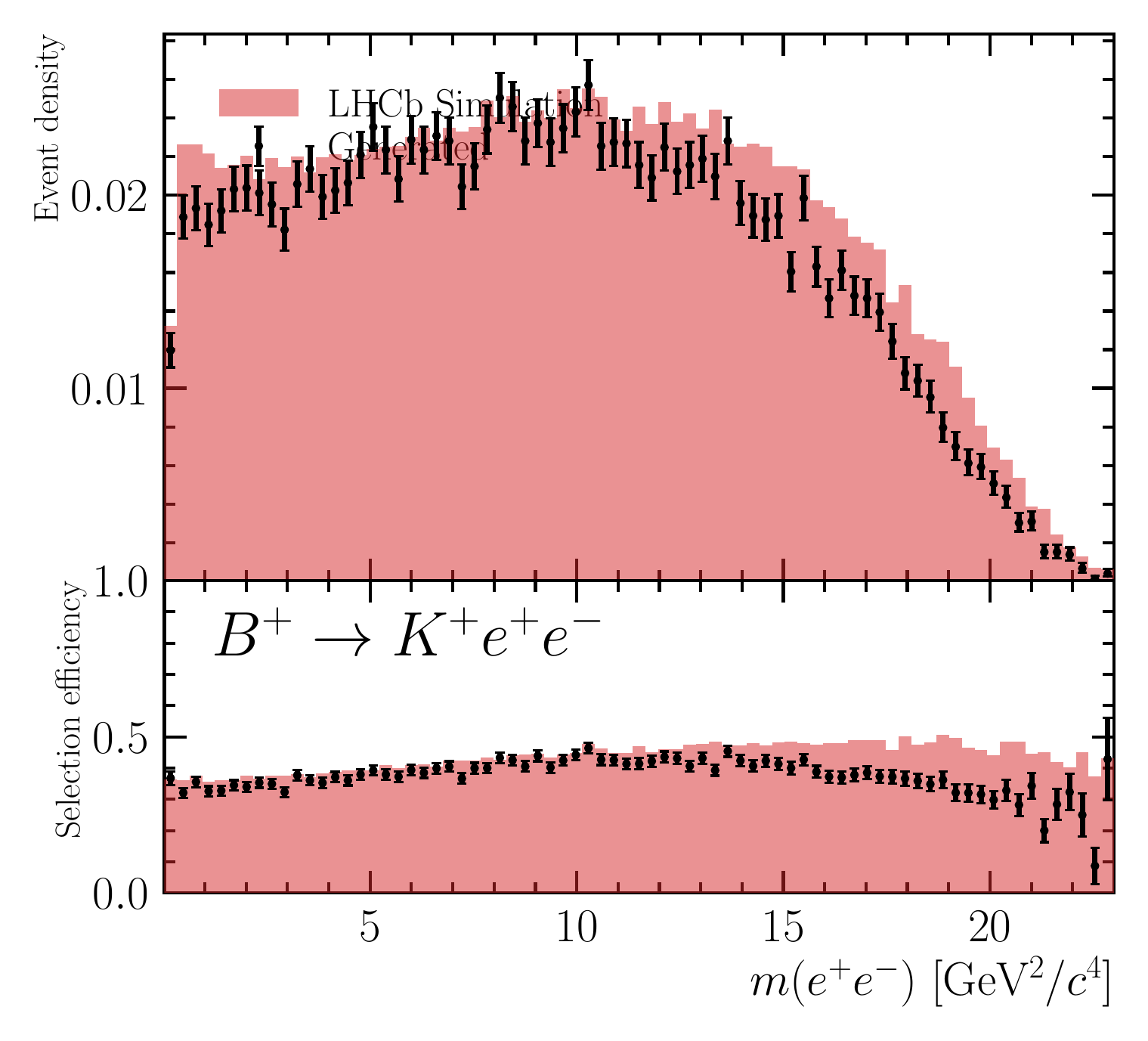}
        \end{minipage}
        \begin{minipage}{0.23\textwidth}
        \includegraphics[page=122, height=0.85\textwidth, keepaspectratio]{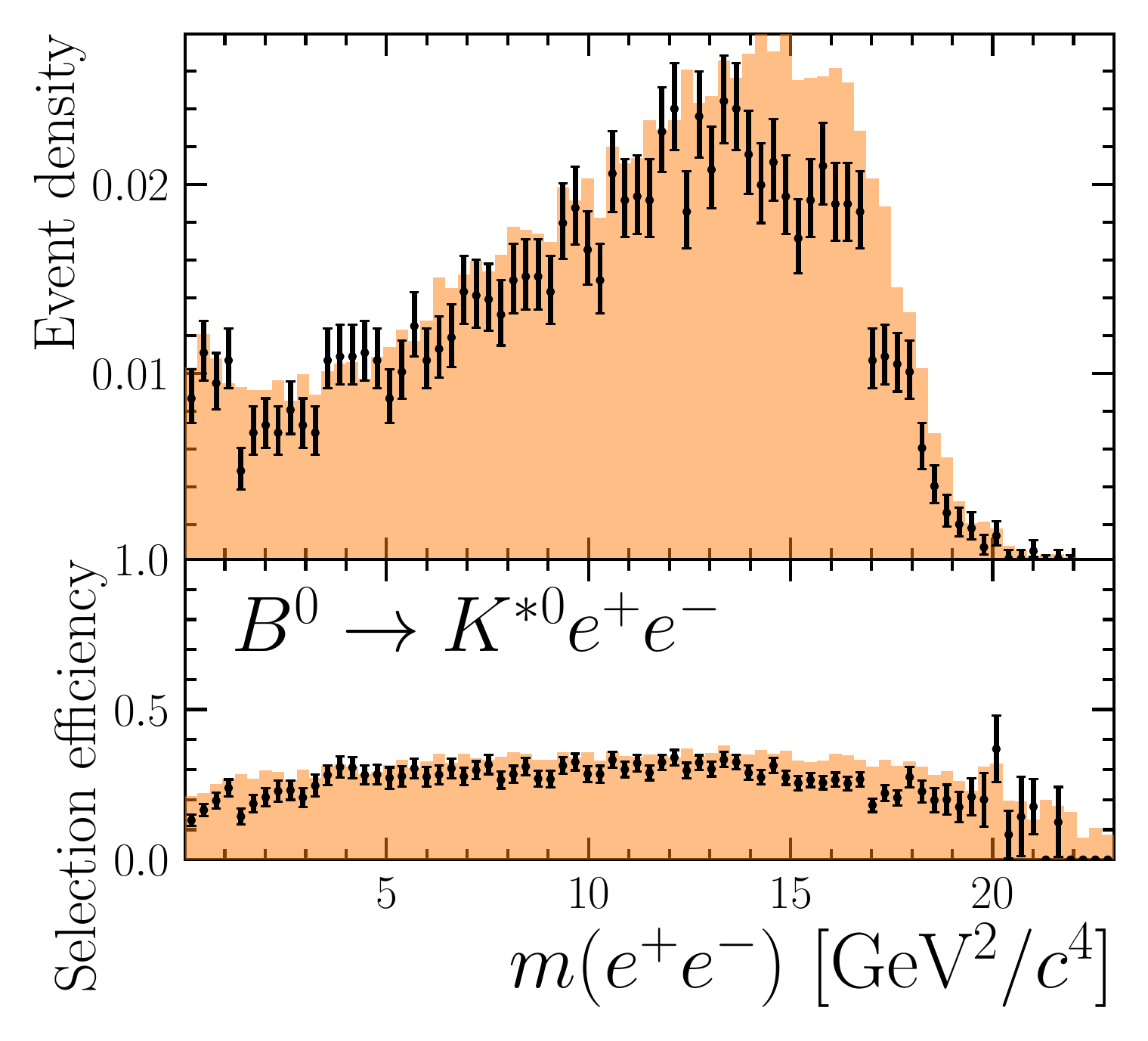}
        \includegraphics[page=129, height=0.85\textwidth, keepaspectratio]{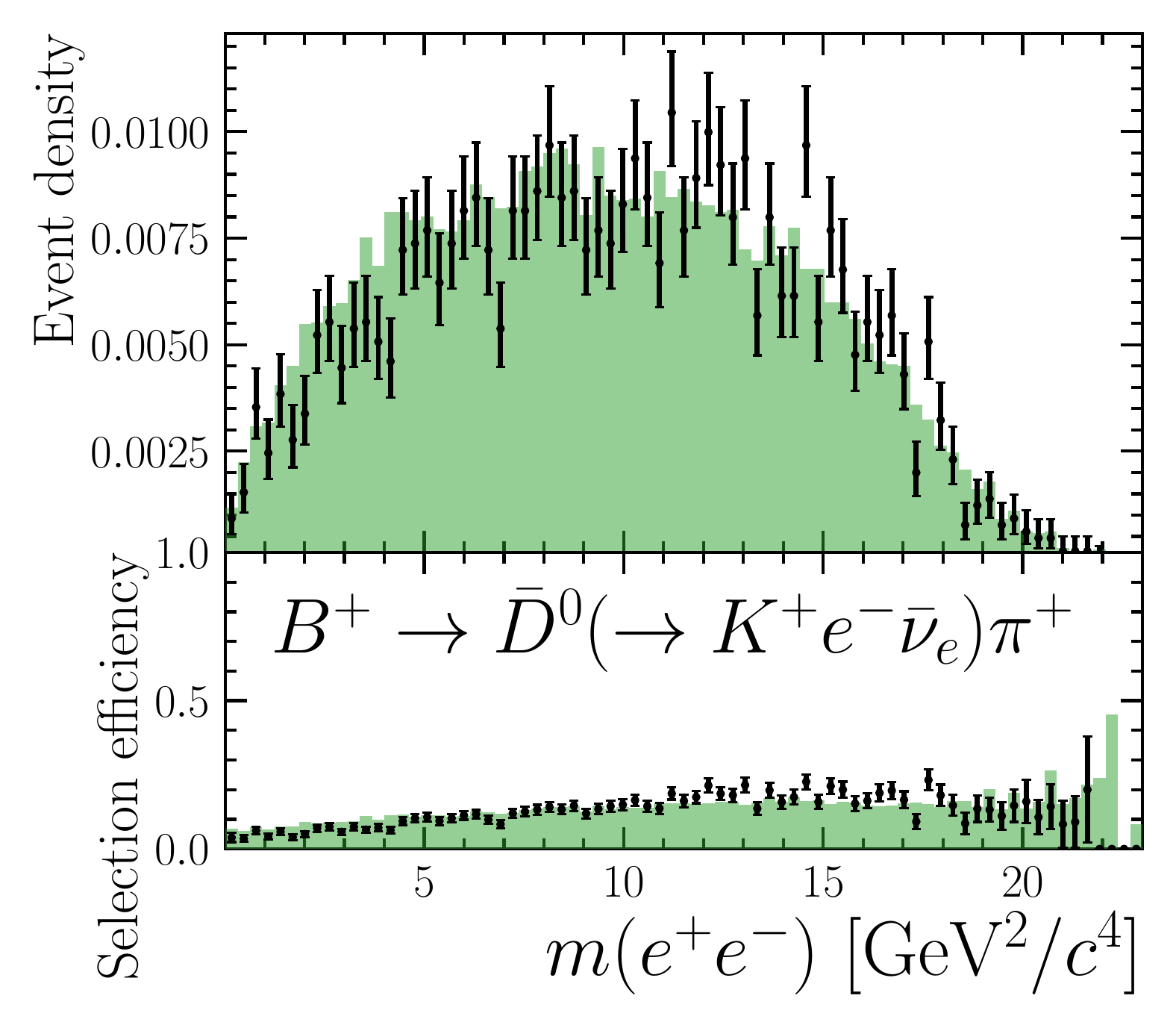}
        \end{minipage}
        \begin{minipage}{0.23\textwidth}
        \includegraphics[page=136, height=0.85\textwidth, keepaspectratio]{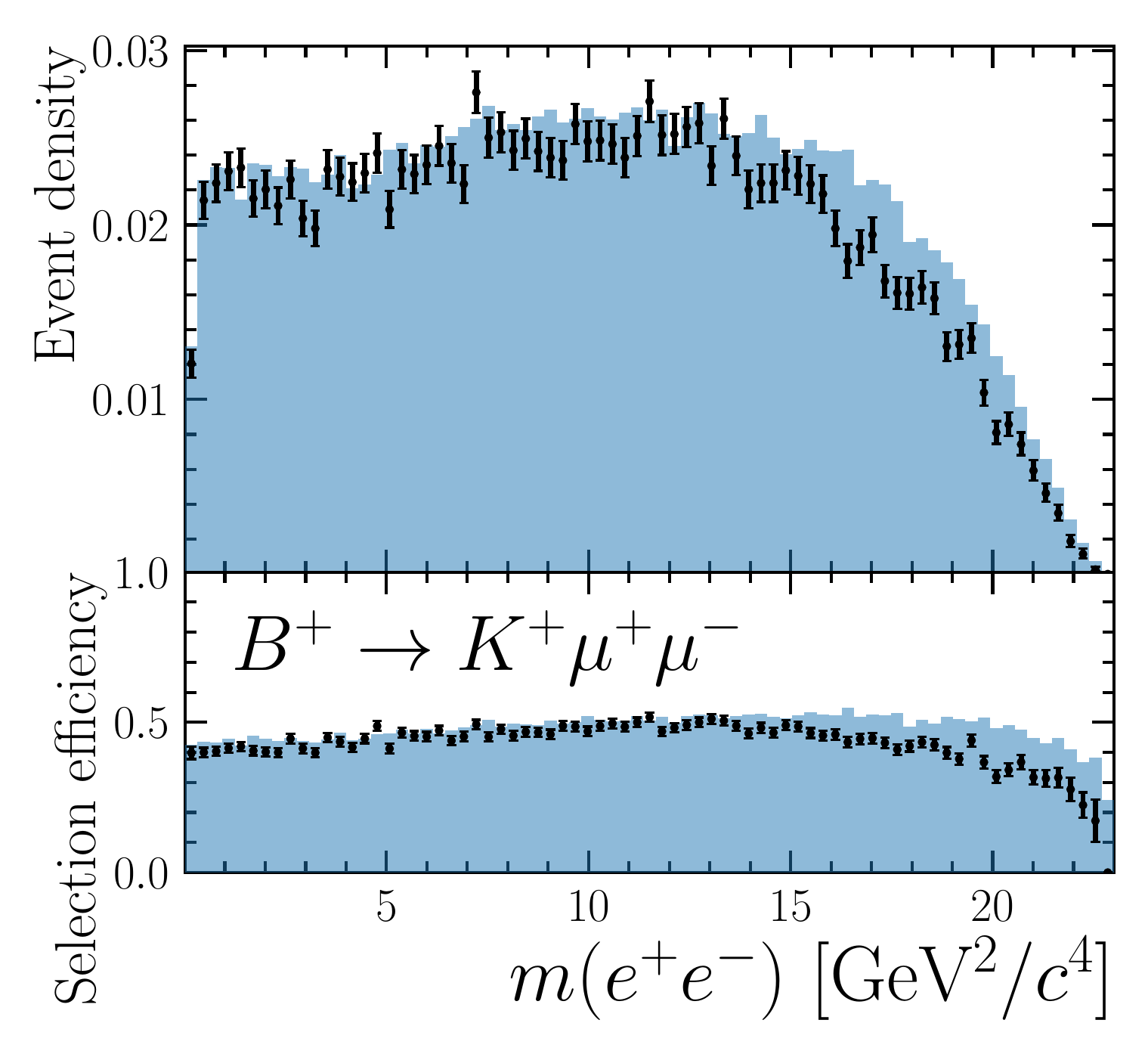}
        \includegraphics[page=143, height=0.85\textwidth, keepaspectratio]{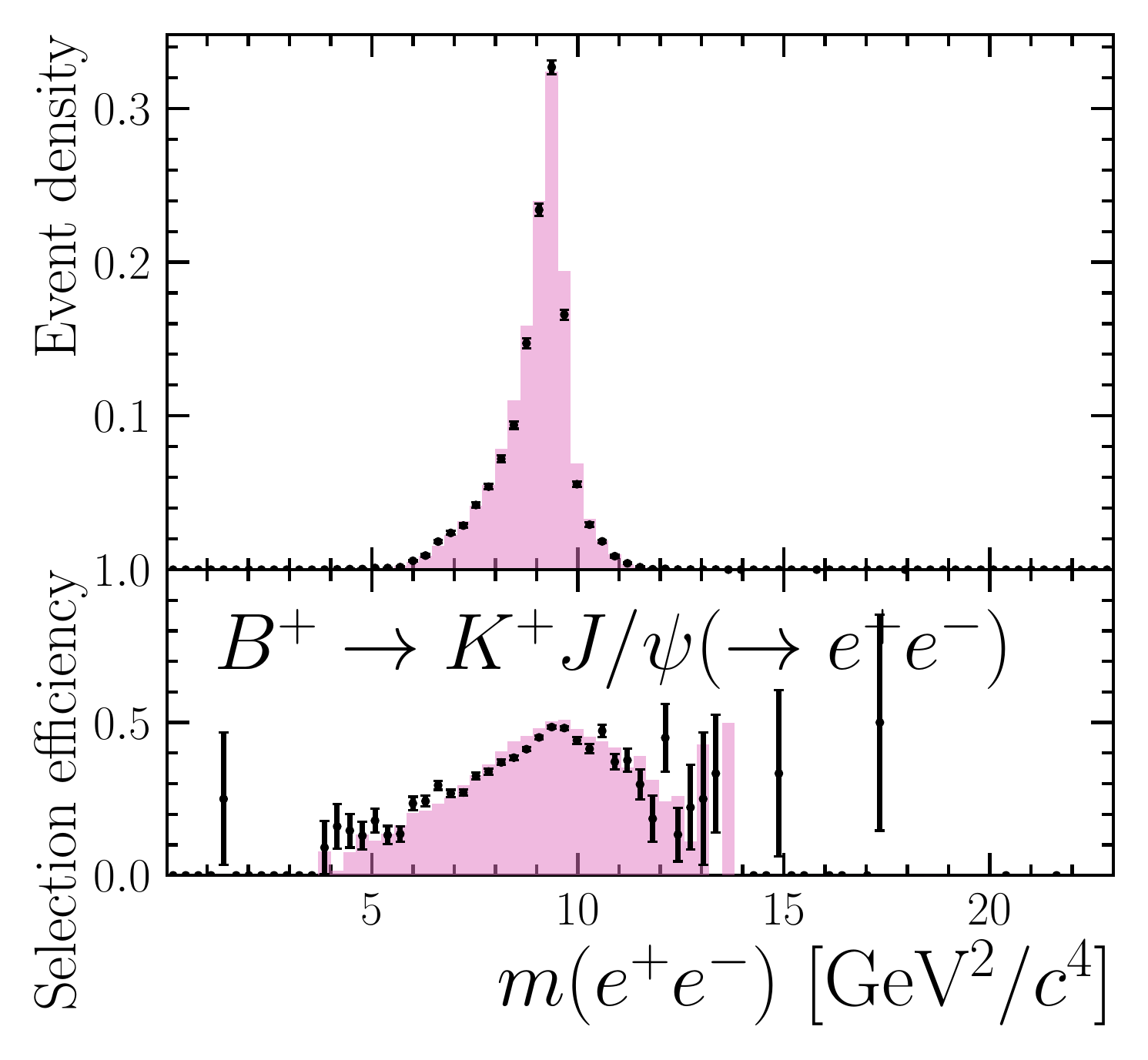}
        \end{minipage}
        
        \begin{minipage}{0.47\textwidth}
        \includegraphics[page=151, width=\textwidth, keepaspectratio]{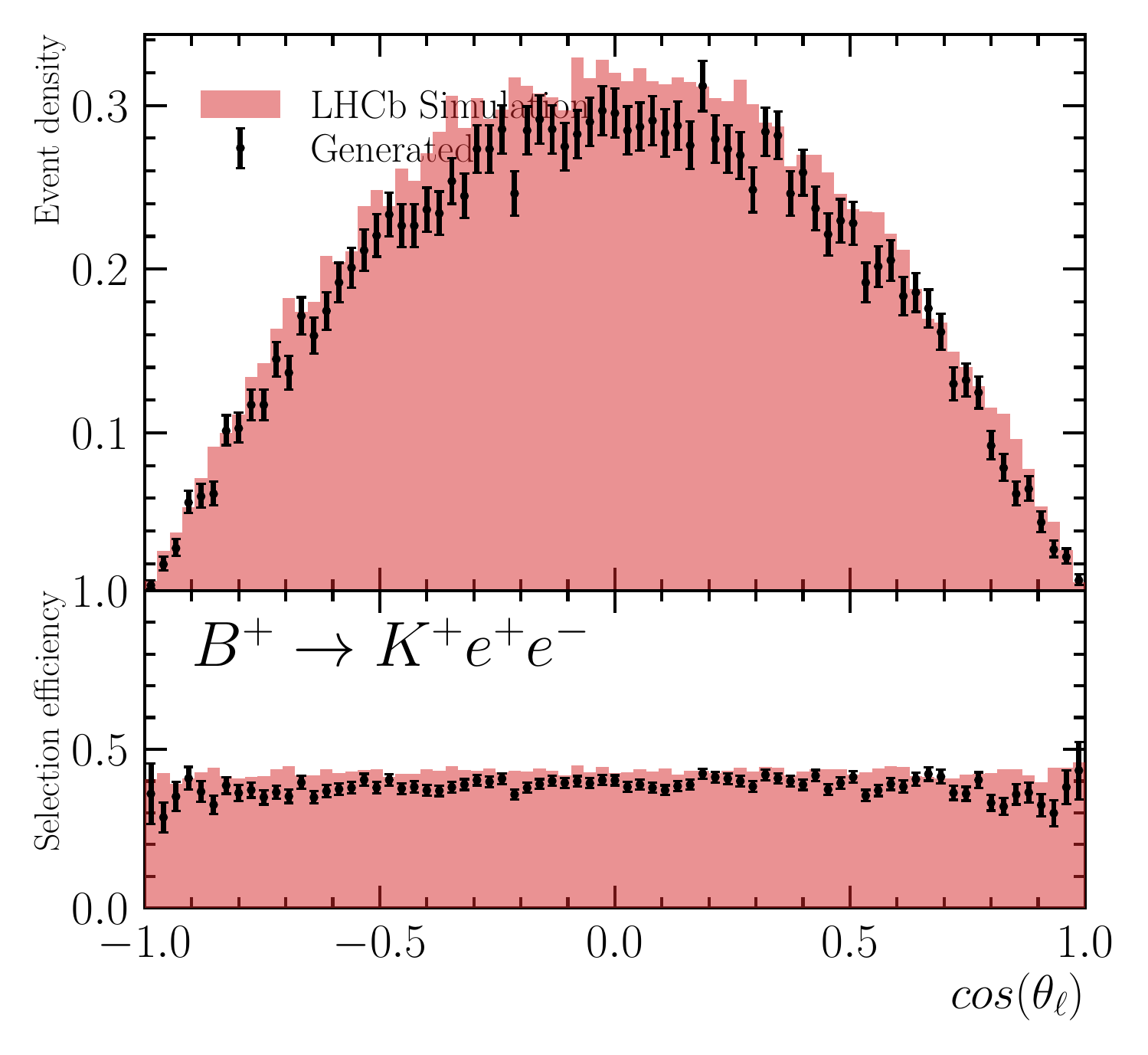}
        \end{minipage}
        \begin{minipage}{0.23\textwidth}
        \includegraphics[page=158, height=0.85\textwidth, keepaspectratio]{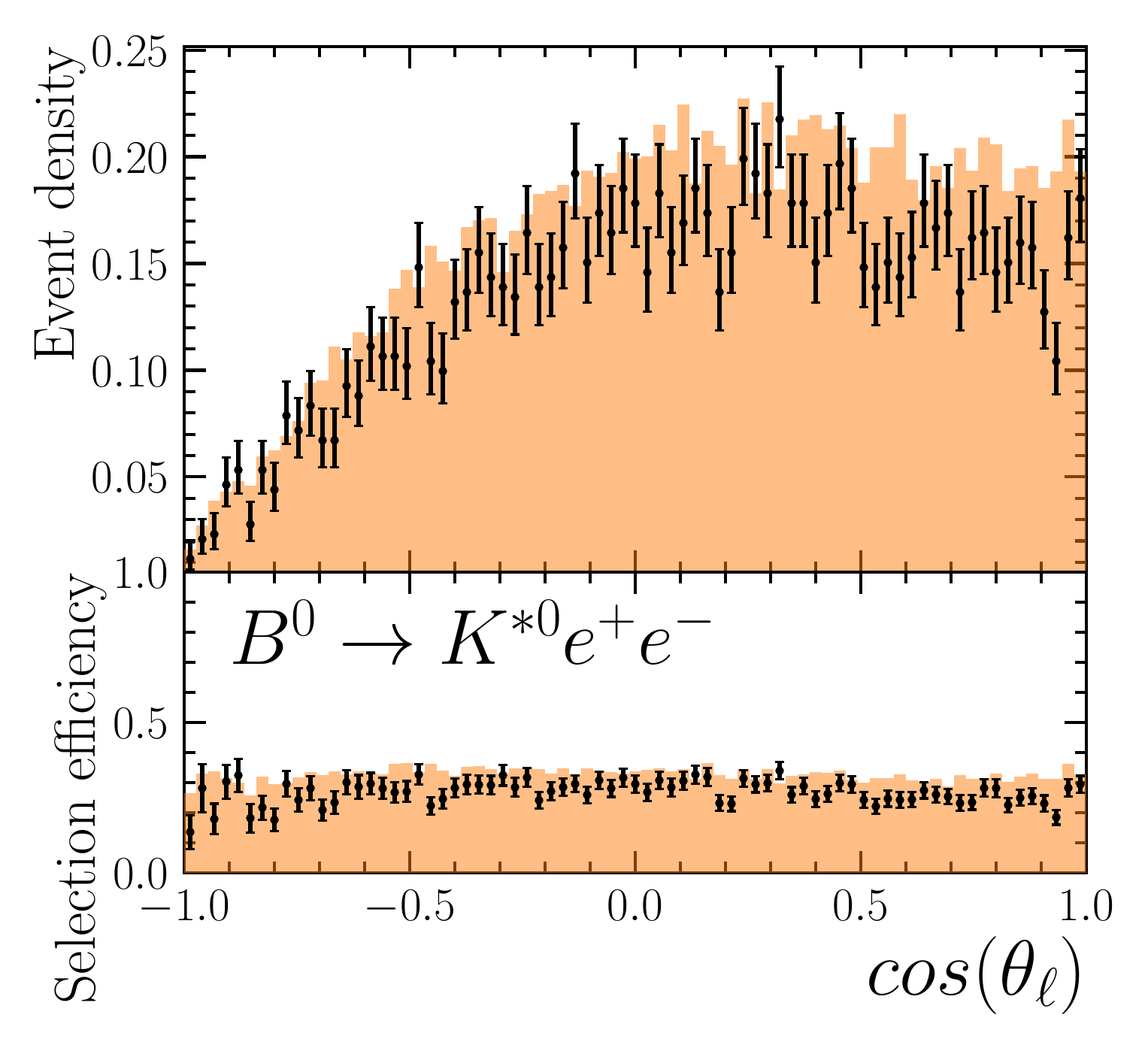}
        \includegraphics[page=165, height=0.85\textwidth, keepaspectratio]{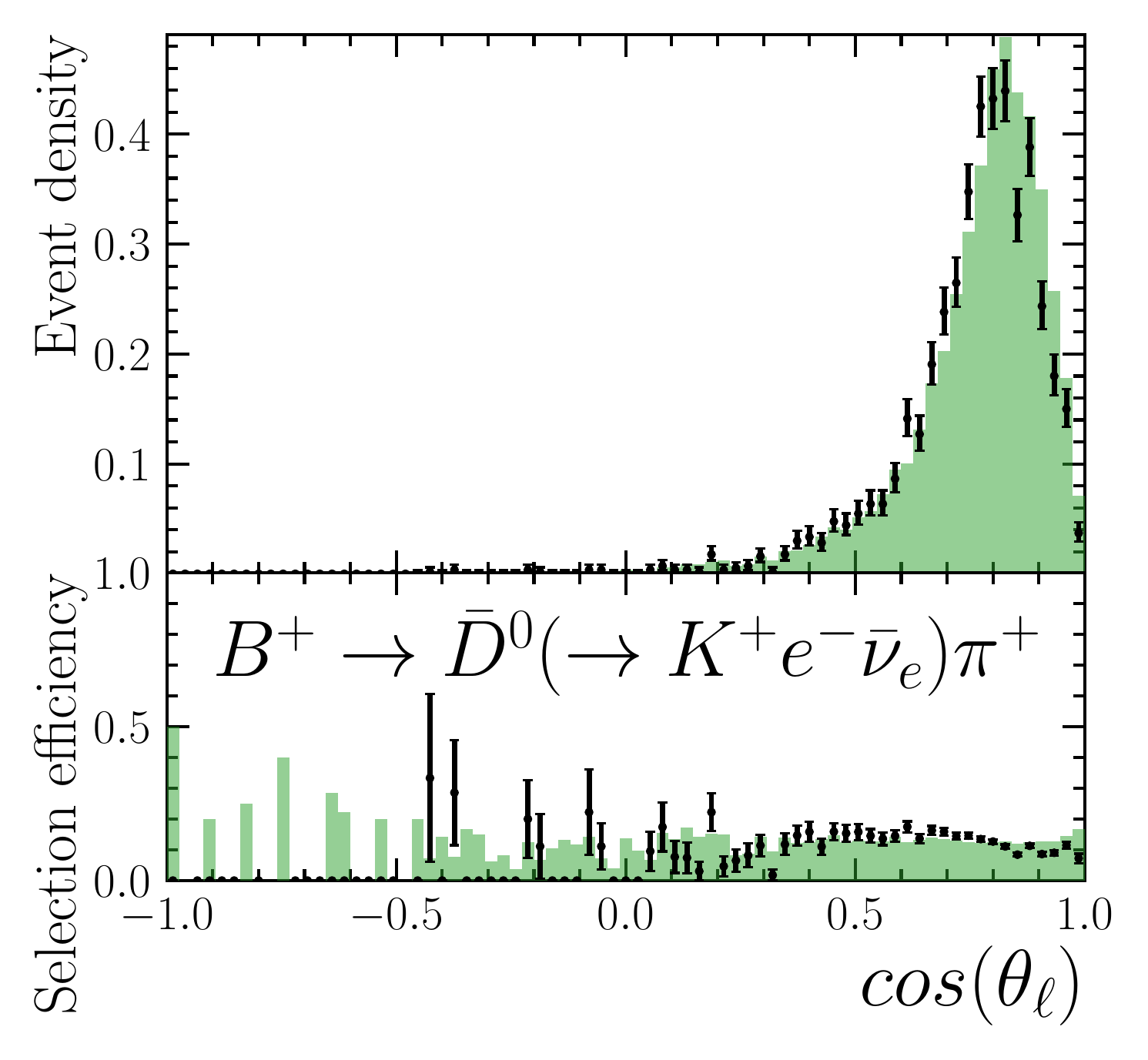}
        \end{minipage}
        \begin{minipage}{0.23\textwidth}
        \includegraphics[page=172, height=0.85\textwidth, keepaspectratio]{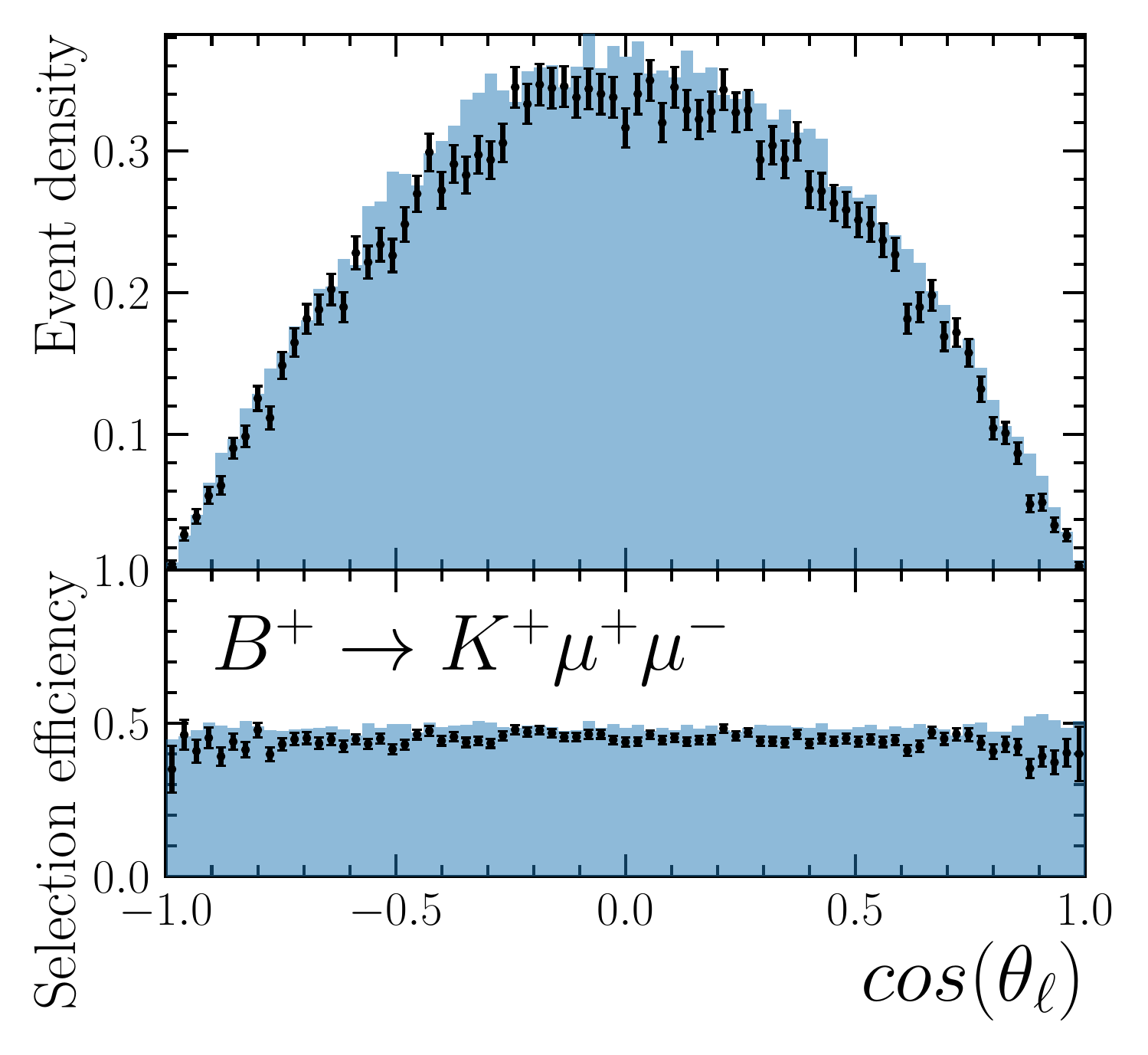}
        \includegraphics[page=179, height=0.85\textwidth, keepaspectratio]{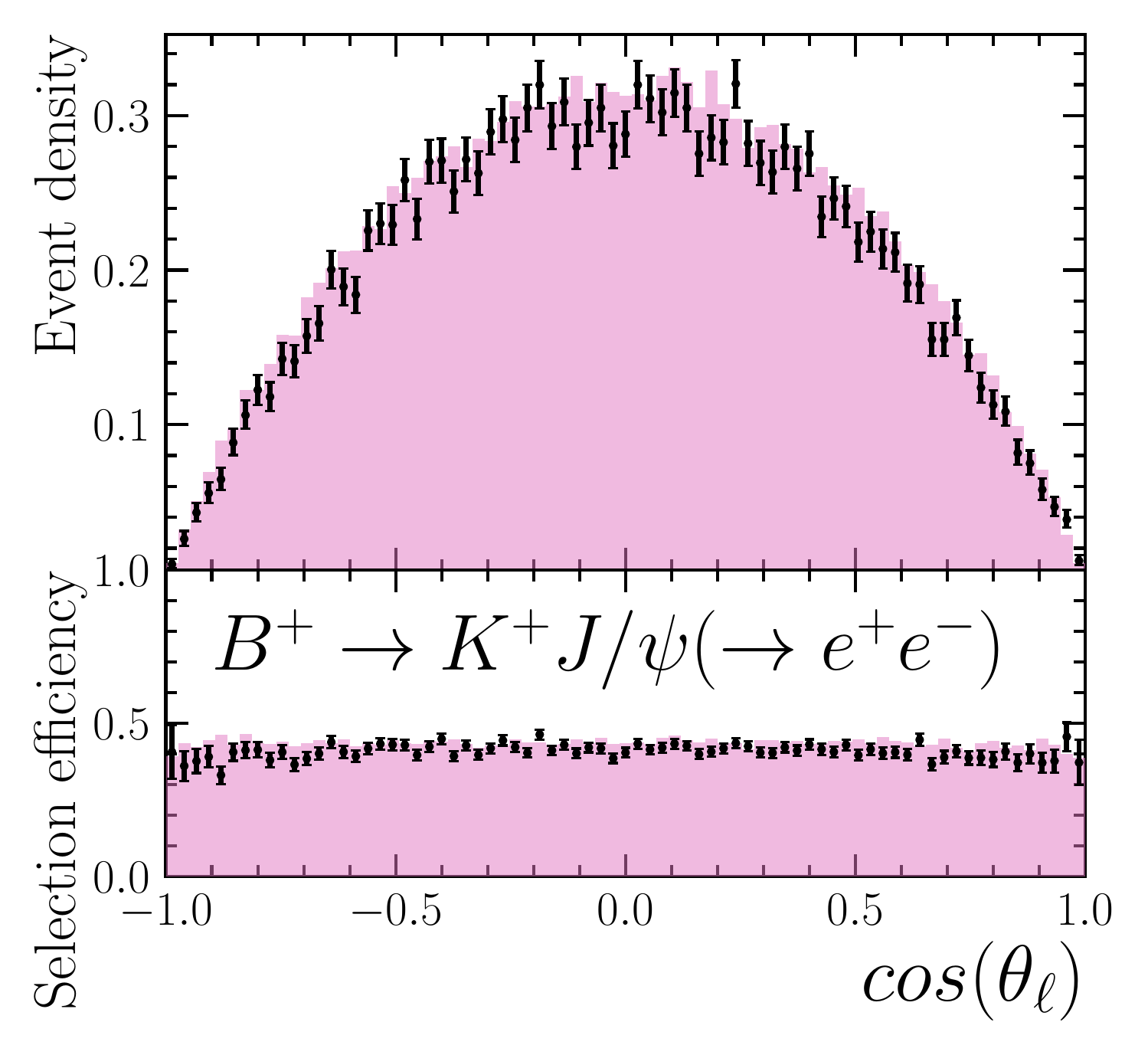}
        \end{minipage}
        \vspace*{-0.5cm}
        \end{center}
        \caption{\small Example response to combinatorial BDT and stripping as a function of various reconstructed kinematic variables.}
        \label{fig:ack:kinematic_variables}
    \end{figure}

\clearpage
\bibliographystyle{JHEP}
\bibliography{main}

\end{document}